\documentclass[usenatbib,useAMS,usedcolumn]{mnras}

\usepackage{epsfig}
\usepackage{times,pdflscape}
\pdfoutput=1

\hypersetup{pdfauthor={E. O'Sullivan},
            pdftitle={The Complete Local Volume Groups Sample - I. Sample Selection and X-ray Properties of the High-Richness Subsample},
            pdfkeywords={galaxies: groups: general, X-rays: galaxies, X-rays: galaxies: clusters, galaxies: clusters: general, galaxies: clusters: intracluster medium, galaxies: active},
            bookmarksnumbered=true}

\newcommand{\arcm}{\hbox{$^\prime$}}

\newcommand{\degree}{\hbox{$^\circ$}}
\newcommand{\rosat}{\emph{ROSAT}}
\newcommand{\chandra}{\emph{Chandra}}
\newcommand{\xmm}{\emph{XMM-Newton}}
\newcommand{\xmms}{\emph{XMM}}
\newcommand{\asca}{\emph{ASCA}}
\newcommand{\einstein}{\emph{Einstein}}
\newcommand{\arcs}{\mbox{\arcm\arcm}}

\newcommand{\Zsol}{\ensuremath{\mathrm{~Z_{\odot}}}}
\newcommand{\Lsol}{\ensuremath{\mathrm{~L_{\odot}}}}
\newcommand{\Msol}{\ensuremath{\mathrm{~M_{\odot}}}}
\newcommand{\Msolpyr}{\ensuremath{\mathrm{~M_{\odot}~yr^{-1}}}}

\newcommand{\LBsol}{\ensuremath{~\mathrm{L_{B\odot}}}}

\newcommand{\s}{\ensuremath{\mbox{~s}}}
\newcommand{\ps}{\ensuremath{\s^{-1}}}
\newcommand{\cm}{\ensuremath{\mbox{~cm}}}
\newcommand{\pcmsq}{\ensuremath{\cm^{-2}}}
\newcommand{\pcmcu}{\ensuremath{\cm^{-3}}}
\newcommand{\kev}{\ensuremath{\mbox{~keV}}}
\newcommand{\kevcmsq}{\ensuremath{\kev\cm^{2}}}
\newcommand{\km}{\ensuremath{\mbox{~km}}}

\newcommand{\erg}{\ensuremath{\mbox{~erg}}}
\newcommand{\ergps}{\ensuremath{\erg \ps}}
\newcommand{\ergpspcmsq}{\ensuremath{\erg \ps \pcmsq}}
\newcommand{\kmps}{\ensuremath{\km \ps}}

\newcommand{\gtsim}{\,\rlap{\raise 0.5ex\hbox{$>$}}{\lower 1.0ex\hbox{$\sim$}}\,} 

\newcommand{\Hi}{H\textsc{i}}

\newcommand{\nh}{\ensuremath{\mathrm{n}_\mathrm{H}}}
\newcommand{\Mdot}{\ensuremath{\dot{\mathrm{M}}}}

\begin{document}

\title[ 
The Complete Local Volume Groups Sample I
] 
{ 
The Complete Local Volume Groups Sample - I. Sample Selection and X-ray Properties of the High--Richness Subsample 
}

\author[E. O'Sullivan et al.] {Ewan O'Sullivan\footnotemark[1]$^{1,2}$, Trevor J. Ponman$^2$, Konstantinos Kolokythas$^{3}$, 
\newauthor Somak Raychaudhury$^{2,3,4}$, Arif Babul$^{5,6}$, Jan M. Vrtilek$^{1}$, Laurence P. David$^{1}$ 
\newauthor Simona Giacintucci$^7$, Myriam Gitti$^{8,9}$, Christopher P. Haines$^{10}$ \\
  $^1$ Harvard-Smithsonian Center for Astrophysics, 60 Garden Street, Cambridge, MA 02138 \\
  $^2$ School of Physics and Astronomy, University of Birmingham, Birmingham, B15 2TT, UK \\
  $^3$ Inter-University Centre for Astronomy and Astrophysics, Pune 411007, India\\ 
  $^4$ Department of Physics, Presidency University, 86/1 College Street, 700073 Kolkata, India \\
  $^5$ Department of Physics and Astronomy, University of Victoria, Victoria, BC V8P 1A1, Canada \\
  $^6$ Center for Theoretical Astrophysics and Cosmology, Institute for Computational Science, University of Zurich, Winterthurerstrasse 190, 8057 Zurich, Switzerland\\
  $^7$ Naval Research Laboratory, 4555 Overlook Avenue SW, Code 7213, Washington, DC 20375, USA \\
  $^8$ Dipartimento di Fisica e Astronomia - Universit\'{a} di Bologna, via Gobetti 93/2, 40129 Bologna, Italy \\
  $^9$ INAF - Instituto di Radioastronomia di Bologna, via Gobetti 101, 40129 Bologna, Italy \\
  $^{10}$ INAF - Osservatorio Astronomico di Brera, via Brera 28, 20122 Milano, via E. Bianchi 46, 23807 Merate, Italy\\
}

\date{Accepted 2017 August 9; Received 2017 August 8; in original form 2017 June 28}

\pagerange{\pageref{firstpage}--\pageref{lastpage}} \pubyear{2015}

\maketitle

\label{firstpage}

\begin{abstract} 
  We present the Complete Local-Volume Groups Sample (CLoGS), a statistically complete optically-selected sample of 53 groups within 80~Mpc. Our goal is to combine X-ray, radio and optical data to investigate the relationship between member galaxies, their active nuclei, and the hot intra-group medium (IGM). We describe sample selection, define a 26-group high-richness subsample of groups containing at least 4 optically bright (log L$_B\ge$10.2\LBsol) galaxies, and report the results of \xmm\ and \chandra\ observations of these systems. We find that 14 of the 26 groups are X-ray bright, possessing a group-scale IGM extending at least 65~kpc and with luminosity $>$10$^{41}$\ergps, while a further 3 groups host smaller galaxy-scale gas haloes. The X-ray bright groups have masses in the range M$_{500}$$\simeq$0.5-5$\times$10$^{13}$\Msol, based on system temperatures of 0.4-1.4~keV, and X-ray luminosities in the range 2-200$\times$10$^{41}$\ergps. We find that $\sim$53-65\% of the X-ray bright groups have cool cores, a somewhat lower fraction than found by previous archival surveys. Approximately 30\% of the X-ray bright groups show evidence of recent dynamical interactions (mergers or sloshing), and $\sim$35\%  of their dominant early-type galaxies host AGN with radio jets. We find no groups with unusually high central entropies, as predicted by some simulations, and confirm that CLoGS is in principle capable of detecting such systems. We identify three previously unrecognized groups, and find that they are either faint (L$_{X,R500}$$<$10$^{42}$\ergps) with no concentrated cool core, or highly disturbed. This leads us to suggest that $\sim$20\% of X-ray bright groups in the local universe may still be unidentified. 
\end{abstract}


\footnotetext[1]{E-mail: eosullivan@cfa.harvard.edu}

\section{Introduction}
\label{sec:intro}

Galaxy groups, systems consisting of a few to a few tens of galaxies bound in a common gravitational potential, are key to our understanding of galaxy evolution and the build-up of large-scale structure. The majority of the matter in the universe \citep{Fukugitaetal98}, including more than half of all galaxies \citep{Ekeetal06}, is thought to reside in group-scale systems. The low velocity dispersions of groups are conducive to galaxy mergers and tidal interactions, driving galaxy evolution \citep{McIntoshetal08,Alonsoetal12}. Many groups are known to possess extensive haloes of ionized, X-ray emitting plasma with temperatures $\sim$1~keV, within which the galaxy population is embedded. These haloes provide proof that groups are indeed gravitationally bound systems dominated by dark matter.

However, the physical properties of groups, and particularly the lowest mass groups, are not well understood, owing to the difficulty of identifying and studying these systems. Optical selection is hampered by the fact that groups typically only contain a handful of bright galaxies. This leads to a significant rate of false detections in optically-selected group samples, caused by chance superpositions along the line of sight, and the difficulty of discriminating fully virialized groups from those which are still in the process of formation. Statistical tests can be devised to improve the effectiveness of selection \citep[see e.g.,][and references therein]{Pearsonetal16}, but these depend on having a statistically meaningful number of galaxies in the system, and therefore push selection toward higher-mass systems and/or volumes covered by spectroscopic surveys capable of probing the dwarf regime.

X-ray selection provides a more reliable way of identifying virialized groups, and has been extensively used to examine their physical properties, but it also has drawbacks. Most nearby X-ray bright groups were first detected by the \einstein\ or \rosat\ observatories, often from relatively shallow surveys \citep[e.g., the \rosat\ All-Sky Survey (RASS) or \einstein\ Slew Survey][]{Vogesetal99,Vogesetal00,Elvisetal92}. Groups are typically at the lower limit of sensitivity for these surveys, and are most easily detected when they possess a bright, highly concentrated core. \citet{Eckertetal11} demonstrated that this introduces a bias toward detection of relaxed, cool-core systems, with the strength of the bias increasing as mass decreases from poor clusters to groups. It is therefore likely that significant numbers of groups with flat or disturbed morphologies, and/or low luminosities are missed by these surveys, leaving important gaps in our knowledge of the group population.

Until recently, most X-ray studies of group properties have been archival in nature; their samples consist of the known systems with available data \citep[e.g.,][]{Ponmanetal96,Mulchaeyetal96,HelsdonPonman00,Mulchaeyetal03,OsmondPonman04,Finoguenovetal06,Finoguenovetal07,RasmussenPonman07,Gastaldelloetal07,Sunetal09}. While providing much of our basic knowledge of group properties, they are subject to biases and may not be representative. 

The biases affecting large-scale X-ray surveys mean that even where groups are selected at another wavelength \citep[e.g., in the optical,][]{Mahdavietal00} we are unlikely to gain an unbiased viewpoint while those surveys are the primary source of X-ray information. Flux-limited, statistically complete samples of nearby groups are less common and understandably focus on higher-mass, high-luminosity groups \citep[e.g.,][]{Eckmilleretal11,Lovisarietal15} where the basic properties of the groups have already been established, and the maximum return from new observations can be guaranteed. The advent of the \xmm\ and \chandra\ observatories has made possible deep surveys of limited areas, from which less biased samples of groups can be selected \citep[e.g.,][]{Jeltemaetal09,Finoguenovetal09,Leauthaudetal10,Adamietal11,Connellyetal12,Erfanianfaretal13,Finoguenovetal15}. However, many of these groups are at moderate redshift, and the observations often provide only limited information (e.g., a luminosity) and cannot support detailed studies of group properties or the interaction between the galaxy population and the X-ray halo.

These selection problems hamper attempts to study the properties of the group population, and determine whether groups are more strongly affected by radiative cooling and feedback processes (star formation, AGN outbursts) than more massive clusters. For example, the 51-group \chandra\ archival sample of \citet{Dongetal10} contains only 8 non-cool-core (NCC) groups, whereas comparable (but statistically complete) X-ray cluster samples find a roughly even split between cool-core (CC) and non-cool-core \citep{Sandersonetal06}. This difference could indicate important mass-dependent physical differences, or could be a product of the biases affecting the RASS, on which the Dong et al. sample is largely based. 

One solution to this problem is to use optical selection to identify groups that can then be observed with sufficient depth in the X-ray to confirm whether a hot halo is present. Starting from an optically-selected sample avoids the X-ray selection bias toward systems with highly concentrated cool-core haloes, while the X-ray follow-up provides information on the gas content and properties, and confirms that the groups are fully collapsed systems. This approach has been used with some success to identify unbiased samples of groups \citep{Rasmussenetal06b,Miniatietal16,Pearsonetal16} and clusters \citep{Baloghetal11}, but these samples have generally targeted more distant systems, either because their goal was to trace the gas halo to large radii, or because they were based on optical surveys whose limited footprint precludes identification of nearby groups with large angular extents. Such samples are well-suited to cosmological studies. However, they are less useful if examination of the detailed interaction between galaxies, AGN, and the intra-group medium (IGM) is the priority. For these purposes, a sample of groups in the local universe has significant advantages. Nearby systems provide the best opportunity to resolve small-scale structure in the ISM, and the greatest sensitivity to faint emission, allowing us to trace haloes down to the scale of individual galaxies, and reducing our vulnerability to X-ray selection effects.

In this paper we describe the Complete Local-Volume Groups Sample (CLoGS), consisting of 53 optically-selected groups in the nearby universe (D$<$80~Mpc). This statistically complete sample is designed to allow the study of the properties and structure of the IGM, using a combination of X-ray and radio observations to examine the role of feedback in balancing radiative cooling. In section~\ref{sec:sample} we describe the optical sample selection and our definition of a high-richness subsample of 26 groups whose X-ray properties we discuss in the remainder of the paper. Section~\ref{sec:obs} describes the new and archival \xmm\ and \chandra\ data available for this subsample, and our data reduction and analysis. The results of the X-ray analysis are presented in section~\ref{sec:res} including, for those systems in which an extended hot halo is detected, estimates of luminosity, system temperature and abundance, and temperature, entropy and cooling time profiles. We discuss our results in section~\ref{sec:disc} and consider the implications for our current knowledge of the local population of groups, and for future surveys. We present our conclusions in section~\ref{sec:conc}. The radio properties of the dominant early-type galaxy in each group are presented in a companion paper (Kolokythas et al., in prep.).

\section{Sample Selection}
\label{sec:sample}

The selection process for CLoGS was driven by a number of factors. The groups were required to be located in the nearby universe, so that relatively short \chandra\ and \xmms\ observations would be capable of detecting a low surface brightness IGM and characterizing its temperature structure. The sample was also limited to Northern hemisphere and equatorial systems visible from the Giant Metrewave Radio Telescope (GMRT) and Very Large Array (VLA), since we intended to study the radio properties of AGN within the groups using new observations from the former, as well as the NVSS \citep{Condonetal93NVSS} and FIRST \citep{Beckeretal95} 1.4~GHz radio surveys performed by the latter.

Examination of group samples drawn from the SDSS and 2dFGRS suggested that these were unsuitable for our purposes. The parent surveys cover significant areas of the sky but only limited volumes of the local universe, and tend to have relatively poor coverage of nearby galaxies with large angular scales. We instead chose to start from the relatively shallow all-sky Lyon Galaxy Group (LGG) catalogue \citep{Garcia93}. The LGG sample contains 485 groups, and is based on an early version of the Lyon Extragalactic Data Archive\footnote{\url{http://leda.univ-lyon1.fr}} \citep[LEDA,][]{Makarovetal14} galaxy catalogue, which at that time contained 23490 galaxies, complete to $m_B$=14 and $v_{rec}$=5500\kmps, equivalent to a distance limit of D$<$80~Mpc after correction for Virgocentric flow. Groups were identified through friends--of--friends and hierarchical clustering algorithms, the final sample consisting of systems identified by both methods.

We selected groups from the LGG catalogue which met the following criteria:
\begin{enumerate}
\item $\geq$~4 member galaxies,
\item $\geq$~1 early-type member \citep[revised morphological type T$<$0;][]{DeVaucouleursetal76},
\item optical luminosity L$_B$$>$3$\times$10$^{10}$\Lsol\ for the brightest early-type member,
\item declination $>$-30\degree.
\end{enumerate}

The motivation for each criterion was to (i) exclude small galaxy associations (pairs and triplets) which may lack a group-scale halo, (ii) and (iii) exclude groups where spirals make up the entire population of massive galaxies, and (iv) ensure visibility from the VLA and GMRT. 

Removing spiral--dominated groups excludes a significant fraction of the group population. However, the presence of a massive elliptical galaxy is a good indicator that an apparent group is a genuinely collapsed system, in which mergers and tidal interactions can drive evolution of the galaxy population. This is supported by the fact that groups containing ellipticals are more X-ray luminous than their spiral--only counterparts \citep{Mulchaeyetal03,Milesetal06}, indicating a superior ability to retain an intra-group medium. While spiral--rich groups with X-ray detected IGMs are known, they are rare, typically extremely X-ray faint \citep[e.g.,][]{Trinchierietal08,OSullivanetal14c}, and in some cases appear to be still in the process of forming their X-ray IGM through transient episodes of shock heating \citep{OSullivanetal09} or starburst winds \citep{OSullivanetal14c}. We therefore considered that the inclusion of spiral-dominated systems would likely lead to an unacceptable fraction of non-detections among our sample, and that those groups which were detected might have X-ray properties determined by short-lived ``special'' events (shocks, starbursts), providing a biased view of halo properties.

In each group, we identified the brightest group-member early-type galaxy (BGE). In relaxed systems, we expect the most massive elliptical galaxies to occupy the group core. In X-ray luminous groups the IGM is typically centred on the most massive elliptical(s). We therefore consider the position of the BGE a reasonable initial indicator of the group centre.

As the LGG sample was drawn from a relatively small galaxy catalogue, some of the groups contain only a handful of galaxies. It is desirable to trace the galaxy populations of the groups to lower luminosities, since this provides more statistically reliable information on their physical properties (e.g., velocity dispersion, projected galaxy distribution). We therefore refined and expanded the group membership by comparison with the current version of the LEDA archive, which contains $\sim$10$^5$ galaxies with measured $m_B$ within 10,000\kmps. For each group, we selected galaxies within 1~Mpc and 2000\kmps\ of the BGE, and iteratively determined the mean group velocity and velocity dispersion using the gapper algorithm \citep{Beersetal90}. Galaxies within 3$\sigma$ of the mean velocity were considered group members, and iteration was continued until the group membership stabilized. 

\begin{figure*}
\includegraphics[width=\textwidth]{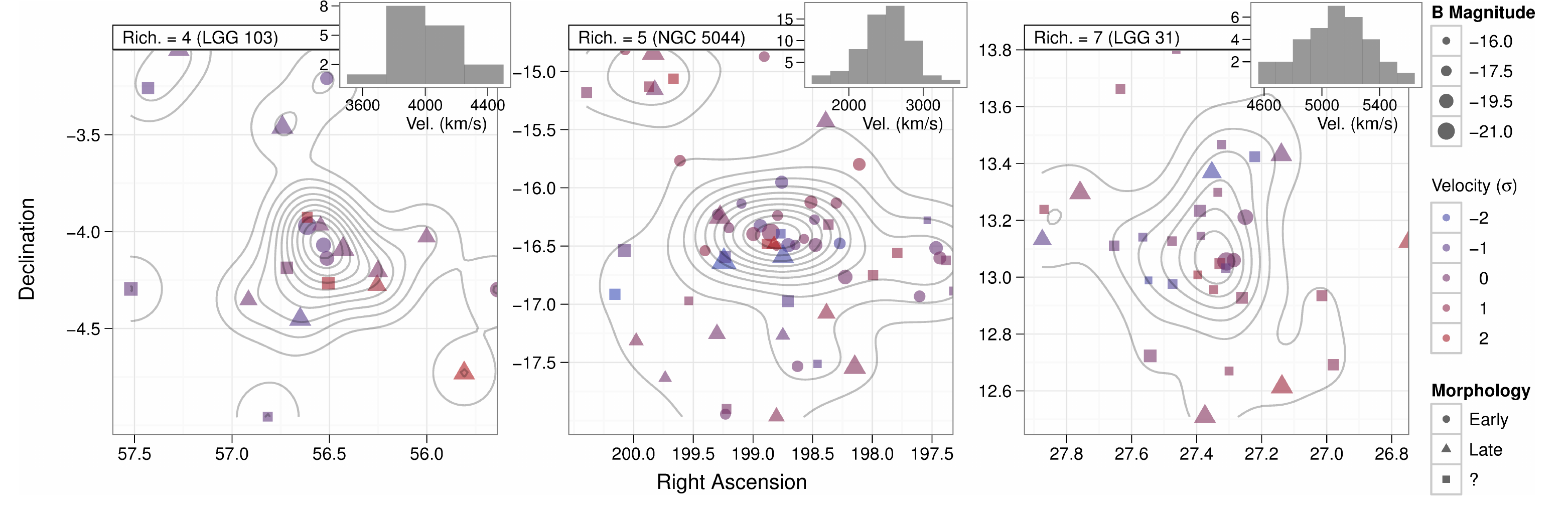}
\vspace{-6mm}
\caption{\label{fig:isopleth}Examples of plots of galaxy position for three CLoGS groups, with galaxy iso-density contours overlaid. Symbol size indicates galaxy $B$-band magnitude, symbol colour indicates galaxy velocity relative to the group mean in units of the velocity dispersion $\sigma$ (blue for lower velocities, red for higher), and symbol shape denotes galaxy morphology, with circles triangles and squares indicating early-type, late-type, and unknown morphologies respectively. Velocity histograms for the group members are shown in the upper right of each panel.}
\end{figure*}

We next examined the spatial distribution of the member galaxies using maps of galaxy iso-density, such as those shown in Figure~\ref{fig:isopleth}. These allowed us to reject systems which lacked a clear galaxy density peak, and sub-clumps of known clusters which had been falsely identified as independent groups. We also identified cases where the BGE was not associated with the main galaxy density peak, and re-evaluated these after selecting a more likely BGE. This process resulted in a set of 67 candidate groups.

We define a richness parameter $R$ as the number of galaxies with log L$_B$$\geq$10.2 (equivalent to the 90\% completeness limit of LEDA at our distance limit) within 1~Mpc and 3$\sigma$ in velocity of the BGE. We excluded all systems with $R$$\geq$10, since the majority of these were found to be subsets of known clusters. No $R$=9 groups were identified, so the maximum richness in our sample is $R$=8. Six groups with $R$=1 were also excluded. While probably bound systems (e.g., HCG~42), they are too poor to allow reliable determination of group properties. This leaves us with a sample of 53 groups. Table~\ref{tab:sample} provides basic information on the sample.

\begin{table*}
\caption{\label{tab:sample}Basic properties of the CLoGS groups, The position and redshift are those of the BGE, as the object most likely to be at the group centre. Alternate group identifications are drawn from the catalogues of \citet[HCG]{Hicksonetal92}, \citet[HG]{HuchraGeller82}, \citet[GH]{GellerHuchra83}, \citet[WBL]{Whiteetal99}, \citet[USGC]{Ramellaetal02}, and the high and low density contrast catalogues of \citet[HDC and LDC]{Crooketal07,Crooketal08}. For LGG~97 and 100, entries marked * indicate cases where neither CLoGS group individually includes 50\% of the comparison system, but the combination of the two includes $>$50\%.}
\begin{center}
\begin{tabular}{lcccccccl}
\hline
LGG & BGE & RA & Dec. & z & D & scale & $R$ & Related group IDs\\
 & & (J2000) & (J2000) & & (Mpc) & (kpc/\arcs) & & \\
\hline
\multicolumn{9}{l}{\textsc{High-Richness subsample}}\\
9   & NGC~193    & 00 39 18.6 & +03 19 52 & 0.0147 & 74 & 0.359 & 7 & HDC~25, GH~6\\
18  & NGC~410    & 01 10 58.9 & +33 09 07 & 0.0177 & 77 & 0.373 & 6 & WBL~31, GH~9\\
27  & NGC~584    & 01 31 20.7 & -06 52 05 & 0.0060 & 25 & 0.121 & 4 & HDC~81, LDC~95, HG~45\\
31  & NGC~677    & 01 49 14.0 & +13 03 19 & 0.0170 & 78 & 0.378 & 7 & USGC~U77, HDC~92, GH~19\\
42  & NGC~777    & 02 00 14.9 & +31 25 46 & 0.0167 & 73 & 0.354 & 5 & USGC~U92, HDC~109\\
58  & NGC~940    & 02 29 27.5 & +31 38 27 & 0.0171 & 74 & 0.359 & 4 & USGC~U127, HDC~143\\
61  & NGC~924    & 02 26 46.8 & +20 29 51 & 0.0149 & 64 & 0.310 & 4 & USGC~U123, HDC~142, LDC~168, GH~29\\
66  & NGC~978    & 02 34 47.6 & +32 50 37 & 0.0158 & 69 & 0.334 & 7 & WBL~77\\
72  & NGC~1060   & 02 43 15.0 & +32 25 30 & 0.0173 & 76 & 0.368 & 8 & WBL~85, USGC~U145, HDC~165\\
80  & NGC~1167   & 03 01 42.4 & +35 12 21 & 0.0165 & 72 & 0.349 & 4 & HDC~202\\
103 & NGC~1453   & 03 46 27.2 & -03 58 08 & 0.0130 & 63 & 0.305 & 4 & USGC~S134, HDC~245, HG~47\\
117 & NGC~1587   & 04 30 39.9 & +00 39 42 & 0.0123 & 51 & 0.247 & 4 & HDC~292, LDC~311, GH~38\\
158 & NGC~2563   & 08 20 35.7 & +21 04 04 & 0.0149 & 65 & 0.315 & 6 & WBL~178, USGC~U173, HDC~480\\
185 & NGC~3078   & 09 58 24.6 & -26 55 36 & 0.0086 & 34 & 0.165 & 6 & HDC~554, HG~29\\
262 & NGC~4008   & 11 58 17.0 & +28 11 33 & 0.0121 & 54 & 0.262 & 4 & USGC~U435, HDC~686, LDC~855, GH~95\\
276 & NGC~4169   & 12 12 18.8 & +29 10 46 & 0.0126 & 45 & 0.218 & 4 & HCG~61, WBL~385, USGC~U469, HDC~699, LDC~875, GH~101\\
278 & NGC~4261   & 12 19 23.2 & +05 49 31 & 0.0075 & 32 & 0.155 & 7 & WBL~392, HG~41, GH~106\\
310 & ESO~507-25 & 12 51 31.8 & -26 27 07 & 0.0108 & 45 & 0.218 & 4 & USGC~S187, HDC~734\\
338 & NGC~5044   & 13 15 24.0 & -16 23 08 & 0.0093 & 38 & 0.184 & 5 & HDC~775\\
345 & NGC~5084   & 13 20 16.9 & -21 49 39 & 0.0057 & 23 & 0.112 & 4 & USGC~S210, HDC~784, HG~35\\
351 & NGC~5153   & 13 27 54.3 & -29 37 05 & 0.0144 & 60 & 0.291 & 7 & HDC~788\\
363 & NGC~5353   & 13 53 26.7 & +40 16 59 & 0.0078 & 35 & 0.170 & 7 & HCG~68, WBL~475, USGC~U578, HDC~827, LDC~1006, HG~69, GH~123\\
393 & NGC~5846   & 15 06 29.3 & +01 36 20 & 0.0057 & 26 & 0.126 & 5 & USGC~U677, HDC~897, HG~50, GH~150\\
402 & NGC~5982   & 15 38 39.8 & +59 21 21 & 0.0101 & 44 & 0.213 & 4 & LDC~1141, GH~158 \\
421 & NGC~6658   & 18 33 55.6 & +22 53 18 & 0.0142 & 63 & 0.305 & 4 & HDC~1043\\
473 & NGC~7619   & 23 20 14.5 & +08 12 22 & 0.0125 & 54 & 0.262 & 8 & WBL~710, USGC~U842, HDC~1240, LDC~1573, GH~166\\
\hline
\multicolumn{9}{l}{\textsc{Low-Richness subsample}}\\
6   & NGC~128    & 00 29 15.0 & +02 51 51 & 0.0141 & 60 & 0.291 & 3 & USGC~U17\\
12  & NGC~252    & 00 48 01.5 & +27 27 25 & 0.0165 & 72 & 0.349 & 3 & USGC~U32, HDC~38\\
14  & NGC~315    & 00 57 48.9 & +30 21 09 & 0.0165 & 73 & 0.354 & 2 & WBL~22, USGC~U39, GH~8\\
23  & NGC~524    & 01 24 47.7 & +09 32 20 & 0.0080 & 34 & 0.165 & 2 & USGC~U60, HDC~71, LDC~85, GH~13\\
78  & NGC~1106   & 02 50 40.5 & +41 40 17 & 0.0145 & 64 & 0.310 & 3 & - \\
97  & NGC~1395   & 03 38 29.7 & -23 01 39 & 0.0057 & 21 & 0.102 & 3 & USGC~S128$^*$, HDC~236$^*$, LDC~251, HG~32\\
100 & NGC~1407   & 03 40 11.8 & -18 34 48 & 0.0059 & 23 & 0.112 & 2 & USGC~S128, HDC~236$^*$, LDC~251$^*$, HG~32\\
113 & NGC~1550   & 04 19 37.9 & +02 24 34 & 0.0124 & 53 & 0.257 & 2 & HDC~280, LDC~297\\
126 & NGC~1779   & 05 05 18.1 & -09 08 50 & 0.0111 & 45 & 0.218 & 3 & - \\
138 & NGC~2292   & 06 47 39.6 & -26 44 46 & 0.0068 & 30 & 0.145 & 3 & - \\
167 & NGC~2768   & 09 11 37.5 & +60 02 14 & 0.0045 & 23 & 0.112 & 2 & HDC~506, HG~80\\
177 & NGC~2911   & 09 33 46.1 & +10 09 09 & 0.0106 & 45 & 0.218 & 3 & WBL~226, USGC~U239, HDC~535, LDC~655, GH~47\\
205 & NGC 3325   & 10 39 20.4 & -00 12 01 & 0.0189 & 80 & 0.388 & 3 & USGC~U315\\
232 & NGC~3613   & 11 18 36.1 & +58 00 00 & 0.0068 & 32 & 0.156 & 3 & HDC~647, LDC~867, GH~94, HG~60\\
236 & NGC~3665   & 11 24 43.7 & +38 45 46 & 0.0069 & 32 & 0.156 & 2 & USGC~U383, HDC~648, LDC~805, GH~79\\
255 & NGC~3923   & 11 51 01.7 & -28 48 22 & 0.0058 & 20 & 0.097 & 2 & HDC~675, LDC~860\\
314 & NGC~4697   & 12 48 35.9 & -05 48 03 & 0.0041 & 18 & 0.087 & 3 & HG~41\\
329 & NGC~4956   & 13 05 00.9 & +35 10 41 & 0.0158 & 71 & 0.344 & 2 & USGC~U514, GH~114\\
341 & NGC~5061   & 13 18 05.1 & -26 50 14 & 0.0069 & 28 & 0.136 & 3 & HDC~782, HG~31\\
350 & NGC~5127   & 13 23 45.0 & +31 33 57 & 0.0162 & 72 & 0.349 & 2 & - \\
360 & NGC~5322   & 13 49 15.3 & +60 11 26 & 0.0059 & 29 & 0.141 & 2 & HG~81, GH~122\\
370 & NGC~5444   & 14 03 24.1 & +35 07 56 & 0.0131 & 60 & 0.291 & 3 & USGC~U593, HDC~845, GH~131\\
376 & NGC~5490   & 14 09 57.3 & +17 32 44 & 0.0162 & 71 & 0.344 & 2 & WBL~493, USGC~U599, LDC~1039, GH~133\\
383 & NGC~5629   & 14 28 16.4 & +25 50 56 & 0.0150 & 67 & 0.325 & 2 & WBL~509, USGC~U629, HDC~875, GH~143\\
398 & NGC~5903   & 15 18 36.5 & -24 04 07 & 0.0086 & 36 & 0.175 & 3 & HDC~904, LDC~1117\\
457 & NGC~7252   & 22 20 44.7 & -24 40 42 & 0.0160 & 66 & 0.320 & 2 & - \\
463 & NGC~7377   & 22 47 47.5 & -22 18 44 & 0.0111 & 46 & 0.223 & 2 & USGC~S278, HDC~1205, LDC~1535\\
\hline
\end{tabular}
\end{center}
\end{table*}

\subsection{Overlap with previous group samples}

A number of other surveys have identified the groups in the sample as probable bound systems, based on their galaxy populations; cross-identifications for some of these are listed in Table~\ref{tab:sample}. The surveys include percolation and density contrast studies based on optical \citep{HuchraGeller82,GellerHuchra83,Whiteetal99} and near-infrared \citep{Crooketal07,Crooketal08} catalogues, as well as friends-of-friends searches \citep{Ramellaetal02}. A few of the groups are dense enough to be classed as \citet{Hicksonetal92} compact groups. \citet{Garcia93} identified cross-matches with the Huchra \& Geller and Geller \& Huchra catalogues. For the other catalogues, we consider the group identifications as matched if they a) both include the BGE, and b) the group membership overlaps by $>$50 per cent. Catalogue comparison was performed using \textsc{topcat} \citep{Taylor05}.

All but five of the groups in our sample have been identified by these
surveys, and most groups have been identified by multiple surveys. The
different selection methods used by these surveys lead to variations
in group membership. In many cases, our selection method results in a
larger number of group members, with previous catalogues identifying
only a subset. However, in some cases CLoGS groups
are subsets of apparent larger structures identified by other
surveys. For example, LGG~97 and 100, centred on NGC~1407 and
NGC~1395, are identified as parts of the Fornax-Eridanus
supercluster in the Huchra \& Geller, Ramella et al. and Crook et
al. samples. In this case, X-ray observation proves its utility as a
group-identification method, showing that each group possesses its own
separate group-scale dark halo. In general, we consider it supportive
of our group selection approach that multiple surveys using different
approaches identify the dominant galaxies of our groups as being
members of larger bound systems.

Many of the groups are also known to be X-ray luminous, primarily from \rosat\ All-Sky Survey and pointed data. The \citet{Mulchaeyetal03} atlas of groups lists \rosat\ X-ray detections for 13 of the 53 groups, as does the GEMS catalogue \citep{Forbesetal06,OsmondPonman04}. Adding systems in which the dominant elliptical was detected by \rosat\ or \textit{Einstein}, often with an X-ray luminosity typical of a galaxy group \citep{OSullivanetal01b} brings the total to 24. Roughly 45\% of the sample had been observed by pointed \xmm\ or \chandra\ X-ray observations at the time of selection, and these had typical temperatures between 0.5 and 1.7~keV, in the expected range for group-scale systems. A number of the groups are well-known systems which have been included in previous X-ray-selected samples \citep[e.g.,][]{OsmondPonman04,Finoguenovetal06,Finoguenovetal07,Sunetal09,Eckmilleretal11,Panagouliaetal14} and in some cases detailed studies of their IGM structure have been made, e.g., NGC~5044 \citep{OSullivanetal14a,Davidetal11,Davidetal09}, NGC~7619 \citep{Randalletal09}, NGC~5846 \citep{Machaceketal11}, NGC~4261 \citep{OSullivanetal11c}, NGC~1407 \citep{Suetal14,Giacintuccietal12}, NGC~1550 \citep{Sunetal03} and NGC~193 \citep{Bogdanetal14}.   

It should be noted that our sample selection excludes a number of well known, X-ray bright, nearby groups. In some cases the original \citet{Garcia93} catalogue includes the group within a larger structure, e.g., NGC~4636 is considered part of the Virgo cluster, and NGC~5813 as part of the NGC~5846 group. Again, both groups are known to possess their own distinct IGM \citep[e.g.,][]{Baldietal09b,Randalletal11} and therefore dark halo, but we accept the definition in the Garcia catalogue so as not to bias our selection. Systems with recession velocities greater than 5500\kmps\ (e.g., NGC~741) are not included in the Garcia catalogue. Our richness criteria also excludes a few systems, the most obvious cases being NGC~507 and NGC~499, which are identified as individual groups by Garcia and which have X-ray temperatures characteristic of groups, but which both exceed our upper richness bound.

\subsection{The High-Richness Subsample}
\label{sec:highR}

We further divide the sample into two subsamples, based on their richness. The high-richness subsample contains the 26 groups with $R$=4-8, while the low-richness subample contains the 27 groups with $R$=2-3. Both subsamples contain well-known X-ray luminous systems, e.g., NGC~777, NGC~4261, NGC~5044, NGC~5846 and NGC~7619 in the high-richness, and NGC~315, NGC~1407 and NGC~1550 in the low-richness subsamples. Both subsamples also contain systems with a variety of radio properties and dynamical states. Each subsample is statistically complete in its own right.

In the remainder of this paper, we focus on the X-ray properties of the high-richness subsample, for which we have acquired complete X-ray coverage of the group cores using a combination of \xmm\ and \chandra\ data.

\section{X-ray Observations and Data Reduction}
\label{sec:obs}

For both \chandra\ and \xmms\ spectral fitting was performed using \textsc{xspec} 12.6.0k \citep{Arnaud96}, with source models including absorbed by a hydrogen column set at the Galactic value \citep[drawn from the the Leiden/Argentine/Bonn survey,][]{Kalberlaetal05}. Abundances were measured relative to the abundance
ratios of \citet{GrevesseSauval98}. 1$\sigma$ uncertainties for one
interesting parameter are reported for all fitted values. 

\subsection{XMM-Newton}
\label{sec:XMM}

At the time of sample selection, 10 of the 26 groups in the high-richness
subsample had been observed by \xmm. Observations of a further 9 groups
were performed during cycles 10 and 11. Table~\ref{tab:Xobs}
provides a summary of these observations. A detailed summary of the \xmm\
mission and instrumentation can be found in \citet[and references
therein]{Jansenetal01}.

\begin{table*}
\caption{\label{tab:Xobs} Summary of X-ray observations. EPIC-pn mode is the readout mode of the \xmms\ EPIC-pn, either Full Frame (F) or Extended Full Frame (EF). All MOS exposures use Full Frame mode. ACIS array indicates position of the focal point, on either the ACIS-I or -S array of \chandra. ACIS mode indicates the telemetry mode used, either Faint (F) or Very Faint (VF). Total exposures and exposure times after flare cleaning are listed in a net/gross format. }
\begin{center}
\begin{tabular}{lcccccccccc}
\hline
LGG & Obs. Date & \multicolumn{3}{c}{XMM-Newton} & \multicolumn{3}{c}{Chandra ACIS} & \multicolumn{3}{c}{Exposure} \\
 & & ObsID & EPIC-pn Mode & Filter & ObsID & Array & Mode & MOS & pn & ACIS \\
 & & & & & & & & (ks) & (ks) & (ks) \\
\hline
9   & 2009-08-21 & - & - & - & 11389 & S & VF & - & - & 93.9/93.9 \\
18  & 2004-01-01 & 0203610201 & F & Thin & - & - & - & 14.2/22.8 & 11.0/19.5 & - \\
27  & 2010-09-07 & - & - & - & 12175 & I & VF & - & - & 9.9/10 \\
31  & 2012-07-02 & 0673770301 & F & Med & - & - & - & 31.3/39.1 & 22/33.6 & - \\ 
42  & 2004-07-18 & 0203610301 & F & Thin & 5001 & I & VF & 7.5/29.3 & 4.4/20.7 & 9.3/10.0\\
58  & 2012-07-31 & 0693970301 & F & Med & - & - & - & 41.5/42.2 & 35.5/36.5 & - \\
61  & 2012-08-14 & 0693970401 & F & Med & - & - & - & 27.7/28.3 & 23.1/24.1 & - \\
66  & 2011-08-17 & 0673770201 & F & Med & - & - & - & 8.4/25.7 & 5.6/21.5 & - \\
    & 2012-01-17 & 0673771001 & F & Med & - & - & - & 8.8/9.5 & 5.5/7.1 &   \\
72  & 2012-02-10 & 0673770101 & F & Med & - & - & - & 36/44.5 & 28/39 & - \\
80  & 2005-08-03 & 0301650101 & F & Thin & - & - & - & 10.0/11.4 & 7.4/8.8 & - \\
103 & 2012-01-27 & 0673770601 & F & Med & - & - & - & 11.5/21.3 & 6.5/17.6 & - \\
117 & 2000-10-03 & - & - & - & 2217 & I & F & - & - & 19.9/21.1 \\
158 & 2001-10-15 & 0108860501 & F & Med & 7925 & I & VF & 20.0/21.4 & 16.2/17.2 & 47.8/48.8\\
185 & 2005-02-02 & - & - & - & 5902 & I & VF & - & - & 8.1/8.6 \\
262 & 2011-06-10 & 0673770501 & F & Med & - & - & - & 5.6/13.7 & $<$4.8/12.4 & - \\ 
276 & 2006-06-20 & 0301651701 & F & Thin & - & - & - & 12.3/12.5 & 9.4/9.7 & - \\
278 & 2007-12-16 & 0502120101 & F & Med & 9569 & S & F & 73/119 & 46.8/103 & 100.9/100.9\\
310 & 2011-03-26 & - & - & - & 12174 & I & VF & - & - & 9.9/10 \\
338 & 2008-12-27 & 0554680101 & F & Thin & 9399 & S & I & 98.1/125 & 71.4/107 & 82.7/82.7\\
345 & 2011-08-22 & - & - & - & 12173 & I & VF & - & - & 9.4/10 \\
351 & 2011-03-27 & - & - & - & 12176 & I & VF & - & - & 19.8/20 \\
363 & 2002-06-02 & 0041180401 & F & Thick & - & - & - & 19.6/22 & 15.7/18 & - \\
393 & 2001-01-25 & 0021540101 & F & Thin & 7923 & I & VF & - & 25.4/25.6 & 90.0/90.0 \\
    & 2001-08-26 & 0021540501 & EF & Thin & - & - & - & 13.9/17.2 & 8.6/12.3 & - \\
402 & 2011-05-18 & 0673770401 & F & Med & - & - & - & 11.8/12.6 & 9.7/10.1 & - \\
421 & 2012-09-17 & 0673970101 & F & Med & - & - & - & 24/31.1 & 19.9/26.9 & - \\
473 & 2003-12-16 & 0149240101 & F & Med/Thin & 2074 & I & VF & 39.5/40.5 & 30.7/35.3 & 26.5/26.7 \\
\hline
\end{tabular}
\end{center}
\end{table*}

\xmm\ data were reduced and analysed using the \xmms\ Science Analysis
System (\textsc{sas v12.0.1}), and reprocessed using the \textsc{emchain}
and \textsc{epchain} tasks. In many of the observations diffuse X-ray
emission from the intra-group medium (IGM) fills the field of view of the
EPIC instruments. This makes accurate scaling and correction of blank-sky
background data to match the observation dataset difficult. We therefore
performed our analysis in two stages. For every dataset we carried out
imaging and spectral analysis using scaled and corrected blank-sky data. In
systems with extended diffuse emission we then used the \xmms-Extended
Source Analysis Software (\textsc{esas}) to carry out a second spectral
analysis, taking advantage of the background modelling approach to improve
the accuracy of spectral fitting and sensitivity to low surface--brightness
emission.

The basic analysis follows the methods described in
\citep{OSullivanetal11c}. Bad pixels and columns were identified and
removed, and the events lists filtered to include only those events with
FLAG = 0 and patterns 0-12 (for the EPIC-MOS cameras) or 0-4 (for the EPIC-pn).
Background lightcurves in hard (10-15~keV), medium (2-5~keV) and soft
(0.3-1~keV) bands were extracted for each dataset, and times when the total
count rate in any band deviated from the mean by more than 3$\sigma$ were
excluded. In systems where the emission is bright enough to produce a significant number of out--of--time (OOT) events on the EPIC-pn detector, OOT event lists were produced and used to provide appropriately scaled correction images and spectra.

Point sources were identified using \textsc{edetect$\_$chain}, and regions
corresponding to the 85 per cent encircled energy radius of each source
(except those potentially associated with the active galactic nuclei of
group member galaxies) were excluded. Imaging analysis was typically
carried out in the 0.5-2~keV band, which provides optimal signal--to--noise
for the spectrally soft diffuse emission, using monoenergetic 1~keV
exposure maps. Spectral analysis was performed in the 0.5-7~keV band, using
responses generated with the \textsc{arfgen} and \textsc{rmfgen} tasks.
Background images and spectra were created using the ``double-subtraction''
technique \citep{Arnaudetal02,Prattetal01}.

For datasets where the diffuse emission fills the field of view, we
repeated the reduction and spectral analysis using ESAS and the
general spectral modelling approach suggested by
\citet{Snowdenetal04}. The \textsc{mos-filter} and \textsc{pn-filter}
tasks were used to filter out periods of high background, and CCDs in
anomalous background states were excluded. Point sources were
identified using the \textsc{cheese-bands} task, and excluded if they
were not associated with the cores of group--member galaxies.  Spectra
and responses for each region were extracted, as well as an additional
RASS spectrum extracted from an annulus typically 2\degree\ from the
group centre using the HEASARC X-ray Background
Tool\footnote{\url{http://heasarc.gsfc.nasa.gov/cgi-bin/Tools/xraybg/xraybg.pl}}.

Spectral analysis was performed using the 0.3-10.0~keV (MOS) and
0.4-7.2~keV (pn) energy bands, with all spectra for a given target
fitted simultaneously.  The particle component of the background was
partially subtracted using particle--only datasets scaled to match the
event rates in regions of the detectors which fall outside the field
of view. Out--of--time (OOT) events in the EPIC-pn data were
statistically subtracted using scaled OOT spectra.  The remainder of
the particle background was modelled with a powerlaw whose index was
linked across all annuli. As this element of the background is not
focused by the telescope mirrors, diagonal Ancillary Response Files
(ARFs) were used. The instrumental Al K$\alpha$ and Si K$\alpha$
fluorescence lines were modelled using Gaussians whose widths and
energies were linked across all annuli, but with independent
normalizations. The X--ray background was modelled with four
components whose normalizations were tied between annuli, scaling to a
normalization per square arcminute as determined by the
\textsc{proton-scale} task.  The cosmic hard X-ray background was
represented by an absorbed powerlaw with index fixed at $\Gamma$=1.46.
Thermal emission from the Galaxy, local hot bubble and/or heliosphere
was represented by one unabsorbed and two absorbed APEC thermal plasma
models with temperatures of 0.1, 0.1 and 0.25 keV respectively. The
normalizations of the APEC models were free to vary relative to one
another. Absorption was represented by the WABS model, fixed at the
Galactic column density taken from the RASS spectrum was fitted using
only the X-ray background components.

Comparison of the two approaches to spectral analysis was carried out
whenever they were applied. The results suggest that the ESAS
spectral-modelling approach is more sensitive to low
surface--brightness emission, but that both methods agree within the
errors where a source--free region is available. We are therefore
confident in the robustness of our approach to \xmms\ spectral
fitting.

\subsection{Chandra}
\label{sec:chandra}

At the time of sample selection, \chandra\ had observed 11 systems in the
high-richness sample (8 of which had also been observed by \xmms). A
further 4 systems were observed as part of the Guaranteed Time allocation
in cyle~12.  A summary of the \chandra\ mission and instrumentation can be
found in \citet{Weisskopfetal02}.

\textit{Chandra} observations were reduced using \textsc{ciao} 4.4.1
\citep{Fruscioneetal06} and CALDB 4.5 following techniques similar to
those described in \citet{OSullivanetal07} and the \chandra\ analysis
threads\footnote{\url{http://asc.harvard.edu/ciao/threads/index.html}}.
The level 1 event files were reprocessed, bad pixels and events with
\asca\ grades 1, 5 and 7 were removed, and the cosmic ray afterglow
correction was applied. Very Faint mode cleaning was performed where
applicable. The data were corrected to the appropriate gain map, the
standard time-dependent gain and charge-transfer inefficiency (CTI)
corrections were made, and background light curves were produced. The
\textsc{lc\_clean} task was used to filter periods of high background,
so as to produce event lists with similar contributions from the
particle component of the background to that of the blank-sky
background files.

Point source identification was performed using the \textsc{ciao} task
\textsc{wavdetect}, with detection thresholds chosen to ensure that
the task detects $\leq$1 false source in the S3 or ACIS-I fields of
view, working from 0.3-7.0 keV images and exposure maps. The resulting
point source regions were then used to remove point sources from all
further analysis, with the exception of sources potentially associated
with active galactic nuclei in group member galaxies. Spectra were
extracted from each dataset using the \textsc{specextract} task. When
examining diffuse emission, background spectra were drawn from
blank-sky event lists, scaled to match the data in the 9.5-12.0 keV
band.

\section{Results}
\label{sec:res}
The results of our X-ray analysis of the high-richness subsample are described below. More detailed comments on each system are given in Appendix~\ref{sec:notes}, and X-ray, radio and optical images of the core of each group can be found in Appendix~\ref{sec:images}.

\subsection{Surface Brightness and Luminosity}
\label{sec:SB}

As an initial step, we examined 0.5-2~keV images of the groups to determine whether extended emission was visible. The \textsc{ciao} \textsc{sherpa} modelling and fitting package \citep{Freemanetal01} was used to characterize the emission distribution. Point sources not associated with the dominant galaxy were generally removed, though in some systems we chose to model a small number of sources associated with group member galaxies. For \chandra\ data we incorporated a monoenergetic exposure map in the model, and use a flat component to model the background. For \xmms\ data we fitted the combined EPIC MOS+pn image, again incorporating a monoenergetic exposure map in the model. The \xmms\ background was modelled using two flat components, one (representing the X-ray background) folded through the exposure map, the other (representing the particle background) only masked to account for chip gaps, bad pixels and columns, and excluded point sources,

Source components were modelled using one or more $\beta$-models \citep{Cavaliere76},

\begin{equation}
S(r) = S_o\bigg(1+\Big(\frac{r}{r_c}\Big)^2\bigg)^{3\beta+0.5}
\label{eqn:SB}
\end{equation}

where $r_c$ is the core radius, $\beta$ governs the slope at large radii, and $S_0$ is the central surface brightness. For \chandra\ images, the radial profile was compared to that of the on-axis point spread function (PSF) to determine whether a central point source was present. For \xmms\ data, all source components were convolved with a monoenergetic on-axis PSF including weighted contributions from the three detectors, and any suspected central point source was modelled using a narrow (0.5-1 pixel FWHM) Gaussian. 

Table~\ref{tab:SB} lists the core radii and $\beta$ parameters for the $\beta$-model fits to the images. The table also lists $r_{det}$, the radius to which emission was detected above the background. In several of the \xmms\ observations, only a central source comparable to the PSF was identified, in which case $r_{det}$ is considered to be $<$30\arcs. It should also be noted that extended emission does not always arise from hot gas. For some of the \chandra\ observations (LGGs~27, 310, 345), spectral fitting shows that diffuse emission extending a few kpc has a powerlaw spectrum and is therefore likely to arise from the unresolved stellar and X-ray binary populations.

\begin{landscape}
\begin{table}
\caption{\label{tab:SB}Detection radii, surface brightness parameters and luminosities for all high-richness groups. Core radius r$_c$ and slope parameter $\beta$ are defined in equation~\ref{eqn:SB}. X-ray luminosities are measured in the 0.5-7~keV band. For any thermal plasma component, luminosity represents the aperture luminosity within the detection radius. For any powerlaw component, the luminosity is measured within the central spectral bin. Values marked with and asterisk (*) represent the predicted flux from low mass X-ray binaries and were held fixed during fitting. The penultimate column gives the luminosity of the thermal plasma scaled to R$_{500}$, a measure of the luminosity of the IGM of the group. X-ray gas morphology is our classification of the X-ray extent as either group-like (GRP, $>$65~kpc), galaxy-like (gal, $<$65~kpc) or point-like (pnt, no extent beyond the PSF). }
\begin{center}
\begin{tabular}{lccccccccccc}
\hline
LGG & \multicolumn{2}{c}{r$_{det}$} & Thermal & \multicolumn{2}{c}{Powerlaw} & \multicolumn{4}{c}{Surface brightness parameters} & L$_{\rm X,R500}$ & X-ray gas\\
 & & & L$_{\rm X,rdet}$ & $\Gamma$ & L$_{\rm X}$ & r$_{\rm c1}$ & $\beta_1$ & r$_{\rm c2}$ & $\beta_2$ & & Morphology\\
 & (\arcs) & (kpc) & (10$^{40}$\ergps) & & (10$^{40}$\ergps) & (\arcs) & & (\arcs) & & (10$^{40}$\ergps) & \\
\hline\\[-3mm]
9   & 380 & 135 & 58.40$\pm$3.95 & 1.51$\pm$0.06 & 3.27$^{+0.15}_{-0.18}$ & 23.65$^{+97.03}_{-23.46}$ & 0.61$^{+0.17}_{-0.05}$ & - & - & 66.8$\pm$0.5 & GRP \\[+0.5mm]
18  & 840 & 310 & 137.56$^{+6.82}_{-6.76}$ & - & - & 1.73$^{+0.04}_{-0.03}$ & 0.612$^{+0.008}_{-0.001}$ & - & - & 141.6$\pm$7.0 & GRP \\[+0.5mm]
27  &  40 & 5   & - & 2.4$^{+0.8}_{-0.6}$ & 0.56$^{+0.79}_{-0.84}$ & 12.40$^{+6.46}_{-4.31}$ & 0.77$^{+0.25}_{-0.13}$ & - & - & $<$11.25 & - \\[+0.5mm]
31  & 350 & 130 & 150.07$^{+10.79}_{-10.37}$ & - & - & 45.4$^{+22.9}_{-17.9}$ & 0.41$^{+0.06}_{-0.05}$ & - & - & 334.8$^{+24.1}_{-23.1}$ & GRP \\[+0.5mm] 
42  & 780 & 275 & 126.28$^{+13.64}_{-14.82}$ & - & - & 6.16$^{+0.83}_{-0.75}$ & 0.575$^{+0.012}_{-0.010}$ & - & - & 129.3$^{+14.0}_{-15.2}$ & GRP \\[+0.5mm]
58  & $<$30 & $<$10 & 1.38$^{+0.08}_{-0.09}$ & 1.69$\pm$0.07 & 8.52$^{+0.43}_{-0.42}$ & - & - & - & - & $<$1.30 & pnt \\[+0.5mm]
61  & $<$30 & $<$9 & - & 1.69$\pm$0.07 & 3.21$\pm$0.20 & - & - & - & - & $<$3.31 & - \\[+0.5mm]
66  & 150 & 50  & 2.08$^{+0.62}_{-1.11}$ & 1.65$^*$ & 2.02$^*$ & 12.78$^{+3.41}_{-2.74}$ & 0.66$^{+0.06}_{-0.05}$ & - & - & 2.26$^{+0.67}_{-1.20}$ & gal \\[+0.5mm]
72  & 840 & 310 & 420.23$^{+19.24}_{-18.87}$ & - & - & 228.2$^{+20.1}_{-17.2}$ & 0.41$^{+0.02}_{-0.18}$ & 11.52$\pm$0.26 & 0.69$\pm$0.01 & 591.2$^{+27.1}_{-26.6}$ & GRP \\[+0.5mm]
80  & $<$30 & $<$10 & - & 3.17$^{+0.26}_{-0.25}$ & 3.14$\pm$0.39 & - & - & - & - & $<$6.64 & - \\[+0.5mm]
103 & 350 & 107 & 28.76$\pm$1.39 & -1.34$^{+2.03}_{-1.05}$ & 2.47$^{+16.82}_{-0.65}$ & 2.09 & 0.365$\pm$0.01 & - & - & 83.2$\pm$4.0 & GRP \\[+0.5mm]
117 & 555 & 137 & 16.01$^{+1.24}_{-1.70}$ & 3.83$^{+0.50}_{-0.69}$ & 4.36$\pm$1.24 & 0.11$^{+0.13}_{-0.10}$ & 0.43$\pm$0.01 & - & - & 21.3$^{+1.6}_{-2.3}$ & GRP \\[+0.5mm]
158 & 950 & 300 & 233.36$^{+16.69}_{-13.38}$ & - & - & 55.39$^{+7.74}_{-7.65}$ & 0.358$\pm$0.006 & - & - & 390.3$^{+27.9}_{-22.4}$ & GRP \\[+0.5mm]
185 & 100 & 16  & 5.61$^{+1.29}_{-2.33}$ & 1.41$^{+0.66}_{-0.73}$ & 4.31$^{+2.06}_{-1.23}$ & 14.35$^{+28.27}_{-0.75}$ & 0.71$^{+0.27}_{-0.15}$ & 0.72$^{+0.51}_{-0.02}$ & 0.82$^{+0.03}_{-0.18}$ & 5.69$^{+1.31}_{-2.36}$ & gal \\[+0.5mm]
262 &  90 & 24  & 4.65$^{+0.49}_{-0.43}$ & 1.65$^*$ & 1.60$^*$ & 9.98$^{+2.33}_{-1.97}$ & 0.61$^{+0.05}_{-0.04}$ & - & - & 5.80$^{+0.61}_{-0.54}$ & gal \\[+0.5mm]
276 & $<$30 & $<$6.5 & 0.65$\pm$0.08 & -1.11$^{+0.21}_{-0.22}$ & 4.67$\pm$0.46 & - & - & - & - & $<$0.99 & pnt \\[+0.5mm]
278 & 550 & 85 & 51.49$^{+2.30}_{-1.95}$ & 0.40$^{+0.06}_{-0.08}$ & 4.97$^{+0.44}_{-0.56}$ & 1.0$^{+1.2}_{-1.0}$ & 0.55$^{+0.01}_{-0.02}$ & - & - & 114.3$^{+5.1}_{-4.3}$ & GRP \\[+0.5mm]
310 &  30 & 6.5 & - & 2.1$^{+0.8}_{-0.6}$ & 1.72$^{+0.27}_{-0.26}$ & 0.99$^{+0.59}_{-0.48}$ & 0.51$\pm$0.03 & - & - & $<$6.03 & - \\[+0.5mm]
338 & 850 & 157 & 1090.24$^{+7.34}_{-11.60}$ & - & - & 51.31$^{+0.75}_{-1.58}$ & 0.538$^{+0.001}_{-0.016}$ & - & - & 1385.5$^{+9.3}_{-14.7}$ & GRP \\[+0.5mm] 
345 & 100 & 11  & - & 2.1$\pm$0.2 & 7.10$^{+0.47}_{-0.44}$ & 0.50$^{+0.02}_{-0.01}$ & 0.89$^{+0.03}_{-0.02}$ & 0.04$\pm$0.01 & 0.40$\pm$0.01 & $<$4.89 & - \\[+0.5mm]
351 &  50 & 15   & - & 1.65$^*$ & 1.50$^*$ & $<$0.01 & 0.44$^{+0.06}_{-0.05}$ & - & - & $<$1.30 & - \\[+0.5mm]
363 & 600 & 100 & 96.39$^{+11.26}_{-11.28}$ & 0.55$^{+0.16}_{-0.14}$ & 0.40$^{+0.26}_{-0.08}$ & 0.61$^{+0.48}_{-0.01}$ & 0.414$^{+0.027}_{-0.073}$ & 4.82$^{+12.16}_{-6.61}$ & 1.38$\pm$0.71 & 186.7$\pm$21.8 & GRP \\[+0.5mm]
393 & 850 & 107 & 112.34$^{+1.75}_{-1.78}$ & 0.06$^{+0.06}_{-0.17}$ & 0.36$^{+0.04}_{-0.07}$ & 16.8$^{+9.1}_{-15.3}$ & 0.69$^{+1.46}_{-0.18}$ & 134.90$^{+60.67}_{-40.50}$ & 1.14$^{+1.30}_{-0.32}$ & 133.4$^{+2.1}_{-2.1}$ & GRP \\[+0.5mm]
402 & 400 & 85  & 26.34$^{+3.08}_{-3.92}$ & 0.91$^{+0.94}_{-0.84}$ & 0.98$^{+2.38}_{-0.71}$ & 10.2$^{+1.1}_{-1.0}$ & 0.54$\pm$0.01 & - & - & 31.7$^{+3.7}_{-4.7}$& GRP  \\[+0.5mm]
421 & $<$30 & $<$9 & 0.39$^{+0.11}_{-0.10}$ & 0.44$^{+0.34}_{-0.37}$ & 1.67$^{+0.35}_{-0.30}$ & - & - & - & - & $<$1.07 & pnt \\[+0.5mm]
473 & 850 & 223 & 256.72$^{+5.48}_{-8.98}$ & - & - & 19.2$^{+14.6}_{-19.1}$ & 0.34$^{+0.01}_{-0.07}$ & 2.478$\pm$0.953 & 0.537$^{+0.001}_{-0.028}$ & 449.3$^{+9.6}_{-15.7}$ & GRP \\
\hline
\end{tabular}
\end{center}
\end{table}
\end{landscape}

For systems where thermal emission was detected, we classified the extent of the gas halo as either group-like (extent $>$65~kpc), galaxy-like (extent $\sim$10-65~kpc) or point-like (unresolved, extent smaller than the XMM PSF). Although somewhat arbitrary, these classifications give a simple picture of the scale of the emission. The 65~kpc cutoff between group and galaxy-scale emission is similar to the 60~kpc boundary used by \citep{OsmondPonman04}. Examination of the luminosities and temperatures of our group-scale and galaxy-scale systems suggests that the 65~kpc cutoff provides a reasonable rule of thumb; selecting groups on the basis of temperature $>$0.6~keV or luminosity $>$10$^{41}$\ergps\ would produce similar results. We discuss the relationship between these X-ray morphological classes and other properties in Section~\ref{sec:disc} and list the classification of each group in Table~\ref{tab:SB}.

In most of the groups where extended emission was detected, the emission was found to be centred on the BGE, confirming its location as the centre of the X-ray halo. For two groups (LGGs~72 and 473) a second X-ray peak was identified, centred on another early-type galaxy, and in LGG~72 the surface brightness distribution between the two peaks is clearly disturbed. We consider both systems to be ongoing mergers, and include additional model components to account for the secondary peaks.

 Where extended emission from hot gas was detected, we estimated the gas luminosity from spectral fits, and used the surface brightness model to extrapolate to a luminosity within R$_{500}$, excluding any contribution from AGN or stellar sources. Estimation of R$_{500}$ is discussed in Section~\ref{sec:mass}. Note that this extrapolation cannot account for changes in gas temperature or abundance at large radius. While a single or double $\beta$-model was sufficient to model most systems with extended emission, a few systems were problematic. In cases where two X-ray peaks are visible in the field of view (merging groups or observations including background groups/clusters) we either excluded the second peak or used an additional $\beta$-model component to model its contribution. Where sloshing affected the surface brightness profile we restricted the fits to the least disturbed parts of the halo. In two of the brightest, closest systems (LGGs~278 and 393) we found it necessary to simultaneously fit the \xmms\ and \rosat\ PSPC data in order to constrain the outer slope of the model. In LGG~9, where a cavity and shock strongly distort the central surface brightness, we limited our fit to the area outside the shock.

Comparison of our best-fitting $\beta$ parameters for six bright groups with those found from \rosat\ analysis shows excellent agreement except in the case of NGC~7619, where \citet{HelsdonPonman00a} found a very steep $\beta$=0.78$\pm$0.08, probably owing to the disturbed state of this merging group. Comparison of our luminosity with that of \citet{OsmondPonman04} for five overlapping groups shows good agreement. 

For systems with no extended emission, or where only a small powerlaw spectrum extended component was identified, we estimated a 3$\sigma$ upper limit on L$_{X,R500}$ as follows. For most systems in the sample, the field of view of the observations extends at least 65~kpc from the dominant galaxy, and we adopt this as the radius within which to estimate an initial upper limit. The exceptions are the closest systems observed with \chandra\ ACIS-I, where we are forced to perform the estimation within 8\arcm\ radius (equivalent to 53.5~kpc for LGG~345 and 58.2~kpc for LGG~18). In each case, we determined the number of 0.5-2~keV counts in the region and a local background value, based on either an annulus outside the 65~kpc circle, or from the corners of the ACIS-I field. The central source and any nearby point sources were excluded, and the area thus lost corrected for assuming a flat flux distribution. We then used the Bayesian Estimate of Hardness Ratios tool \citep[\textsc{behr},][]{Parketal06} to estimate the 3$\sigma$ upper limit on the number of detected counts in the region, scaling the background using the monoenergetic exposure map.

These limits were converted from counts to luminosity assuming a 0.5~keV, 0.5\Zsol\ APEC plasma model with Galactic absorption. This produced limits in the range $\sim$0.2-2.5$\times$10$^{40}$\ergps\ for the 65~kpc radius region. This can be considered as a limiting sensitivity of the sample. The sensitivity is driven by the distance of the group, the length of the exposure, the collecting area of the telescope and the background level of the observation. Our observation planning took this into account, using short ACIS-I observations to target the closest groups, and scaling the requested \xmm\ exposures with distance. We note that we adopted a temperature and abundance typical of the smaller detected groups. Increasing to 0.8-1~keV and solar abundance would increase the expected luminosity by a factor $\sim$2, and decreasing to 0.3~keV and 0.3\Zsol\ would decrease expected luminosity by a factor $\sim$2.5.

To extrapolate to R$_{500}$, we estimate the fiducial radius from the assumed temperature, and assume a $\beta$-model surface brightness distribution with core radius 0.1R$_{500}$ (31.2~kpc) and $\beta$=0.4, a flat distribution being more likely to be undetected than a centrally-peaked source. The resulting limits are listed in Table~\ref{tab:SB}. We note that using $\beta$=0.54 and core radius 10~kpc (similar to LGG~398/NGC~5044) would decrease luminosity by a factor $\sim$3. Decreasing $\beta$ to 0.3, an exceptionally flat distribution, would increase luminosity by a factor $\sim$1.9.

\subsection{Spectral fits and temperature profiles}
\label{sec:tprofs}

For each system, we use an annular adaptive binning algorithm to select regions with a fixed number of net counts in the 0.5-7~keV band. Annuli are centred on the BGE, and we exclude the secondary X-ray peaks identified in the merging groups LGG~72 and LGG~473. We initially require 2000 counts per region, and increase or decrease this requirement in steps of factor two for bright or faint systems. For the faintest objects, the minimum acceptable for spectral fitting is 500 net counts in a circle centred on the core of the dominant galaxy. For the brightest groups we use annuli containing up to 16000 net counts. For \xmms\ the count requirement is applied to all three detectors combined.

Spectra for each region are initially fitted separately to determine the basic properties of the system. For the faintest systems with only a single spectral region, we determine whether the emission is best fitted by an APEC thermal model \citep{Smithetal01}, a power-law, or a combination of the two. Where multiple spectral regions can be used, we test whether a power-law component is needed in any region which overlaps the dominant galaxy. We estimate the expected powerlaw flux from low-mass X-ray binaries (LMXBs) in the dominant galaxy based on its $K$-band luminosity, using the L$_{LMXB}$:L$_K$ of \citet{Borosonetal11}. Where necessary, the expected flux in each annular bin is estimated based on a \citet{Sersic68} profile fit to the 2MASS $K$-band image \citep{Skrutskieetal06}. In most systems, either the expected flux is negligible compared to the thermal component, or is less than or comparable to that of the fitted powerlaw component. However, in three systems (LGGs~66, 262, 351) we expect a significant powerlaw contribution but find that an APEC-only model is the best fit. In these cases we fit APEC+powerlaw models with the photon index fixed at $\Gamma$=1.65 and normalization set to reproduce the expected LMXB flux. In the case of LGG~351, the resulting fit has an APEC normalization consistent with zero. We treat this galaxy as gas-poor, but note that deeper observations might allow the separation of gas and powerlaw components in the spectrum. Table~\ref{tab:SB} lists the luminosities derived from the spectral fits, and where powerlaw emission was detected or assumed, the photon indices. Note that fitted powerlaw contributions often include both LMXB and AGN emission, which we do not attempt to separate.

\begin{figure*}
\includegraphics[width=\textwidth,bb=60 110 560 730]{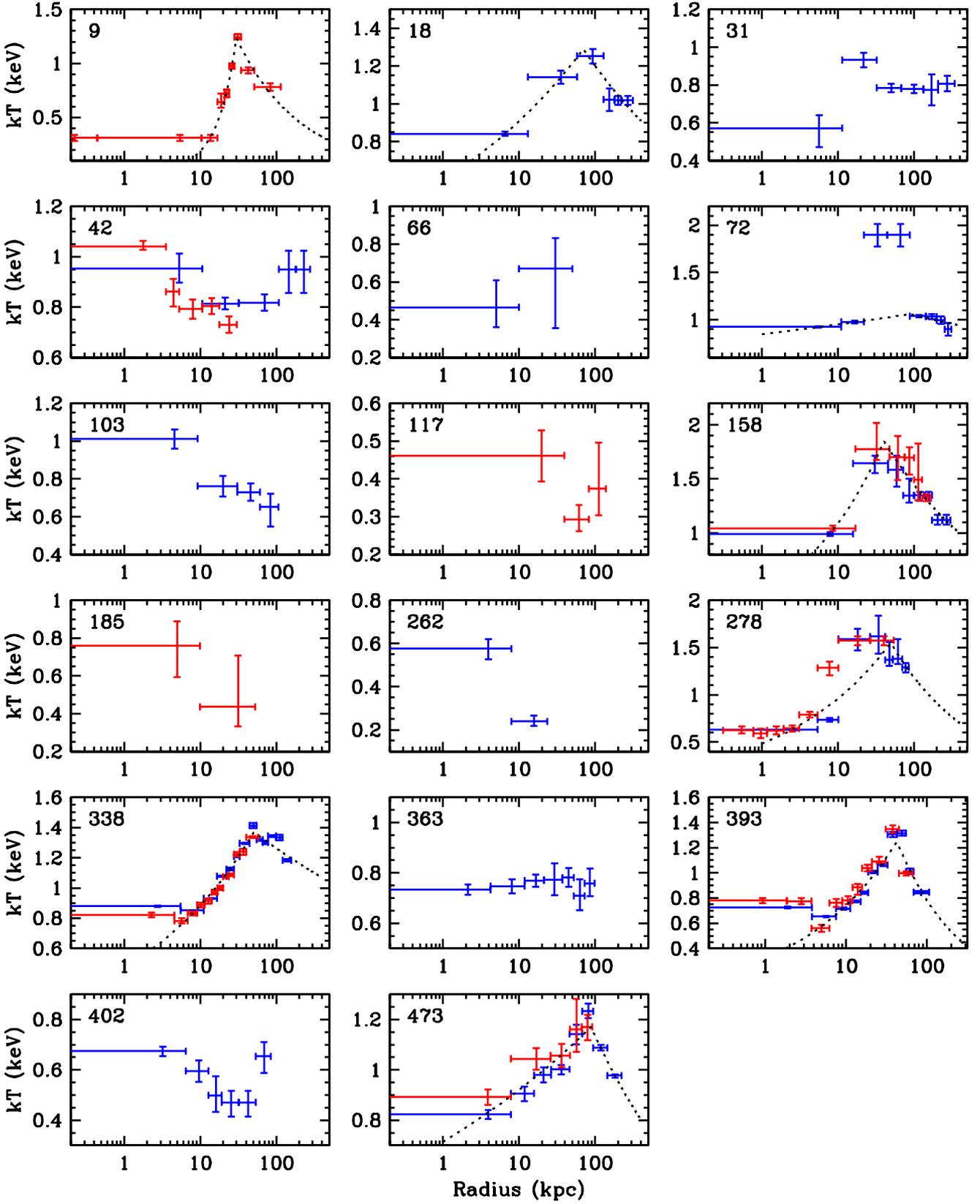}
\caption{\label{fig:kTprofs} Temperature profiles for the groups of the high-richness subsample for which at least two gas temperatures can be measured. Red profiles are derived from \chandra\ observations, blue profiles from \xmms. Group number is indicated in the top left corner of each plot. Deprojected profiles are shown for all systems with $>$3 bins. Dotted lines indicate the broken powerlaw fits used to determine the system temperatures of the cool-core groups.}
\end{figure*}

Where possible, we performed deprojected fits to the full spectral profile (using the \textsc{xspec} PROJCT model), again allowing a powerlaw component in the central bins where necessary. The deprojection approach was similar to that used in \citep{OSullivanetal10}, but in some systems with lower signal-to-noise spectra we tied the abundance values of adjacent radial bins to suppress unphysical bin-to-bin variations. Figure~\ref{fig:kTprofs} shows radial temperature profiles for every group with sufficient numbers of counts to allow two or more spectra to be extracted. In most cases these are deprojected profiles. Projected profiles are shown for the datasets with the lowest signal-to-noise ratios; those with only two or three spectral bins (LGG~66, 117, 185, 262), and the \chandra\ profile of LGG~42 (NGC~777).

\subsection{System temperature and Mass}
\label{sec:mass}
Table~\ref{tab:mass} lists our estimates of the mass within a volume in which the overdensity is 500 times the critical density of the universe (M$_{500}$), and the corresponding fiducial radius (R$_{500}$) for each group, estimated from the X-ray temperature. Owing to the wide range of data quality, we elected to use the mass-temperature relation to estimate the mass of each group. To determine the characteristic system temperature (and abundance) of each system, we follow a tiered approach similar to that of \citet{Hudsonetal10} and \citet{Eckmilleretal11}. We consider systems detected to $<$65~kpc to be dominated by galactic gas. If only a single temperature measurement can be made, we take this as the system temperature; if two spectral bins can be fitted, we take the system temperature to be that of the outer bin, to avoid any emission from the stellar population.

In group-scale systems, we determine whether the temperature profile is best described as flat, centrally peaked, or cool-core. For flat-profile systems (LGG~117, 363) and those which are centrally peaked (LGG~103, 402), we fit a single projected temperature and abundance across all spectral bins, freezing the background components at their best fit values when using \xmms\ \textsc{esas}. 

For systems with a central temperature decline (cool-core systems) we wish to exclude the cooling region, where the temperature is not determined by the gravitational potential. We therefore fit the temperature profiles for each system with a broken powerlaw model:

\begin{equation}
T(r)=\left\{
  \begin{array}{lr}
    T_a \times (r/R_{\rm break})^{i1} & \textrm{for $r<R_{\rm break}$} \\
    T_a \times (r/R_{\rm break})^{i2} & \textrm{for $r\ge R_{\rm break}$}
  \end{array}
  \right.
\end{equation}

where $T_a$ is a normalization factor, $i1$ and $i2$ are the slopes of the two parts of the broken powerlaw, $r$ is radius and $R_{\rm break}$ is  the radius of the turnover in the temperature profile, beyond which the temperature is either flat or declining. We define the system temperature of these systems by fitting a single temperature and abundance to all bins outside $R_{\rm break}$, excluding the bin in which the break radius is found. Where \xmms\ observations are available, we use them in preference to the \chandra\ data, since in many cases the \chandra\ field of view does not extend far beyond the turnover radius.

The broken powerlaw fits are shown as dotted lines in Figure~\ref{fig:kTprofs}. The system temperature and abundances, break radii of the broken powerlaw fits, and resulting fiducial mass and radius estimates, M$_{500}$ and R$_{500}$, are listed in Table~\ref{tab:mass}. Note that in the case of LGG~31, where both cool core and temperature peak are represented by only single bins, the uncertainty of the break radius is taken to be the radial limits of the bin in which the peak occurs.

\begin{table*}
\caption{\label{tab:mass} System temperatures and abundances, temperature profile break radii, R$_{500}$ and the total mass within that radius, for those groups where emission from a thermal plasma is detected. Columns 2 and 3 indicate the number of radial bins used in the \chandra\ and \xmms\ temperature profiles. For systems with only one or two measured temperatures, T$_{sys}$ and Z$_{sys}$ are derived directly from a spectral fit. Systems with no break radius have either flat or declining temperature profiles. R$_{500}$ and M$_{500}$ are estimated from T$_{sys}$ using the \citet{Sunetal09} scaling relations for their Tier 1+2 groups.}
\begin{center}
\begin{tabular}{lccccccc}
\hline
LGG & \multicolumn{2}{c}{No. of bins} & T$_{sys}$ & Z$_{sys}$   & $R_{\rm break}$ &  M$_{500}$        & R$_{500}$ \\
    & Chandra & XMM    & (keV)    & (Z$_\odot$) & (kpc)         & (10$^{13}$\Msol) & (kpc)     \\
\hline\\[-3mm]
9   & 9 & - & 0.88$^{+0.02}_{-0.03}$ & 0.68$\pm$0.27            & 29.9$\pm$0.5            & 2.33$^{+0.09}_{-0.13}$ & 432$^{+5}_{-8}$ \\[+0.5mm]
18  & - & 6 & 0.98$\pm$0.02          & 0.42$^{+0.12}_{-0.08}$   & 70.0$^{+11.5}_{-10.2}$    & 2.78$^{+0.10}_{-0.09}$ & 458$\pm$5 \\[+0.5mm]
31  & - & 6 & 0.79$\pm$0.01          & 0.38$\pm$0.04          & 21.7$\pm$10.4                    & 1.94$\pm$0.04          & 406$\pm$3 \\[+0.5mm]
42  & 5 & 5 & 0.89$\pm$0.02          & 0.63$^{+0.15}_{-0.11}$   & -                       & 2.37$\pm$0.09          & 434$\pm$5 \\[+0.5mm]
58  & - & 1 & 0.41$^{+0.04}_{-0.03}$ & 0.06$^{+0.03}_{-0.02}$ & -                       & 0.65$^{+0.11}_{-0.08}$ & 282$^{+15}_{-12}$ \\[+0.5mm]
66  & - & 2 & 0.49$^{+0.15}_{-0.11}$ & $>$0.29 & -                       & 0.87$^{+0.49}_{-0.30}$ & 312$^{+50}_{-41}$ \\[+0.5mm]
72  & - & 8 & 1.02$\pm$0.01          & 0.28$\pm$0.02          & 88.5$^{+134.1}_{-28.1}$ & 2.97$^{+0.15}_{-0.14}$ & 468$\pm$8 \\[+0.5mm]
103 & - & 4 & 0.74$\pm$0.03          & 0.42$^{+0.22}_{-0.12}$ & -                       & 1.74$\pm$0.12          & 392$\pm$9 \\[+0.5mm]
117 & 3 & - & 0.37$^{+0.08}_{-0.06}$ & 0.03$^{+0.03}_{-0.02}$ & -                       & 0.55$^{+0.21}_{-0.14}$ & 267$^{+31}_{-25}$ \\[+0.5mm]
158 & 6 & 8 & 1.25$\pm$0.01          & 0.24$\pm$0.02          & 40.3$^{+5.0}_{-4.5}$    & 4.18$\pm$0.06          & 525$\pm$2 \\[+0.5mm]
185 & 2 & - & 0.44$^{+0.27}_{-0.11}$ & 0.03$^{+0.10}_{-0.03}$ & -                       & 0.73$^{+0.90}_{-0.28}$ & 294$^{+90}_{-44}$ \\[+0.5mm]
262 & - & 2 & 0.56$^{+0.04}_{-0.07}$ & 0.32$^{+0.42}_{-0.13}$ & -                       & 1.09$^{+0.13}_{-0.22}$ & 336$^{+13}_{-24}$ \\[+0.5mm]
276 & - & 1 & 0.71$^{+0.08}_{-0.10}$ & 0.11$^{+0.08}_{-0.05}$ & - & 1.63$^{+0.32}_{-0.36}$ & 384$^{+23}_{-31}$ \\[+0.5mm]
278 & 8 & 7 & 1.36$^{+0.03}_{-0.02}$ & 0.23$\pm$0.02          & 49.4$^{+3.3}_{-3.6}$    & 4.83$^{+0.18}_{-0.12}$ & 552$^{+7}_{-5}$ \\[+0.5mm]
338 & 12 & 13 & 1.28$\pm$0.01          & 0.39$\pm$0.01          & 50.6$\pm$0.7            & 4.36$\pm$0.06          & 533$\pm$2 \\[+0.5mm]
363 & - & 8 & 0.72$\pm$0.01          & 0.17$\pm$0.01          & -                       & 1.67$\pm$0.04          & 387$\pm$3 \\[+0.5mm]
393 & 10 & 12 & 0.95$\pm$0.01          & 0.27$\pm$0.12          & 42.7$^{+0.80}_{-0.83}$  & 2.65$\pm$0.05          & 452$\pm$3 \\[+0.5mm]
402 & - & 6 & 0.59$\pm$0.01          & 0.38$^{+0.06}_{-0.05}$ & -                       & 1.20$\pm$0.03          & 346$\pm$3 \\[+0.5mm]
421 & - & 1 & 0.29$^{+0.08}_{-0.06}$ & $<$0.18                & - & 0.36$^{+0.18}_{-0.12}$ & 233$^{+34}_{-28}$ \\[+0.5mm]
473 & 5 & 8 & 1.00$\pm$0.01          & 0.54$\pm$0.03          & 88.5$\pm$5.8            & 2.88$\pm$0.05          & 464$\pm$3 \\
\hline
\end{tabular}
\end{center}
\end{table*}

We use the relations for Tier 1+2 groups from \citet{Sunetal09} to estimate M$_{500}$ and R$_{500}$ from the system temperatures. Sun et al. find very similar relations for groups alone, or groups plus clusters, and these relations are similar to those found for groups and HIFLUGCS clusters by \citet{Eckmilleretal11}. The scalings should therefore be reliable across a wide mass range, at least for those systems with sufficiently well-determined system temperatures. 

\begin{figure*}
\includegraphics[width=\textwidth,bb=40 380 565 760,clip=]{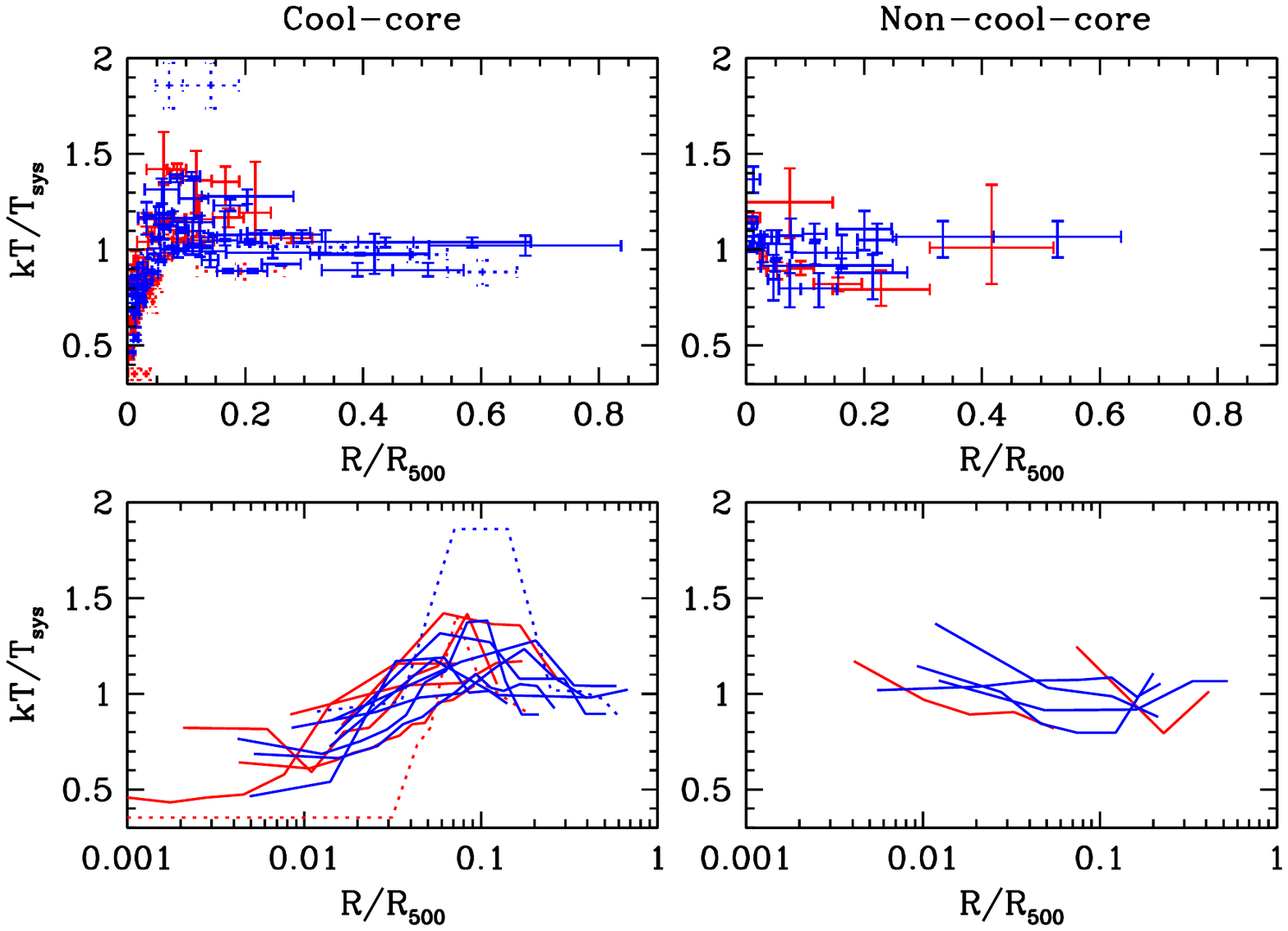}
\caption{\label{fig:kTscaled}Scaled temperature profiles (kT/T$_{sys}$ vs. R/R$_{500}$) for the group-scale X-ray haloes in the high-richness subsample. The upper row shows data points with errors on a linear radius scale, while the lower row shows lines linking the best fitting temperatures, on a log radius scale. The two columns show cool core and non-cool-core groups. Chandra profiles are shown in red, XMM-Newton in blue. Dashed lines indicate LGG~72 and LGG~9, which we believe to be affected by recent shock heating, and whose profiles deviate from the trends followed by the other NCC groups.}
\end{figure*}

Figure~\ref{fig:kTscaled} shows the radial temperature profiles of the groups scaled by T$_{sys}$ and R$_{500}$. For the systems with central temperature declines, the size of the cooling region varies over the range $R_{\rm break}$=0.05-0.2$\times$R$_{500}$, with a mean value of 0.11$\times$R$_{500}$.
\citet{RasmussenPonman07} found that, for a sample of groups observed with \chandra\ most of which had large well-resolved cool cores, the size of the CC was $\sim$0.1$\times$R$_{500}$. \citet{Sandersonetal06} find a somewhat larger scaled size, 0.1-0.2$\times$R$_{500}$ for their sample, which is dominated by more massive clusters.

\subsection{Entropy profiles}

\begin{figure*}
\includegraphics[width=\textwidth,bb=35 440 560 740,clip=]{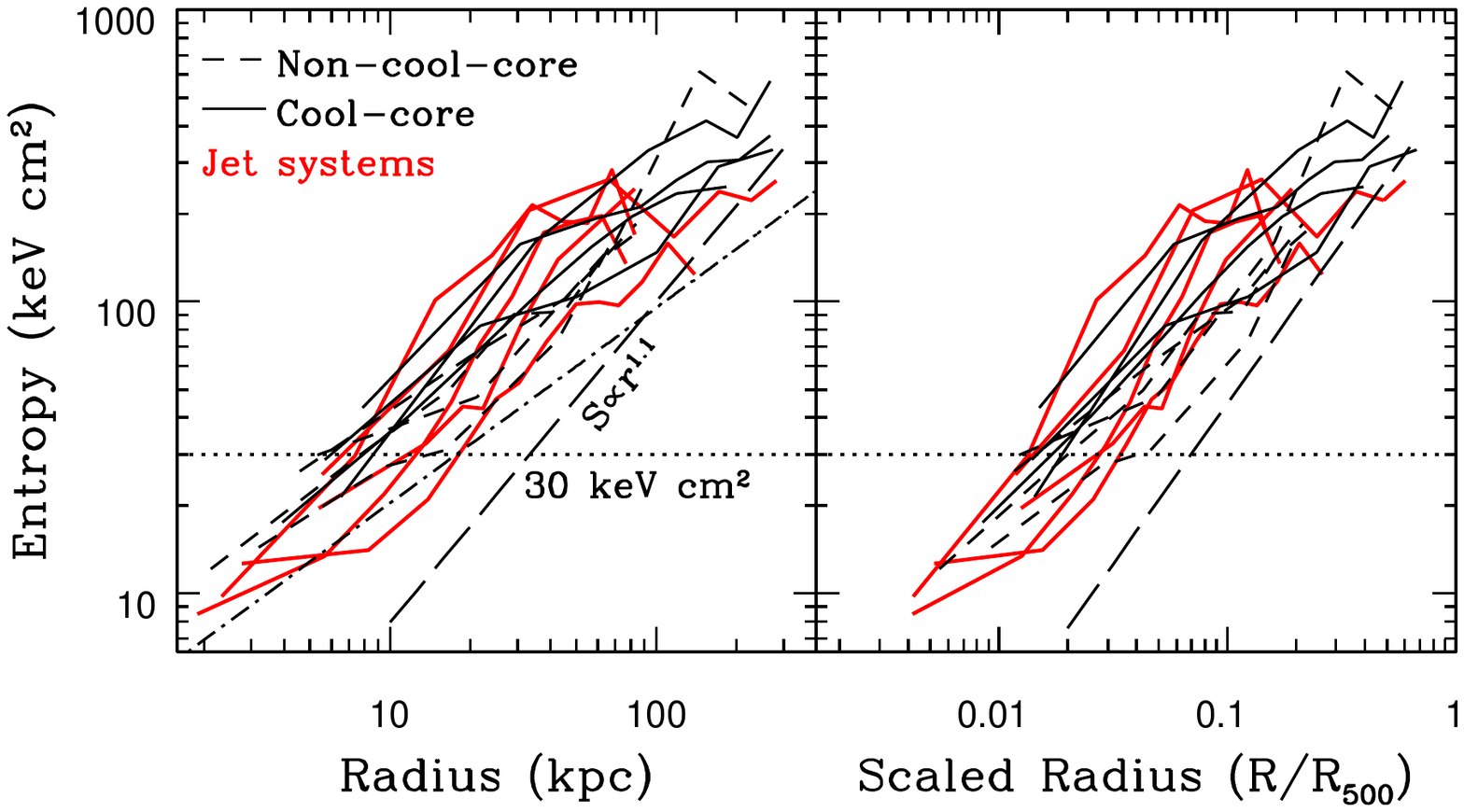}
\caption{\label{fig:entropy} Entropy profiles in unscaled (\textit{left}) and scaled (\textit{right}) radius for the groups of the high-richness subsample. Solid lines indicate cool-core systems, short-dashed lines non-cool-core, and red lines indicate systems whose dominant early-type galaxy hosts radio jets. The straight long-dashed line shows an r$^{1.1}$ scaling, which most of the profiles approximate in their outer parts. The dot-dashed line indicates the best-fitting power law entropy profile from \protect\citet{Panagouliaetal14}. The dotted line indicates the 30\kevcmsq\ entropy level below which ionized gas and other cooling products are often observed in galaxy clusters.}
\end{figure*}

We use the definition of entropy commonly adopted in X-ray astronomy, K = $kT/n_{e}^{2/3}$, where $kT$ is the temperature of the gas in keV and $n_e$ the electron number density in cm$^{-3}$. For those systems where we are able to fit deprojected spectral profiles, we calculate entropy directly from the fitted temperature and electron number density profiles. These resulting entropy profiles are shown in Figure~\ref{fig:entropy}.

\citet{Cavagnoloetal09} show that for the majority of the clusters and groups in their ACCEPT sample, the entropy profile can be described by a power-law plus a minimum `floor' level, K$_0$. Fitting our deprojected profiles with this model, we find that 5 of the 13 systems are best described by models including an entropy floor, with values of 5-25\kevcmsq. The remaining systems have profiles consistent with a simple powerlaw. If we use the entropy value in the innermost bins of the 13 profiles as an estimate of K$_0$ we find that they cover the range $\sim$2-50\kevcmsq. Only one group has K$_0>30$\kevcmsq, LGG~158 / NGC~2563. Its central bin extends beyond 10~kpc, so it is possible that we may miss a fall to lower entropies on smaller scales. We note that \citet{Panagouliaetal14} and other authors have argued that the entropy floor is an artefact caused by poor resolution in the central parts of spectral profiles. This may be the case in our systems, since we have not optimized the \chandra\ profiles for maximum resolution, and the resolution of the \xmms\ profiles are limited by the size of the PSF. 

From self-similar scaling, entropy profiles are expected to rise with radius following an R$^{1.1}$ relation. However, \citet{Panagouliaetal14} report that for a volume limited sample of groups and clusters, the entropy profile within $\sim$100~kpc of the system core deviates from this relation, flattening to a slope of R$^{0.7}$. Figure~\ref{fig:entropy} shows that the behaviour of our CLoGS groups is generally consistent with that found by Panagoulia et al., with a central slope consistent with their best-fit relation, and some indication of steepening to R$^{1.1}$ at large radii. The scatter among our profiles is fairly large, a factor $\sim$3 at 10~kpc. This is also consistent with the Panagoulia et al. sample.

\subsection{Cooling time profiles}
\begin{figure*}
\includegraphics[width=\textwidth,bb=35 440 560 740,clip=]{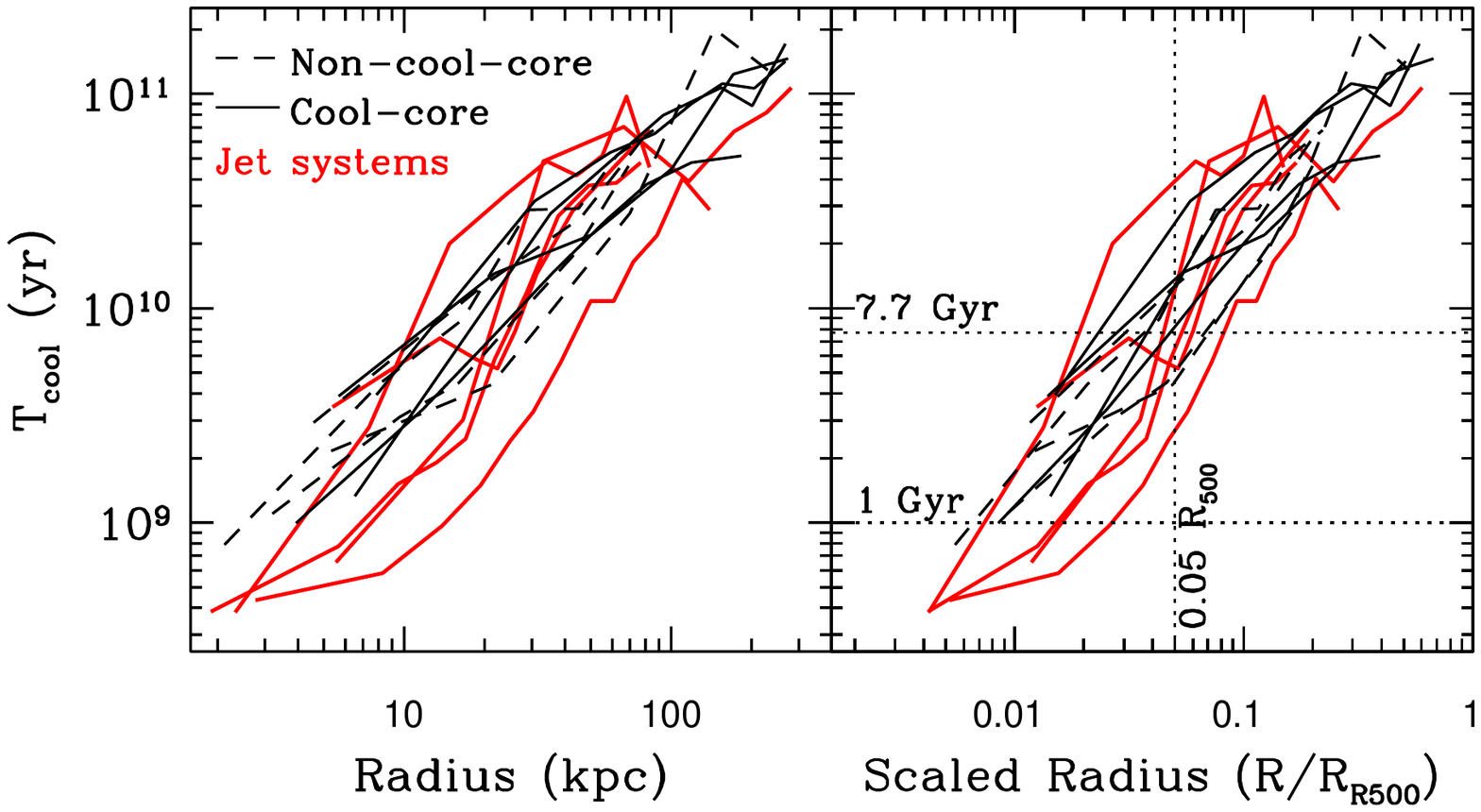}
\caption{\label{fig:tcool} Cooling time profiles in unscaled (\textit{left}) and scaled (\textit{right}) radius for the groups of the high-richness subsample. Solid lines indicate cool-core systems, short-dashed lines non-cool-core, and red lines indicate systems whose dominant early-type galaxy hosts radio jets. The dotted lines indicate 0.05R$_{500}$, and the 1 and 7.7~Gyr cooling time thresholds used by \citet{Hudsonetal10} to define their cool-core classification.}
\end{figure*}

We estimate the isobaric cooling time of the gas as the time required for the gas to lose its thermal energy if it were to continue to radiate at its current rate, approximated as

\begin{equation}
t_{cool} = 5.076\times 10^{-17} \times \frac{5kTn_eV\mu_e}{2\mu L_x},
\end{equation}

where the units of $t_{cool}$ are years, $V$ is the volume of the gas in cm$^{3}$, $L_X$ its bolometric luminosity in \ergps, $\rho$ is the electron number density in \pcmcu\ and $\mu$ and $\mu_e$ are the mean molecular weight (0.593) and the mean mass per electron (1.167) in the gas, respectively. All measured parameters are determined from deprojected fits, and the cooling time is calculated in each shell. The factor of 5/2 allows for work done on the shell as it cools at constant pressure. Figure~\ref{fig:tcool} shows the cooling time profiles, with line colour and style indicating CC and NCC systems, and systems with central radio jet sources.

Considering our cooling time profiles as a sample, we see a fairly large scatter in the profiles, a factor $\sim$10 at 10~kpc. This is comparable to the range of values found by \citet{Panagouliaetal14} for their sample of groups and clusters. Taking the innermost bin as a measure of the central cooling time (CCT), we find $t_{cool}$ in the range $\sim$0.4-4~Gyr, suggesting that all systems will be significantly affected by radiative cooling.

\section{Discussion}
\label{sec:disc}

\subsection{Detection fraction}

As a simple first step, we consider the fraction of our optically-selected groups detected in the X-ray band. Of the 26 groups in the high-richness subsample, 14 ($\sim$54\%) are found to possess extended, group-scale gaseous haloes, with a further 3 ($\sim$12\%) hosting more compact galaxy-scale haloes associated with the dominant early-type galaxy. The remaining 9 groups contain little or no hot gas, at least with properties detectable by our observations.

Comparison of detection fraction with other samples is not straightforward; the details of the optical selection process, the mass and redshift range targeted and the X-ray data available all influence detection efficiency. Among nearby samples, our detection rate is comparable to that of the handful of groups observed in the XMM/IMACS sample \citep[$XI$, 50\%][]{Rasmussenetal06b} and somewhat greater than that for the X-ZENS sample \citep[21\%][]{Miniatietal16}, but the latter survey focused on the smallest groups, with masses in the range 1-5$\times$10$^{13}$\Msol, equivalent to temperatures of $\sim$0.4-0.9~keV. Our detection rate is smaller than the 80\% achieved by \citet{Pearsonetal16} for a sample of groups drawn from the Galaxy And Mass Assembly (GAMA) survey. Their sample includes higher mass groups than ours (kT=0.6-2.8~keV), but the key to their high detection rate appears to be the very high quality optical data available from GAMA, which supports the use of detailed substructure tests to exclude unrelaxed systems. The Pearson et al. detection rate is actually superior to the 70\% achieved by \citet{Baloghetal11} for a sample of low-mass clusters (3-6$\times$10$^{14}$\Msol, equivalent to $\sim$3-5~keV).  

\begin{figure}
\includegraphics[width=\columnwidth,bb=25 210 560 740, clip=]{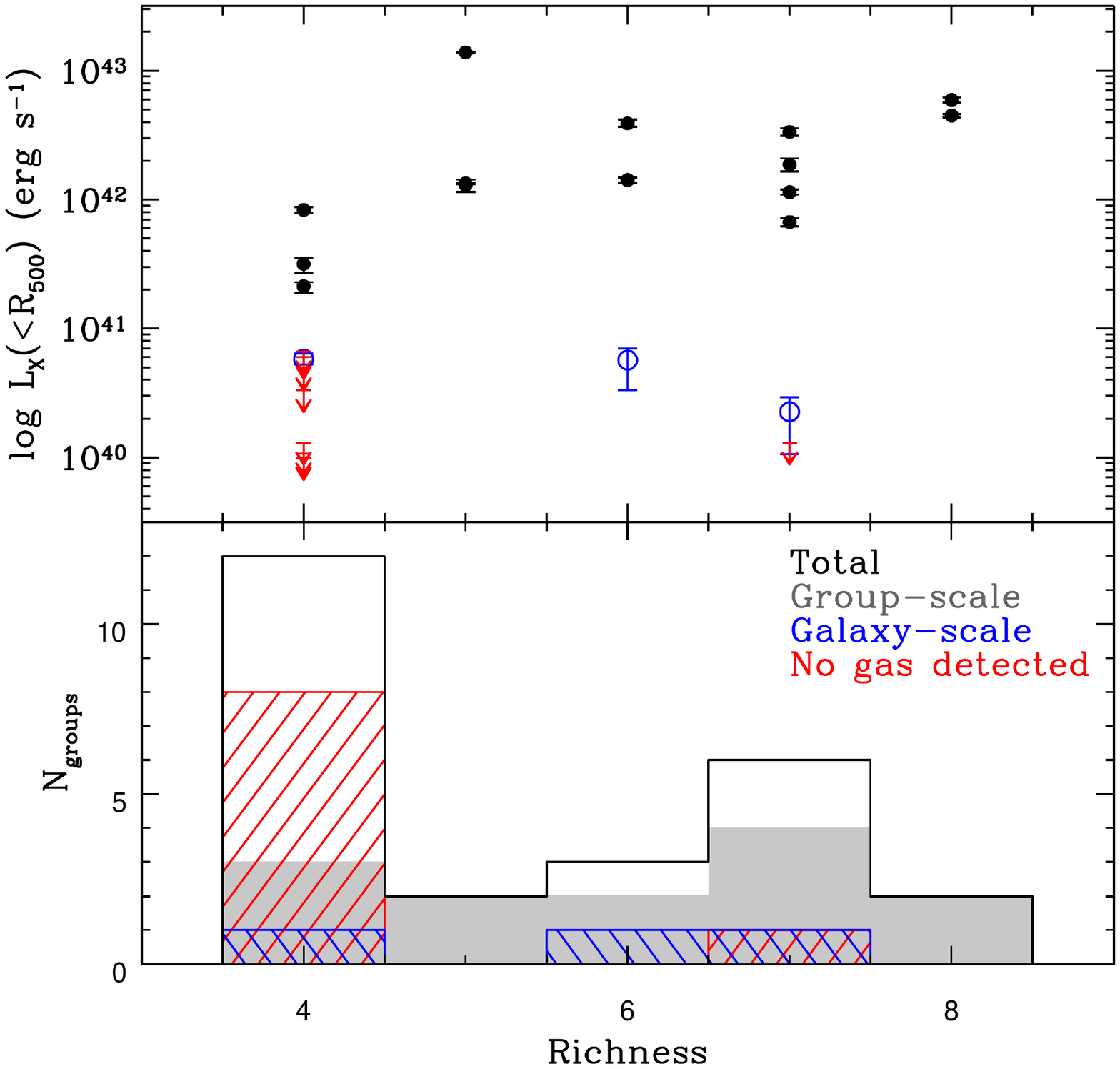}
\vspace{-5mm}
\caption{\label{fig:LxR}Number of groups detected and 0.5-7~keV X-ray luminosity within R$_{500}$ vs. optical richness ($R$) for the high-richness sample. In the upper panel black points indicate X-ray bright groups with emission extending $>$65~kpc, open blue circles galaxy scale haloes (extent $<$65~kpc) and red arrows 3$\sigma$ upper limits. In the lower panel the black histogram shows the total number of optically-selected groups, while the grey, blue and red shaded regions indicate group-scale emission, galaxy-scale and non-detections.}
\end{figure}

Figure~\ref{fig:LxR} shows a histogram of the numbers of X-ray detections against our optical richness estimator $R$, and a comparison with X-ray luminosity. The most obvious result is that 8 of the 9 gas-poor groups are in the lowest richness band, $R$=4. The systems with group- or galaxy-scale haloes are spread fairly evenly across the richness range, with galaxy-scale haloes seen in apparently quite rich ($R$=6-7) systems. There is a weak indication (2$\sigma$ significance) of a mild trend for higher X-ray luminosity with higher richness among the group-scale systems, but this is largely driven by the low values in the $R$=4, and high values in the $R$=8 bins. Our most luminous group (LGG~338 / NGC~5044) has $R$=5. The galaxy-scale haloes have uniformly low luminosities, with a luminosity of 10$^{41}$\ergps\ providing a clean discriminator between group and galaxy-scale haloes.

Our sample is quite dynamically active. Among the X-ray bright systems with group-scale haloes we have two ongoing mergers (LGGs~72 and 473) and two systems with sloshing features indicating recent gravitational interactions (LGG~338 and 393). Considering only the group-scale haloes where the indicators of such events are detectable, this suggests that $\sim$30\% of our sample has undergone a significant interaction within the past few hundred Myr.

AGN are detected via their radio emission in all but two (92\%) of the dominant early-type galaxies. The properties of these radio sources are described in detail by Kolokythas et al. (in prep.), but we summarize their classification here. We define the morphology of the sources based on their appearance in our GMRT observations, NVSS and FIRST survey data, or in some cases based on the literature. Sources which are unresolved at all frequencies are classed as \textit{point-like}. Those which are clearly extended are classified either as \textit{jets}, if the extension is linear or if lobes or plumes are observed, or as \textit{diffuse} if the morphology is amorphous and has no preferred axis. Jet sources are further split into large- and small-scale jets depending on whether they extend $>$20~kpc. This division is intended to distinguish between jets which primarily affect the gas in the central galaxy, and those which extend out into the IGM. In two groups (LGG~80 and LGG~338) previous studies identified large-scale jet/lobe systems, but showed that these structures were produced by past AGN outbursts, and are now passively aging. We refer to these as \textit{remnant} systems, to distinguish them from cases where jets appear to be currently active. We generally class these remnant sources with the currently active jets in our later analysis. The majority of the detections are point-like radio sources, with jet activity (past or present) found in only 6 of the 26 groups ($\sim$23\%). Considering only the systems with group-scale gas haloes, 5 of the 14 host jets ($\sim$36\%), suggesting a duty cycle of approximately one third. We consider the relation between AGN and gas properties in Section~\ref{sec:AGN}.

\subsection{Luminosity--temperature comparison}
\label{sec:LT}

\begin{figure}
\includegraphics[width=\columnwidth,bb=35 300 560 740, clip=]{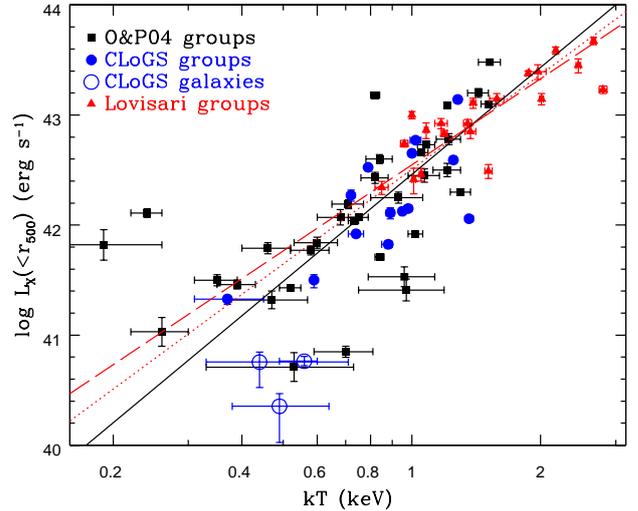}
\vspace{-5mm}
\caption{\label{fig:LT}0.5-7~keV Luminosity within R$_{500}$ versus system temperature for the detected high-richness CLoGS groups (filled blue circles) and galaxies (open blue circles), compared with the samples of \protect\citet{OsmondPonman04} (black squares) and \protect\citet{Lovisarietal15} (red triangles). Lines indicate luminosity--temperature relations for Osmond \& Ponman groups+clusters (solid black), Lovisari groups + HIFLUGCS clusters (red dashed) and Lovisari completeness-corrected groups (red dotted).}
\end{figure}

Figure~\ref{fig:LT} shows the luminosity and temperature of CLoGS high-richness groups and galaxies compared with the samples from \citet{OsmondPonman04} and \citet{Lovisarietal15}. The Lovisari sample is a statistically complete, X-ray flux-limited sample selected from RASS and observed with \xmms\, while the Osmond \& Ponman sample is drawn from the \rosat\ PSPC archive of pointed observations. The Lovisari sample probes a somewhat higher temperature band, $\sim$0.8-3~keV, compared to $\sim$0.2-1.5~keV for Osmond \& Ponman, and $\sim$0.3-1.4~keV for the CLoGS. In general we consider the Lovisari sample as best representing the underlying luminosity--temperature relation for relaxed groups and poor clusters, while the Osmond \& Ponman sample includes less relaxed systems and gives a better idea of the scatter across the population.

The CLoGS groups have a comparable distribution to the other two samples, and fall close to the fitted luminosity--temperature relations. With only 14 group--scale systems, our sample cannot strongly constrain the luminosity--temperature relation, but for comparison we performed linear regression fits to our data using the bivariate correlated error and intrinsic scatter (BCES) algorithm \citep{AkritasBershady96}. We found a normalization at 1~keV which is equal (within uncertainties) to the Osmond \& Ponman and Lovisari et al. relations (42.50$\pm$0.11). The slope of the relation was poorly constrained and dependent on the regression method used (e.g., orthogonal vs. bisector) but generally comparable to the results of prior studies. A more constrained fit can only be achieved by expanding our sample; observations of the low-richness subsample may allow this in future.

The dynamically active systems all fall relatively close to the relation. The group furthest from the relation is LGG~278 / NGC~4261, whose luminosity ($\sim$1.1$\times$10$^{42}$\ergps) is rather low for its temperature (1.36~keV). Osmond \& Ponman report a slightly higher luminosity and lower temperature, but \rosat\ PSPC data are incapable of resolving the complex structures in the core of the group \citep[including an X-ray bright AGN and jets, e.g.,][]{Worralletal10} and some degree of luminosity overestimation must be expected. 

Of the three CLoGS systems in which only galaxy--scale thermal emission is detected, LGG~262 falls closest to the luminosity--temperature relation, at its extreme low end. LGG~66 falls further from it, with a luminosity $\sim$1~dex below that expected for its temperature. Gas temperature in these small systems is more likely to be affected by AGN, supernovae (SNe) and stellar winds, so such deviations are not unexpected. These systems are similar to the Osmond \& Ponman H-sample of galaxy-scale haloes.

\subsection{High entropy groups}
One goal of the CLoGS project is to search for groups with properties which differ from those of the population observed in X-ray selected surveys. The OverWhelmingly Large Simulations project \citep[OWLS,][]{Schayeetal10} uses simulations of cosmological volumes which incorporate radiative cooling and both stellar and AGN feedback to examine the formation history of groups. The simulations fairly accurately reproduce many of the observed properties of groups, including the radial entropy distribution \citep{McCarthyetal10, McCarthyetal11}. However, they also predict a significant number of groups with higher central entropies than have been observed. These high entropies are caused by AGN feedback both preceding and early in the process of group formation. Quasar-driven winds in the simulations heat and expel low-entropy gas from progenitor haloes at moderate redshift ($z$=2-4), leaving only higher entropy material to contribute to the build up of the group. The question of whether quasar winds can effectively drive gas out of haloes is still open, but there is considerable evidence for groups being baryon deficient \citep[see e.g., Fig.~8 of][]{Lichenetal16}, and a number of studies show that in this redshift range, quasar hosts have typical masses of a few 10$^{12}$\Msol\ \citep[e.g.,][]{Whiteetal12,Shenetal13,Wangetal15}, or perhaps as much as 1-2$\times$10$^{13}$\Msol\ \citep{Richardsonetal12,TrainorSteidel12,Verdieretal16}. It therefore seems plausible that the progenitor haloes which merged to form nearby groups could have hosted powerful quasars at $z$=2-4.

Such high-entropy groups would lack the bright centrally-peaked cool cores by which groups were typically identified in RASS, and we might therefore expect them to be missing from RASS-based X-ray selected samples. Optically selected samples such as CLoGS are more likely to find such groups if they exist in the local universe. 

\begin{figure}
\includegraphics[width=\columnwidth,bb=25 260 560 740, clip=]{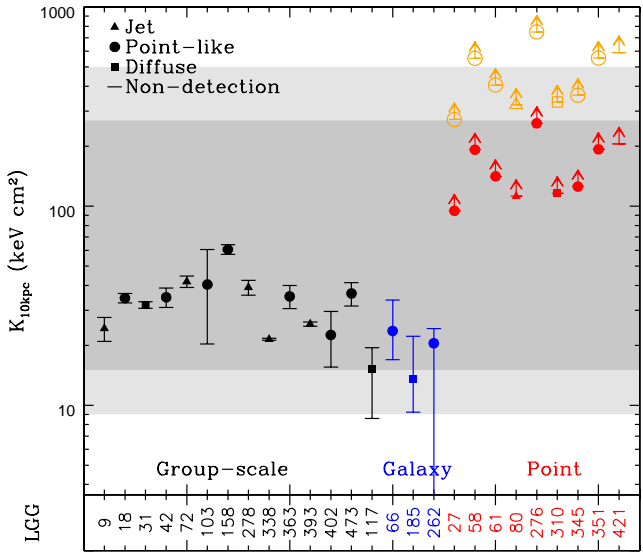}
\vspace{-5mm}
\caption{\label{fig:K10} Entropy at 10~kpc for each high-richness CLoGS group, or the 3$\sigma$ lower limit on entropy for systems where no extended gas component is detected. Colours indicate systems with group-scale (black) or galaxy-scale haloes (blue), or systems where no extended gas was detected (red and orange, representing limits assuming 0.5 keV and 1 keV haloes respectively). Symbols indicate the radio morphology of the AGN in the dominant early-type galaxy (from Kolokythas et al., in prep.) either point-like (circles), diffuse (squares), or jet-lobe (stars). Filled symbols are used where the system is detected, or for lower limits assuming a 0.5~keV halo (red), while open orange symbols show the lower limits for 1~keV haloes. The dark (light) grey shaded region indicates the 1$\sigma$ (2$\sigma$) range of entropies expected from the OWLS simulations.}
\end{figure}

Figure~\ref{fig:K10} shows the entropy at 10~kpc (K$_{10}$) predicted by the OWLS simulations \citep[][and references therein]{Pearsonetal16}, and measurements or lower limits for our CLoGS high-richness groups. Where deprojection was possible, entropies are taken directly from the powerlaw+constant fits to the deprojected entropy profiles. Where gas emission extends to 10~kpc but only a projected temperature profile (or single temperature) could be measured, K$_{10}$ is determined from the projected temperature in the bin covering 10~kpc, and a density profile derived from the $\beta$-model surface brightness model, normalized to the emission measure in that bin. Among the systems with only projected temperatures, only LGG~117 has a group-scale halo. The presence of a cool core has a significant effect on K$_{10}$. For our CC systems, the ratio between the temperature at 10~kpc and the system temperature is $\sim$0.7, with individual systems having ratio $\sim$0.3-0.9. This emphasizes the importance of studying nearby groups where temperature structure can be resolved.

For systems where gas emission was not detected, or does not extend to 10~kpc, we estimate an lower limit on K$_{10}$. We assume a density profile with $\beta$=0.4 and core radius r$_c$=0.1$\times$R$_{500}$ for every system, and normalize the density of each group so that the emission measure matches that expected at the 3$\sigma$ upper limit on the luminosity, for system temperatures of 0.5 and 1~keV. This provides 3$\sigma$ lower limits on the entropy profile, from which we determine limits on K$_{10}$. The choice of temperature leads to a factor $\sim$3 difference in lower limit, with higher temperature groups detectable to higher entropies. Increasing core radius to 0.2$\times$R$_{500}$ (To match Pearson et al.) increases K$_{10}$ by $\sim$12\%.

From Figure~\ref{fig:K10} we can see that all detected groups have K$_{10}$$\simeq$10-60\kevcmsq, comparable to previous X-ray selected samples. We detect no group or galaxy-scale haloes in the upper part of the range expected from the OWLS simulations. Our lower limits for 0.5~keV groups are in the range $>$90-260\kevcmsq, suggesting that a large fraction of the predicted high-entropy groups would be detectable in our observations. For 1~keV groups the lower limits are all higher than the upper bound of the 1$\sigma$ range, and four systems have lower limits above the upper bound of the 2$\sigma$ range. A system temperature of 1~keV corresponds to M$_{500}$ of $\sim$3$\times$10$^{13}$\Msol, so this strongly suggests that, at least for medium-mass groups, CLoGS is in principle capable of detecting the high-entropy IGM predicted by OWLS.

A high-entropy IGM provides one possible explanation for those groups in which no group-scale X-ray halo is detected, particularly if those groups have low masses and system temperatures. However, there are other possibilities. Without the evidence of a hot IGM, we cannot be sure that the groups are virialized systems; some or all of them may still be in the process of collapsing. A related issue is that of lower system temperatures. For system temperatures below 0.5~keV, our data become increasingly insensitive, e.g., at 0.1~keV, the lower limit on K$_{10}$ would be $\sim$3-15\kevcmsq. However, systems with such low temperatures should not be considered as groups. A system temperature of 0.5~keV corresponds to a total mass $\sim$10$^{13}$\Msol. Less massive systems are typically considered as individual galaxies rather than groups, with different halo properties, and this is supported by our own findings for the galaxy-scale systems. At 0.1~keV, the expected virial radius of the BGE in many of our groups would be too small to overlap the other supposed group members, and the total mass would in fact be smaller than the BGE stellar mass. Very few early-type galaxies are found to have temperatures below $\sim$0.3~keV, particularly among those with large K-band luminosities typical of group-dominant galaxies \citep[e.g.,][]{KimFabbiano15}. Comparing such low mass systems to the OWLS predictions for groups is incorrect, given the probable differences in their accretion and merging histories. The system temperatures may in fact be below 0.5~keV, but if that is so, the undetected systems are probably not collapsed groups, and cannot be considered as part of this discussion.

Given the number of undetected groups, and our sensitivity to high-entropy gas, it seems likely that if our sample contained a significant number of high entropy groups, we would have detected some of them. This raises the question of whether the predicted high-entropy groups actually exist. Since the mechanism of quasar-mode AGN feedback is poorly understood and occurs on scales below the resolution of simulations, the energy injection from AGN feedback in the OWLS simulations is by necessity relatively simple. Accretion on to the central super-massive black hole in the dominant galaxy of the group produces heating in randomly selected nearby particles. Heating is only allowed to occur when enough accretion energy has built up to heat these particles by 10$^8$~K ($\sim$8 keV). This threshold is necessary to prevent the particles rapidly radiating away the injected energy. It is possible that this formula for energy injection, while allowing many group properties to be reproduced, overestimates the ability of quasar-mode accretion to affect the IGM. Further observations of optically-selected groups, and the extensive X-ray surveys planned in the next decade, should resolve this question.

\subsection{Cool core fraction}
We have defined groups as cool-core or non-cool-core based on their temperature profile, with groups showing a 3$\sigma$ significant decline between their peak and central temperatures classed as cool-core. This has the advantage of simplicity and highlights systems where cooling has a clear effect on the properties of the hot IGM. Table~\ref{tab:morph} lists the cool-core status of our groups.

\begin{table*}
\caption{\label{tab:morph} Gas morphology, central isochoric cooling time, entropy at 10~kpc (K$_{10}$), minimum isochoric cooling time / free fall time ratio (min($t_{c}/t_{ff}$)), cool-core status and radio characteristics of the groups and their dominant galaxies. Lower limits on K$_{10}$ assume a 1~keV halo as described in the text. min($t_{c}/T_{ff}$) values for systems where no deprojected profile was available are marked with an asterisk and were calculated at 10~kpc. Core type indicates either (T profile) our classification of groups as cool-core/non-cool-core based on their temperature profiles, or (Hudson) as strong-, weak- or non-cool-core (SCC, WCC, NCC) based on the scheme of \protect\citep{Hudsonetal10}. Entries in brackets indicate systems where only a projected temperature profile with $<$3 bins is available. For systems where radio sources are associated with the BGE, we indicate whether they are point-like, diffuse, have small-scale jets (lower-case ``jet'') or large-scale jets (upper-case ``JET''). Aging radio structures observed at low frequency and apparently no longer powered by the AGN are listed in the remnant column. }
\begin{center}
\begin{tabular}{lccccccccc}
\hline
LGG & BGE & CCT & K$_{10}$ & min($t_{c}/t_{ff}$) & \multicolumn{2}{c}{Core type} & \multicolumn{2}{c}{Radio Morphology} & Notes \\
    &     & (Gyr) & (keV cm$^{-2}$) & & T profile  & Hudson & Current & Remnant & \\
\hline
9   & NGC~193    & -                       & 24.3$\pm$3.4       & - & CC    & - & JET & - & shock \\
18  & NGC~410    & 0.102$\pm$0.007         & 34.5$\pm$1.9       & 18.2$\pm$1.3 & CC    & SCC & pnt & - & \\
27  & NGC~584    & -                       & $>$272.7            & - & -     & -   & pnt & - & \\
31  & NGC~677    & 2.042$^{+0.639}_{-0.630}$ & 31.9$\pm$1.3        & 47.9$\pm$15.4 & CC    & WCC & diffuse & - & \\[+0.5mm]
42  & NGC~777    & 0.314$^{+0.044}_{-0.039}$ & 34.8$\pm$3.9        & 39.2$\pm$5.2 & NCC   & SCC & pnt & - & \\[+0.5mm]
58  & NGC~940    & -                       & $>$552.9             & - & -     & -   & pnt & - & \\
61  & NGC~924    & -                       & $>$406.1             & - & -     & -   & pnt & - & \\
66  & NGC~978    & 1.421$^{+0.629}_{-1.365}$ & 23.6$^{+10.1}_{-6.7}$ & 80.6$^{+35.7*}_{-77.5}$ & (NCC) & WCC & pnt & - & \\[+0.5mm]
72  & NGC~1060   & 0.132$^{+0.018}_{-0.022}$ & 41.8$\pm$2.7        & 10.4$\pm$1.7 & CC    & SCC & jet & - & merger \\
80  & NGC~1167   & -                       & $>$322.8             & - & -     & -   & jet & JET & \\
103 & NGC~1453   & 0.390$\pm$0.036         & 40.4$\pm$20.1       & 64.4$\pm$6.7 & NCC   & SCC & pnt & - & \\
117 & NGC~1587   & 0.101$^{+0.064}_{-0.038}$ & 15.3$^{+4.2}_{-6.7}$ & 31.6$^{+20.2*}_{-12.2}$ & NCC & SCC & diffuse & - & \\[+0.5mm]
158 & NGC~2563   & 1.954$^{+0.389}_{-0.396}$ & 60.6$\pm$3.4        & 41.0$\pm$8.3 & CC    & WCC & pnt & - \\[+0.5mm]
185 & NGC~3078   & 0.647$^{+0.475}_{-0.444}$ & 13.5$^{+8.7}_{-4.3}$ & 67.9$^{+133.7*}_{-67.9}$& (NCC) & SCC & diffuse & - & \\[+0.5mm]
262 & NGC~4008   & 0.199$^{+1.394}_{-0.033}$ & 20.5$^{+3.8}_{-20.5}$ & 16.9$^{+118.4*}_{-3.1}$ & (NCC) & SCC & pnt & - & \\[+0.5mm]
276 & NGC~4169   & -                       & $>$750.0             & - & -     & -   & pnt & - & compact group\\
278 & NGC~4261   & 0.097$^{+0.017}_{-0.010}$ & 39.1$\pm$3.3        & 15.0$\pm$2.1 & CC    & SCC & JET & - & \\[+0.5mm]
310 & ESO~507-25 & -                       & $>$333.9             & - & -     & -   & diffuse & - & \\
338 & NGC~5044   & 0.232$\pm$0.003         & 21.4$\pm$0.3        & 7.4$\pm$0.4 & CC    & SCC & pnt & JET & sloshing\\
345 & NGC~5084   & -                         & $>$361.1             & - & -     & -   & pnt & - & \\
351 & NGC~5153   & -                       & $>$555.1             & - & -     & -   & - & - & \\[+0.5mm]
363 & NGC~5353   & 0.288$^{+0.037}_{-0.024}$ & 35.2$\pm$4.7        & 32.4$\pm$3.5 & NCC   & SCC & pnt & - & compact group\\[+0.5mm]
393 & NGC~5846   & 0.138$^{+0.002}_{-0.003}$ & 25.6$\pm$0.6        & 14.7$\pm$0.4 & CC    & SCC & jet & - & sloshing\\[+0.5mm]
402 & NGC~5982   & 0.254$^{+0.069}_{-0.428}$ & 22.6$\pm$7.0        & 25.6$\pm$11.5 & NCC   & SCC & pnt & - & \\[+0.5mm]
421 & NGC~6658   & -                       & $>$591.3             & - & -     & -   & - & - & \\
473 & NGC~7619   & 0.135$^{+0.011}_{-0.008}$ & 36.4$\pm$4.9        & 24.5$\pm$1.8 & CC    & SCC & pnt & - & merger \\
\hline
\end{tabular}
\end{center}
\end{table*}

Of the 17 systems where a temperature profile can be measured, 9 show a central cool core (53\%). This rises to 9/14 (64\%) if we insist on group-scale emission and profiles with at least three temperature bins. Our cool-core fraction is apparently smaller than the fraction found in past X-ray selected group samples \citep[e.g., 85\%][]{Dongetal10}, and is comparable with the $\sim$50\% fraction found in clusters \citep[e.g.,][]{Sandersonetal06}. Unfortunately, the small size of the sample makes the result uncertain.

The spatial resolution of our temperature profiles is generally sufficient to identify central temperature declines on scales $\sim$10~kpc, so we may be missing some small scale cool cores \citep[or galaxy coronae,][]{Sunetal07}. These would also have been missed in the great majority of prior surveys.

Alternative methods for defining cool-core status have been suggested, including definitions based on the central entropy and central cooling time. \citet{Hudsonetal10} used the HIFLUGCS cluster sample to try to determine an optimal method for identifying cool core systems, and suggested the central cooling time as the best solution. They found that their sample was clearly separated into three subsets: strong cool cores (SCC) with isochoric central cooling time (CCT, note the factor 3/5 difference from our isobaric measurements) $<$1~Gyr, weak cool cores (WCC) with CCT=1-7.7~Gyr, and non-cool cores with CCT$>$7.7~Gyr. They found that CCT was closely correlated with central entropy and the temperature profile, with SCC systems showing a central temperature decline and central entropy $<$30\kevcmsq. 

\citet{Bharadwajetal14} study the cool core status of the 26 groups and poor clusters of the \citet{Eckmilleretal11} sample. They note that the correlation between CCT and temperature profile fails for this sample, with some CC systems possessing rising central temperature profiles.

\citet{Hudsonetal10} define CCT based on the density at 0.004$\times$R$_{500}$, and mean temperature within 0.048$\times$R$_{500}$. For systems where we have deprojected profiles, we estimate CCT using the temperature of the innermost bin (which may extend beyond 0.048$\times$R$_{500}$) and gas mass and luminosity values extrapolated in to 0.004$\times$R$_{500}$ using the surface brightness profile. For systems where only a projected profile is available, we follow a similar procedure using the gas mass profile derived in our calculation of K$_{10}$. Table~\ref{tab:morph} lists CCT and Hudson cool core classification for our groups. The centre of LGG~9 / NGC~193 is too disturbed for a meaningful surface brightness fit so we exclude it from consideration.

Using the Hudson classification, every group in our sample is identified as CC, with the majority (13/16) SCC. Groups with central temperature peaks (e.g., LGG~42 / NGC~777, LGG~402, NGC~5982) and flat profiles (e.g., LGG~363, NGC~5353) have CCT$<$1~Gyr and would be classed as SCC. This CC fraction is much higher than found for HIFLUGCS (44\% SCC, 28\% WCC, 28\% WCC) or for the Eckmiller sample \citep[50\% SCC, 27\% WCC, 23\% NCC;][]{Bharadwajetal14}. However, the Eckmiller sample covers a broader temperature range than CLoGS, $\sim$0.6-3~keV. If we restrict that sample to match CLoGS (kT$<$1.5~keV) we find that the CC fraction increases dramatically: 8/12 groups in this range (67\%) are SCC and 4/12 (33\%) are WCC, with no NCC systems at all. Varying the temperature threshold, it appears that there is a correlation between system temperature and CC status. 

Such a correlation is understandable given the nature of radiative cooling in the X-ray regime. In clusters, continuum emission dominates, but at temperatures below $\sim$2~keV line emission (primarily from the Fe-L complex) grows increasingly important. For 0.5-1.5~keV groups, line emission is many times more effective at radiating away thermal energy than the continuum, leading to dramatically shorter cooling times. Centrally peaked abundance profiles likely emphasize the effect even further in group cores. Although this does not invalidate cooling time as an indicator of the likelihood of gas cooling in group cores, it does suggest that the CCT boundary values suggested by Hudson et al. are less helpful at low masses. In clusters, CCT separates the population into three relatively well-defined classes. In groups with kT$<$2~keV, it appears that a short cooling time is almost inevitable, and the great majority of systems will always be classed as SCC.

\subsection{Gas properties and central AGN jet activity}
\label{sec:AGN}
One obvious question is whether we can link the different measures of CC status to other indicators of cooling in our groups. We can begin to answer this question by considering the state of the central AGN. \citet{Mittaletal09} find that for the HIFLUGCS clusters, the presence of a radio source is strongly correlated with CC state, with 100\% of SCC, 67\% of WCC and 45\% of NCC clusters hosting a central radio source. \citet{Bharadwajetal14} extend this study to groups, but find a conflicting result. All their NCC systems host a central radio source, but only 77\% of SCC and 57\% of WCC. For the CLoGS groups, we have the advantage of targeted, high-resolution and low-frequency radio observations for every group, allowing us to go beyond identification of central radio sources to consider morphology, i.e, whether the radio source has current or recently active jets.       

Kolokythas et al. (in prep.) describe the radio properties of the dominant early-type galaxies of each of our high-richness groups. Table~\ref{tab:morph} shows the radio morphology found for each galaxy. As noted above, all but two of the dominant galaxies show evidence of AGN activity. Current jet activity, or evidence of activity recent enough to still be visible in radio emission, is observed in six systems. Of these, five are X-ray confirmed groups. Jet activity in the sixth, LGG~80 / NGC~1167, is probably the result of a cold-gas-rich merger \citep{Shulevskietal12}.

All five X-ray bright groups which host central jet sources are classed as CC based on their temperature profiles. Since there are 9 CC groups, jet systems make up 56\% of the CC groups. None of the non-cool-core systems host central jet sources, though two contain diffuse radio structures. If we instead classify cool cores based on the CCT, we find that radio jets are found only in SCC systems, but this is unsurprising since all but three groups are SCC. The fraction of CCT-selected SCC groups with jets is 5/13 (38\%). We note that systems with diffuse radio emission include SCC, WCC and X-ray undetected groups.

Almost all the confirmed groups have central entropies $<$30\kevcmsq, the boundary below which radio activity and ionized gas emission becomes common in the cores of galaxy clusters \citep{Cavagnoloetal08}. Three of our jet-hosting groups have K$_{10}$$<$30\kevcmsq, indicating a relatively large region of low entropy. However, we see other group and galaxy-scale haloes with low K$_{10}$ which do not host jets. There is no obvious separation between the entropy profiles of the jet hosting groups and the rest of the population, suggesting that while the jets are certainly injecting energy into the IGM, they have not dramatically increased the entropy in their immediate surroundings \citep[see also][]{Jethaetal07}. 

Examining the CCT values, the jet-hosting systems have some of the shortest CCTs in the sample, but there are systems with comparable CCTs which host only point sources. Examining the cooling time profiles, three of the jet-hosting groups have the lowest cooling times at 10~kpc, with a fourth falling to comparable values ($\sim$400-600~Myr) inside $\sim$5~kpc. LGG~9 / NGC~193 has one of the longest cooling times at 5~kpc, a product of the AGN outburst, which has inflated a large cocoon or system of cavities, driving a radial shock and reducing the density of gas in the group core. However, we note that at least three other groups (two of them non-cool-core) have cooling times in their central spectral bin $\la$1~Gyr without showing signs of jet activity.

Numerical simulations have been used to probe the thermal stability of hydrostatic haloes, and suggest that clouds of cool material can precipitate out of the IGM when heating and cooling are in approximate balance \citep{Sharmaetal12,McCourtetal12,Gasparietal12,Gasparietal13,LiBryan14a,LiBryan14b}. In the simulations, precipitation occurs when the ratio of the isochoric cooling time ($t_{c}$) to the free fall time ($t_{ff}$) is $\le$10. Observationally, warm or cool gas which could be the product of such precipitation is observed in systems where the minimum value of $t_{c}/t_{ff}\la$ 20 \citep[][for isochoric cooling time]{VoitDonahue15,Voitetal15}. 

\citet{Voitetal15} noted that in large ellipticals, including those at the centres of groups and clusters, $t_{ff}$ can be approximated (to within $\sim$10\%) as r$\sigma^{-1}$, where $\sigma$ is the stellar velocity dispersion of the galaxy, and r is radius. Their sample of 10 galaxies includes three of the BGEs in the CLoGS high-richness subsample, NGC~5044, NGC~5846 and NGC~4261, in which they find minimum values of $t_{c}/t_{ff}$$\simeq$10. Following their methodology, we use $\sigma$ value drawn from the HyperLEDA catalogue to estimate $t_{c}/t_{ff}$ in our groups. For systems where deprojection was possible, we estimate the value in all spectral bins in the central 20~kpc of each group and determine the minimum value. Where deprojection was not possible, we estimate the value within 10~kpc. This provides a representative value at a radius where resolution effects are likely to be minimal. The resulting values are listed in Table~\ref{tab:morph}. Note that to match previous studies, we use isochoric cooling times rather than the isobaric values used in our cooling time profiles.

Among the X-ray bright groups, the lowest values of $t_{c}/t_{ff}$ are found in four of the five systems which host central jet sources; LGGs 72, 278, 338 and 393 all have min($t_{c}/t_{ff}$)$\le$15. Our values are derived from the \xmms\ profiles; for LGGs 278, 338 and 393 \chandra\ produces similar results \citep{Voitetal15,Davidetal17}. LGG~18 has the next highest value, min($t_{c}/t_{ff}$)$\sim$18, but only hosts a radio point source. Owing to its strongly disturbed state, we do not consider it possible to determine a reliable value for the fifth jet-hosting system, LGG~9. The other groups have values in the range $\sim$25-100, with no obvious indication that diffuse radio sources are found in groups with particularly low (or high) values. 

For the three systems with only galaxy-scale gas haloes, the values of $t_c/t_{ff}$ are poorly constrained. Deeper observations would be needed to investigate the thermal stability of these relatively gas-poor galaxies.

While the limited size of our sample, and the limited spatial resolution of some of our data, makes it difficult to draw definite conclusions, it is noteworthy that the precipitation criterion ($t_{c}/t_{ff}$) is strongly correlated with, and is a good indicator of, jet activity. \citet{McNamaraetal16} have argued, using data from galaxy clusters, that $t_{c}$ is a superior indicator of cooling and that buoyantly rising radio lobes are actually necessary to trigger precipitation. LGG~18 might be an example of a system which has the potential for thermal instability but lacks the uplift necessary to trigger precipitation, but its higher value of $t_{c}/t_{ff}$ makes this uncertain. As noted in above, groups naturally have short cooling times compared to clusters, and thus cooling time may not be as meaningful an indicator of cooling in these lower mass systems. Examining the other indicators, while jet-hosting systems have low entropy and CCT, a central temperature decline is a better indicator of current or recent jet activity for our groups.

\subsection{Centrally peaked temperature profiles}
It is notable that among the groups in our sample, those with cool cores (as defined by a central temperature decline) tend to be the hotter systems. Splitting the sample at 0.8~keV gives nine hotter groups of which eight are CC, and eight cooler groups of which only one is a clear CC. The poor quality of the temperature profiles of some of the cooler groups may disguise some cool cores, and the cooler half of the sample includes three galaxy-scale haloes, but the divide is intriguing nonetheless. If we consider groups with centrally-peaked temperature profiles, two (LGGs~103 and 402) have clear, well resolved temperature peaks in their cores; both are in the cooler half of the sample. Two other cooler groups (LGG~117 and 262, the latter a galaxy-scale halo) have profiles that may suggest a central temperature peak, but both have poor quality data that only supports projected profiles with a small number of bins. The only system in the hotter half of the sample which has a centrally-peaked profile is LGG~42 (see Figure~\ref{fig:kTprofs}). This divergence between the profiles of hotter and cooler groups is intriguing, particularly given that all the merger systems in our sample are CC, and raises the question of how such central peaks form. 

\subsubsection{Galactic coronae and unresolved cool cores}
\begin{figure}
\includegraphics[width=\columnwidth,bb=30 220 570 750]{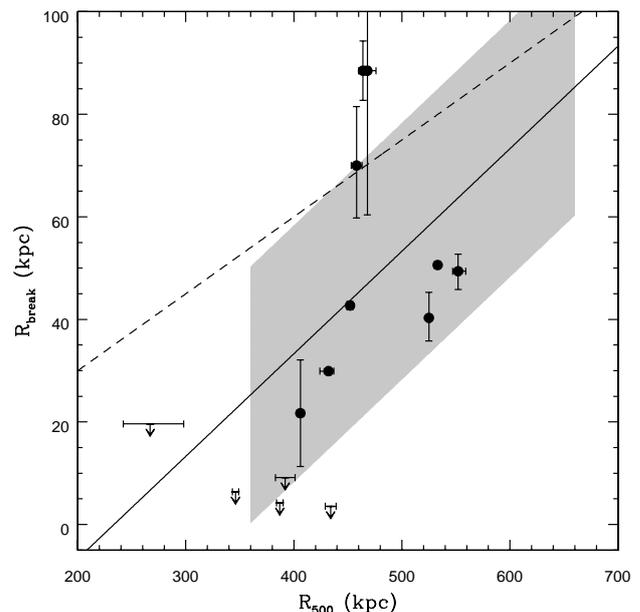}
\vspace{-5mm}
\caption{\label{fig:Rbreak}Cool core radius ($R_{\rm break}$) compared with fiducial radius (R$_{500}$) for high-richness groups with at least three radial temperature bins. CC systems are marked by circles and error bars, NCC systems by upper limits, set by the size of the innermost temperature bin. The black profile shows the best fit to the groups of \citet{RasmussenPonman07}, with the grey region showing an approximation of the range of values in their sample. The dashed line marks R$_{\rm break}$=0.15R$_{500}$, the typical value for clusters \citep{Vikhlininetal05}.}
\end{figure}

Before discussing methods of heating the core, we consider the possibility that the NCC systems might simply have very small cool cores, unresolved by our temperature profiles. Cool cores of with radii $<$10~kpc, sometimes referred to as galactic coronae, are known to exist in a number of systems \citep[e.g.,][]{Sunetal07,Sun09}. LGG~421 provides a possible example in our sample \citep{OSullivanetal11c}. Some of our temperature profiles probe radii $<$10~kpc, but there are cases where a small galactic corona could be missed (e.g., LGG~103). Figure~\ref{fig:Rbreak} shows a comparison of R$_{500}$ with with CC size ($R_{\rm break}$) for our groups.

Most of our CC groups are consistent with the range of values found for groups by \citet{RasmussenPonman07}. There are three high outliers. LGG~31 falls at the upper margin of expected values, but its break radius is uncertain since both cool core and temperature peak are only represented by single radial bins. The two merging groups, LGG~72 and LGG~473 fall above the relation with almost identical values. It is possible that heating of the IGM outside the cool core by the merger in each system produces a large break radius, and LGG~72 also has large uncertainties. Among the NCC systems, LGG~117 provides a clear case of a system where a CC could have gone undetected. However, the other systems have very low upper limits on the break radius (set by the size of the innermost temperature bin), at the lower end of the range of values found by Rasmussen \& Ponman. This suggests that if they contained typical cool cores, we ought to resolve them. LGG~42 is the strongest outlier, since the \chandra\ data rule out any CC down to $\sim$3.5~kpc. We therefore conclude that, while we cannot rule out small galactic coronae in some of our NCC systems, we are probably not missing significant numbers of full-scale cool cores.

\subsubsection{AGN heating and supernovae}
Considering sources of heating which might produce the centrally-peaked temperature profiles, AGN jets are an obvious possibility. Evidence of reheating by AGN is seen in systems such as NGC~3411 \citep[also known as NGC~3402 or SS2b~153,][]{OSullivanetal07}. Its temperature profile shows a central temperature plateau, surrounded by a shell of cooler material, as if the central part of a cool core had been heated. Although only very small-scale jets are observed, the size of the plateau closely matches that of diffuse radio emission in the group core, suggesting a link between the higher temperatures and AGN.

Radiative cooling will erase central temperature peaks relatively rapidly. Taking LGG~402 as an example, the volume represented by the central three bins of the temperature profile has a luminosity $\sim$2.5$\times$10$^{41}$\ergps. Using the temperature and gas mass to estimate the energy of the gas in each bin, we find that this is sufficient to cool the central temperature peak in $\sim$40~Myr. Studies of jet-inflated cavities in groups find likely jet powers in the range $\sim$10$^{41}$-10$^{44}$\ergps\ \citep[e.g.,][]{OSullivanetal11b}, easily sufficient to heat the core of one of our groups. We see no clear evidence of recent or ongoing AGN outbursts in the cooler systems. However, outbursts are difficult to detect after a few tens of Myr, once radio emission from the lobes fades and cavities in the IGM move out to large radii. The absence of cool cores, and of evidence of AGN outbursts suggests we would need to be observing the groups at the midpoint of the cooling cycle, but given this caveat, AGN heating can explain the presence of hot cores in individual groups.

To produce a systematic difference in temperature profiles between hotter and cooler systems is more difficult. While all the groups have central cooling times which are short compared to typical galaxy clusters, the hotter groups have shorter central cooling times on average than the cooler systems in our sample, suggesting that outbursts occur more often. This might explain some of the difference; the hotter systems cool faster, so the period in which we might see them with a hot core caused by recent feedback is shorter. However, the difference in cooling times is not so large as might be expected given the predominance of CCs among the hotter groups. It also raises the question of why these hotter groups are more effective at cooling gas to fuel their AGN. The longer cooling times in the cooler groups suggest that the gas in their cores is proportionately less dense than the cores of the hotter groups. This is difficult to explain if outburst power is, as expected, proportional to cooling rate.

There is good evidence that this proportionality holds on average, in galaxy clusters, over long timescales \citep{HlavacekLarrondoetal12,Mainetal17}. However, if this relationship is imperfect for individual outbursts, or if outbursts can be triggered by other means (e.g., gas-rich mergers), it is worth noting that powerful outbursts could have a disproportionate impact on the least massive groups, since their shallow potentials make them less effective at retaining gas in their central regions. If a single powerful outburst can effectively heat and reduce the density in the group core, this will lead to longer cooling times, delaying the reformation of a CC. \citet{Gittietal07} estimate that $\sim$10\% of outbursts in clusters are high--powered, and if such events are similarly common in groups, we might then expect to see a systematic difference in properties with group mass, since the lowest mass groups will spend the longest period recovering from these over-powered outbursts. However, it should also be noted that we do not see examples of such outbursts. In the two most powerful jet systems in our sample, LGG~278 and LGG~9, the outbursts have not erased the cool cores, despite having quite dramatic impacts on their structure. 

Type~Ia supernovae (SN~Ia) in the group-central galaxy could also contribute to heating a group core. The dominant galaxy of LGG~402, NGC~5982, has a luminosity-weighted stellar population age $\sim$9~Gyr \citep{Kuntschneretal10}. Adopting the SN~Ia rate of \citet{Rodneyetal14}, which for old stellar populations declines with age$^{-1}$, we expect a rate of $\sim$1.8$\times$10$^{-4}$~SN~yr${^-1}$ for every 10$^{10}$\Msol\ of stars. Assuming a stellar mass-to-light ratio M/L$_K$$\sim$1 \citep{LonghettiSaracco09} and an energy release of 10$^{51}$~erg for each supernova, this suggests a maximum heating rate $\sim$1.3$\times$10$^{41}$\ergps. This is comparable to the X-ray luminosity, and we would expect the heating rate to be enhanced if any significant younger stellar sub-population is present. However, the efficiency with which supernovae heat the IGM is likely to be low \citep[][]{KravtsovYepes00}, suggesting that they are unlikely to be capable of producing a central temperature peak. 

\subsubsection{Gravitational heating}
\citet{Khosroshahietal04}, in their study of NGC~6482, suggested a third possibility, gravitational heating. As gas cools and flows inward to the group core, it will be subject to PdV work which, if it exceeds radiative losses, will naturally produce a central temperature peak. NGC~6482 is a fossil group with an exceptionally concentrated mass profile \citep{Buote17}. Since gravitational heating could be unusually effective in such a system, it is necessary to test the model suggested by Khosroshahi et al. on our groups to see how well it applies to the group population in general.

Khosroshahi et al. assume a steady-state cooling flow (see their equations~12-14) in which the temperature of the gas at a given radius is determined by i) the starting temperature at some outer radius, ii) the change in gravitational potential energy between the outer radius and its current position, and iii) the energy lost through radiation during the time taken to flow in to that position. The rate of inflow is varied to obtain a model profile approximating the observed temperature profile. Khosroshahi et al. found that a rate \Mdot=1.5\Msolpyr\ was sufficient to reproduce the observed temperature profile in NGC~6482.

\begin{figure}
\includegraphics[width=\columnwidth,bb=40 210 330 730]{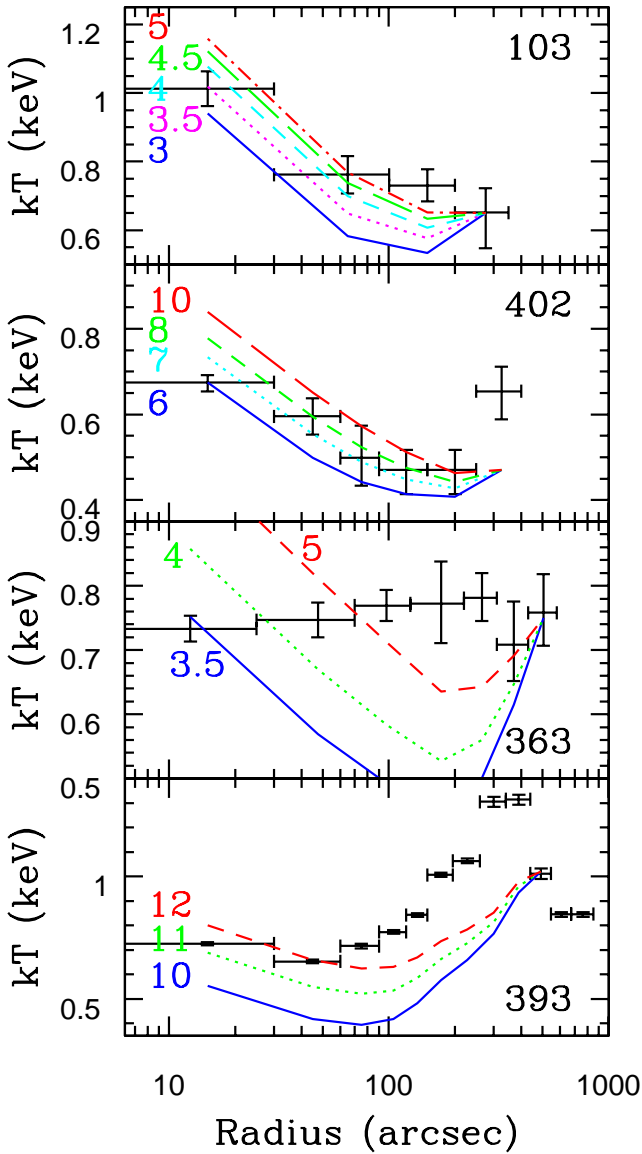}
\caption{\label{fig:pdv}Deprojected temperature profiles for four CLoGS groups with PdV gravitational heating models overlaid. Different line colours and styles indicate different inflow rates, with the \Mdot\ for each line labeled.}
\end{figure}

We estimate the change in gravitational potential from total mass profiles calculated directly from the temperature and density profiles of our groups, assuming hydrostatic equilibrium. We use the measured luminosities to estimate energy losses, and compare a range of inflow rates to our measured temperature profiles. Figure~\ref{fig:pdv} shows the results for four of our groups.

For the two groups with centrally-peaked profiles, we find that the model can produce reasonable temperature profiles. In LGG~103, normalising the model in the outermost bin, \Mdot$\simeq$4-5\Msolpyr\ provides a reasonable match to the two inner bins, though the model consistently underestimates bin 3. In LGG~402 we assume the high outermost temperature is spurious, and find that \Mdot$\simeq$7\Msolpyr\ provides a good match to all five inner bins. These results suggest that the model is generally applicable and does not require an extreme mass profile such as that in NGC~6482. However, the inflow rates are rather large, particularly given that by definition the model requires all of the mass to be deposited in the centre of the flow.

 In LGG~402, we can compare the observed mass of warm and cool gas in the system to the predicted cooling rate. 2.3$\times$10$^4$\Msol\ of H$\alpha$ emitting ionized gas is seen in the core of NGC~5982 \citep{Sarzietal06}, and a 3.4$\times$10$^7$\Msol\ cloud of \Hi\ is located $\sim$6~kpc and 200\kmps\ from the nucleus \citep{Morgantietal06}. CO observations place an upper limit of $<$2.5$\times$10$^7$\Msol\ on the molecular gas content of the galaxy \citep{OSullivanetal15}. The inflow rate suggested by the model would thus be incompatible with observations in $<$10~Myr, whereas the lack of evidence of AGN jet activity suggests that cooling is likely to have been undisturbed for at least a few tens of Myr.

The gravitational heating model predictions for groups with flat or centrally declining temperature profiles highlight another problem. For the observed mass and luminosity profiles, and physically plausible inflow rates, the model always predicts a temperature decline at intermediate radii, with a rise at small radii. This means it can never reproduce the flat profile observed in LGG~363, or the intermediate radius temperature peak and central decline characteristic of cool core systems such as LGG~393. This is particularly troubling in that the model fails in systems where cooling is most clearly occurring; those with ongoing or recent jet activity all have central temperature declines. It is notable that the shape of the model profile is closest to that of the CC systems in the cool core itself, but fails in their outer parts. We might have expected the opposite, since it is at large radii that the model assumptions are most plausible. At small radii other heating mechanisms (AGN, SNe) complicate the situation. 

Given the short cooling times we find in our groups, the indication from this model that radiative losses need not always lead to a central temperature decline is interesting. The failure of the model in systems where cooling seems to be most effective suggests that its assumptions are flawed, but it is striking that it comes close to reproducing some of the profile shapes we observe. However, none of the processes we have considered seem to provide an explanation for the intriguing difference in temperature profiles between our hotter and cooler systems. Further exploration of the low-mass, low-temperature group population is clearly needed if this issue is to be resolved, but given their low luminosities, this is a difficult undertaking for the current generation of X-ray observatories.

\subsection{Previously undetected groups and prospects for eROSITA}
By selecting groups from an optical catalogue, the CLoGS sample aims to avoid the biases associated with X-ray selection. It is important to consider how well we achieve this goal with the high-richness half of the sample.

Of the 14 groups where we confirm the presence of a hot IGM, 11 had previously been identified as X-ray bright groups \citep[e.g.,][]{Bohringeretal00,Mahdavietal00,Mulchaeyetal03,Giacintuccietal11,Panagouliaetal14}, and in all 11 the BGE is associated with a source in the RASS bright or faint source catalogues \citep{Vogesetal99,Vogesetal00}. Our luminosity measurements for these systems (within R$_{500}$) cover a wide range $\sim$2-200$\times$10$^{41}$\ergps, though in some of the faintest cases the RASS detection may be aided by the presence of a bright AGN in the dominant galaxy. 

In one of the remaining three, LGG~402, the dominant early type galaxy, NGC~5982, is correlated with a source in the RASS Bright Source Catalogue, leading to its inclusion in catalogues of X-ray bright galaxies \citep[e.g.,][]{Beuingetal99,OSullivanetal01b}. The group has a relatively low luminosity ($\sim$3.2$\times$10$^{41}$\ergps), so it is plausible that its more extended component was not identifiable in the RASS.

The remaining two groups had not been previously identified as X-ray bright systems. Their dominant galaxies were not detected in the RASS Bright or Faint Source Catalogues, nor in the updated Second ROSAT All-Sky Survey source catalogue \citep{Bolleretal16}. LGG~103 is a relatively low-luminosity system ($\sim$8.3$\times$10$^{41}$\ergps) with a flat surface brightness profile and a hot core. LGG~72 is one of the brighter groups in our sample ($\sim$5.9$\times$10$^{42}$\ergps) but is highly disturbed owing to an ongoing merger. While it does host a small cool core, the more extended emission is not centred on the dominant galaxy, NGC~1060, and is not symmetrically distributed.

Among the systems we identify as hosting galaxy-scale gas haloes, only LGG~185 appears in the RASS source catalogues, owing to the bright AGN of NGC~3078.

These results conform to the known limitations of the RASS; the systems which were previously undetected or miscategorized are either faint and lacking a cool core, or disturbed and asymmetrical, with a cool core not associated with a larger-scale bright central concentration. If we consider only the ability to identify group-scale haloes, RASS identifies $\sim$80\% of the X-ray confirmed galaxy groups in our high-richness sample. 

With the launch of the Spectrum-Roentgen-Gamma mission in 2017, the eROSITA instrument will begin mapping the X-ray sky, eventually providing surveys of groups and clusters a factor $\sim$20 more sensitive than the RASS \citep{Merlonietal12}. We estimate, based on the limits presented by Merloni et al., that at the end of the 4-year all-sky survey (eRASS:8), the limiting flux for a S/N=7 detection of the central 3\arcm\ diameter core of a 0.5~keV group with 0.3\Zsol\ metallicity and Galactic absorption 3$\times$10$^{20}$\pcmsq\ at redshift $\sim$0.01 will be $\sim$3$\times$10$^{-14}$\ergpspcmsq. This is easily sufficient to detect any of the X-ray bright groups in our sample, including those undetected or misidentified in RASS. If we consider the merger system LGG~72, such sensitivity is also sufficient to detect both the bright arc of emission linking the two cores, and the fainter emission inside the arc. It is therefore clear that eROSITA will detect such systems, and be able to identify them as possessing an extended IGM. 

Although we have focused in this paper on the local universe, eROSITA will also trace the population of groups out to moderate redshifts. If we again require S/N=7 and a 3\arcm\ diameter detection region, we find that our group-scale systems are detectable to redshifts $\sim$0.22. The resolution of eROSITA, while not suited to study of complex structure in the IGM, is certainly sufficient to identify extended emission in these systems. Even the larger galaxy-scale systems, with gas extending $>$25~kpc, will be resolvable out to redshifts $\sim$1.

CLoGS provides at least a glimpse of the kinds of systems we might expect to see in the eROSITA surveys. As well as extending our knowledge of the group population to a much larger volume and to lower luminosities, eROSITA is likely to detect more disturbed systems than were identified by RASS, and more non-cool-core groups. The inclusion of these systems will provide a much clearer picture of the gas properties of the group population as a whole, and opens the door to a fuller understanding of the physics of group formation and evolution.

\section{Summary and Conclusions}
\label{sec:conc}

In this paper we have presented the Complete Local-Volume Groups Sample, an optically-selected, statistically complete set of nearby galaxy groups chosen to allow observation in the X-ray and radio bands, and the investigation of the relationship between the group member galaxies, their AGN, and the IGM. As the typical environment of most galaxies in the universe, groups are key to our understanding of galaxy evolution, the build-up of the hot intra-group (and intra-cluster) medium, and the regulation of radiative cooling. The biases affecting prior X-ray selected samples of nearby groups have made it difficult to address these issues. While the CLoGS project cannot hope to resolve issues of this scale, it can at least throw light on them and give us some hints as to the effects of the known biases.

We have described the sample selection of the 53-group CLoGS sample, the definition of a 26-group high-richness subsample, and the analysis of the X-ray observations of that subsample. Our results are summarized below, and Table~\ref{tab:frac} provides an overview of the high-richness groups, classified by X-ray and radio morphology. 

\begin{table}
\caption{\label{tab:frac}Summary of our classification of CLoGS high-richness groups by their X-ray properties, and by those of the AGN in their BGE.}
\begin{center}
\begin{tabular}{llc}
\hline
Category & & No. of groups \\
\hline
\multicolumn{2}{l}{CLoGS sample size}              & 53 \\
\multicolumn{2}{l}{Groups in high-richness subsample} & 26/53 \\
X-ray morphology: & Group-scale halo  & 14/26 \\
 & Galaxy-scale halo & 3/26 \\
 & Point-like  & 9/26 \\
Dynamically active: & & 4/14 \\
 & Merger & 2/14 \\
 & Sloshing & 2/14 \\
Central radio AGN &  & 24/26 \\
 & Jet & 6/24 \\
 & Diffuse & 4/24 \\
 & Point source& 14/24 \\
Cool core fraction: & Temperature decline & 9/14 \\
 & Hudson SCC & 13/16 \\
 & Hudson WCC & 3/16 \\ 
\hline
\end{tabular}
\end{center}
\end{table}

\begin{itemize}
\item Of the 26 groups in the high-richness subsample, $\sim$54\% (14 groups) are confirmed to possess a group-scale hot IGM, with a further $\sim$12\% (3 groups) hosting a smaller galaxy-scale halo associated with the dominant early-type galaxy. The typical temperatures of the detected groups cover the range $\sim$0.4-1.4~keV, corresponding to masses in the range M$_{500}$$\sim$0.5-5$\times$10$^{13}$\Msol, and the systems have IGM luminosities in the range L$_{X,R500}$$\sim$2-200$\times$10$^{41}$\ergps. The galaxy-scale haloes have temperatures 0.4-0.6~keV and gas luminosities L$_{X,R500}\sim$2-6$\times$10$^{40}$\ergps. While extended hot gas is detected in groups across the richness range of the subsample, all but one of the nine groups for which no IGM or galaxy-scale halo was detected are all in the lowest richness class, containing only 4 bright galaxies. It is notable that even quite rich groups ($R$=6-7) may contain only a galaxy-scale halo. The X-ray detected groups have luminosities and temperatures consistent with prior X-ray selected samples, while the galaxy-scale systems generally fall below the luminosity--temperature relation, again consistent with previous studies.

\item Of the groups in which extended gas emission is detected, $\sim$53\% (9/17 groups) possess cool cores, defined as a significant central temperature decline, rising to $\sim$64\% (9/14) if we restrict the sample to systems with at least three radial temperature bins. This is a smaller fraction than that found in some prior X-ray selected samples of groups and poor clusters, and is closer to the roughly even split between CC and NCC systems seen in more massive clusters. The size of the central bin of our temperature profiles is typically $\sim$10~kpc, raising the possibility that some very small cores or galactic coronae may be missed, but similar caveats apply to the prior samples which found higher cool core fractions. 

\item We note that the scheme of cool core classification based on central cooling time, commonly used to divide clusters into strong-, weak-, and non-cool-core systems, fails in galaxy groups. The efficiency of line emission in $\sim$1~keV plasma naturally results in short cooling times in groups, leading to most groups being classed as strong cool cores, regardless of their temperature structure. A redefinition of the boundaries between cool core classes is needed if this scheme is to be usefully applied to groups.
  
\item The X-ray detected groups include a fairly high fraction of dynamically active systems. Two groups are currently undergoing mergers, and a further two show evidence of sloshing, indicating recent gravitational disturbance by infalling galaxies or subgroups. This suggests that $\sim$30\% of our X-ray bright groups (4/14 systems) have undergone a significant interaction within the past few hundred Myr.

\item Our radio analysis of the groups (Kolokythas et al., in prep.) finds AGN in all but two (92\%) of the 26 group-dominant early-type galaxies. Most are point-like, but $\sim$25\% possess radio jets, or show evidence of jet activity in the recent past. Only one of the jet systems is in an X-ray undetected group, leading to a jet fraction $\sim$36\% (5/14 systems) for groups with a full-scale IGM. This implies a duty cycle of $\sim$1/3. All five jet systems are found in groups with cool cores. Jet activity appears to be more closely correlated with short central cooling times rather than low central entropies. Central entropies are not enhanced in systems which host jets, and in only one group, LGG~9 / NGC~193, does the central jet source appear to have had a significant impact on the central cooling time. Where it can be calculated, the thermal instability criterion $t_{c}/t_{ff}$ is strongly correlated with jet activity, and all four groups with values $\le$15 host currently or recently active jets.

\item We estimate the entropy at 10~kpc (K$_{10}$) or lower limits on that value for all 26 groups. For the systems with extended gas emission we find K$_{10}$$\simeq$10-60\kevcmsq, comparable to previous X-ray selected samples. The lower limits are dependent on the temperature assumed for the group, with typical values for 1~keV groups in the range $\sim$300-700\kevcmsq, and values for 0.5~keV groups a factor $\sim$3 lower. Comparing this with the range of entropies expected in groups from simulations, we find that we only detect systems at the lower end of the predicted range of entropies, and our limits suggest that we are not failing to detect a population of high entropy groups unless those groups are predominantly low-temperature systems.

\item We find an interesting suggestion of a separation in core properties between hotter and cooler groups, with almost all systems with T$_{sys}$$>$0.8~keV possessing CCs, while those with T$_{sys}$$<$0.8~keV are more likely to be NCC. We consider possible heating mechanisms that could have produced the centrally-peaked temperature profiles we see in some of the cooler systems, including AGN, supernovae, and gravitational work done on inflowing gas. None of these provides a clear explanation, but we note the ability of gravitational heating to produce central temperature rises, and the possibility that exceptional AGN outbursts may disproportionately heat the cores of low-mass groups, offsetting radiative cooling for long periods. 

\item Of the 14 groups in which we detect a hot IGM, only 11 had previously been identified as X-ray bright groups. The dominant early-type galaxy of one additional system was detected in the RASS, but had been classed as galaxy rather than a group. Of these three previously unidentified groups, two are relatively faint systems (L$_{X,R500}<$10$^{42}$\ergps) which lack cool cores, while the third is more luminous but highly disturbed by an ongoing merger. These properties are consistent with the expectation that RASS-based group samples are likely to miss groups which lack a strong central surface brightness peak. The fraction of groups missed by RASS in our samples is $\sim$20\%. Surveys made by eROSITA will likely resolve this problem for the nearby universe, and CLoGS suggests that groups drawn from eROSITA surveys are likely to contain a higher fraction of disturbed, low-luminosity, non-cool-core systems than previous surveys.
  
\end{itemize}

\medskip
\noindent{\textbf{Acknowledgments}}\\
We thank N. Jetha, M.  Weisskopf and the late S.~S. Murray for their help
in the early stages of the project, and A.J.R.  Sanderson for his
contributions to the sample selection and optical characterization, and for
many helpful discussions. We also thank the anonymous referee for
  their useful comments and suggestions. We acknowledge
support from the European Union's FP7-PEOPLE-2009-IRSES (Marie Curie)
programme, through their grant to the CAFEGroups project. EOS acknowledges
support from the National Aeronautics and Space Administration (NASA)
through through Astrophysical Data Analysis Program grant number NNX13AE71G
and through Chandra Award Number AR3-14014X, issued by the Chandra X-ray
Observatory Center, which is operated by the Smithsonian Astrophysical
Observatory for and on behalf of NASA under contract NAS8-03060. AB
acknowledges support from NSERC, Institut Lagrange de Paris, and Pauli
Center for Theoretical Studies ETH UZH, and thanks the Institut
Astrophysique de Paris and the Centre for Theoretical Astrophysics and
Cosmology (University of Zurich) for hosting him. MG acknowledges partial
support from PRIN-INAF 2014.  Basic research in radio astronomy at the
Naval Research Laboratory is supported by 6.1 Base funding. This work makes
use of observations obtained XMM-Newton, a European Space Agency (ESA)
science mission with instruments and contributions directly funded by ESA
Member States and NASA. This publication also makes use of NASA's
Astrophysics Data System Bibliographic Services, and of data products from
the Two Micron All Sky Survey, which is a joint project of the University
of Massachusetts and the Infrared Processing and Analysis Center/California
Institute of Technology, funded by NASA and the National Science
Foundation. We acknowledge usage of the HyperLeda database, and the
NASA/IPAC Extragalactic Database (NED), which is operated by the Jet
Propulsion Laboratory, California Institute of Technology, under contract
with the NASA.

\bibliographystyle{mnras}
\bibliography{../paper}

\clearpage
\appendix

\section{Notes on Individual Groups}
\label{sec:notes}

\subsection{LGG~9 / NGC~193}
LGG~9 is an X-ray luminous system, and its BGE NGC~193 hosts one of the brightest FR-I radio sources in our sample, 4C~+03.01 \citep{Giacintuccietal11}.The \chandra\ observations show that the AGN jets have inflated a large cavity or cavities in the centre of the group, with a ring of compressed gas surrounding the radio cocoon \citep{Bogdanetal14}. 

\subsection{LGG~18 / NGC410}
LGG~18 is an X-ray luminous system, with an IGM centred on the BGE NGC~410 and extending beyond the edge of the \xmms\ field of view (at least 14\arcm\ or 310~kpc). The central ($\sim$10~kpc radius) bin of the temperature profile shows a significant temperature drop, indicating a cool core. The BGE is detected as a point source in our 235 and 610~MHz GMRT observations, and at 1.4~GHz in the NVSS survey \citep{Condonetal98}. Another group member, the edge-on S0 NGC~407 located east of NGC~410, is also detected in both X-ray and 235~MHz emission.

\subsection{LGG~27 / NGC584}
LGG~27 is an X-ray faint system. \rosat\ observations showed some extended emission around the BGE, NGC~584, but our short \chandra\ observation detects only powerlaw emission with extension comparable to the size of the stellar population. The galaxy is not detected in our GMRT data or previous radio surveys.

\subsection{LGG~31 / NGC~677}
LGG~31 was identified as an X-ray luminous group by \citet[SRGb~115 in their catalogue]{Mahdavietal00}. Our \xmms\ observation confirms the presence of an extended IGM centred on the BGE NGC~677, but also shows a second X-ray peak to the northwest, at approximately 01$^h$48$^m$55\fs 2, +13$\degree$07$^\prime$30$^{\prime\prime}$. This X-ray source, hereafter named XMMU~J014855.2+130730, appears to be a background cluster, and we exclude a 200\arcs-radius circle from all analysis of the LGG~31 IGM to avoid contamination.

Examination of Sloan Digital Sky Survey (SDSS) imaging reveals a number of faint galaxies around XMMU~J014855.2+130730. 2MASX~J01485523+1307285 is located 3.5\arcs\ from the X-ray centroid, and has a measured redshift $z$=0.2269$\pm$0.0001. Spectra extracted from a circular 100\arcs\ radius region around the X-ray peak are reasonably well fitted (red. $\chi^2$=1.0519 for 353 degrees of freedom) by an absorbed APEC model with Galactic hydrogen column (\nh=5.09$\times$10$^{20}$\pcmsq), redshift $z$=0.218$\pm$0.002, temperature kT=3.09$^{+0.10}_{-0.09}$~keV, and abundance 0.57$^{+0.08}_{-0.07}$\Zsol. A NED search finds no previous cluster identifications at this position, and we therefore consider this a newly identified poor cluster.

NGC~677 was previously detected in the NVSS, and is identified as a diffuse source in our GMRT 235 and 610~MHz observations. 2MASX~J01485523+1307285 is detected as a point source at both frequencies, with flux densities of 181~mJy at 235~MHz and 24~mJy at 610~MHz.  

\subsection{LGG~42 / NGC~777}
LGG~42 is an X-ray luminous group, first identified as such from the RASS \citep{Bohringeretal00}. The available short \xmms\ and \chandra\ ACIS-I pointings allow us to trace the hot IGM to a radius of $\sim$275~kpc (13\arcm) from the BGE NGC~777. The \chandra\ data are too shallow to support spectral deprojection, but do provide a more detailed view of the centrally peaked temperature profile, confirming this as a non-cool-core system. NGC~777 hosts a radio point source detected in our 235 and 610~MHz GMRT observations and at 1.4~GHz in the NVSS.

\subsection{LGG~58 / NGC~940}
Our \xmms\ observation of this system does not detect an IGM, revealing only an X-ray point source at the position of the BGE NGC~940. The galaxy hosts a radio point source and, unusually for an early-type galaxy, $\sim$6$\times$10$^9$\Msol\ of molecular gas \citep{OSullivanetal15}, with the CO(1-0) spectrum suggesting a double-peaked profile typical of a rotating disk.

An extended X-ray source is observed to the northwest of NGC~940, with a peak at 2$^h$28$^m$54\fs0, +31$\degree$44\arcm57\arcs. Extracting spectra from a 110\arcs-radius circle and fitting them with an absorbed APEC model, we find that with fixed Galactic hydrogen column the temperature is kT=1.32$^{+0.02}_{-0.05}$~keV, abundance 0.60$\pm$0.12\Zsol\ and redshift $z$=0.081$^{+0.010}_{-0.014}$. We therefore conclude that XMMU~J022854.0+314457 is previously unknown galaxy group in the background of LGG~58. A NED search finds no previous group or cluster identifications at this position, and no galaxies with measured redshift within the extent of the emission. 2MASX~J02285384+3144572 is located $\sim$2.5\arcs\ from the X-ray peak, and hosts a weak (3.9~mJy at 1.4~GHz) radio source. It is possible that it is the group-dominant galaxy, a redshift survey of the region would be needed to confirm this.

\subsection{LGG~61 / NGC~924}
Our \xmms\ observation detects only point-like X-ray emission associated with the BGE of LGG~61, NGC~924. Spectra extracted from the point source are adequately described by a simple absorbed powerlaw, suggesting the galaxy lacks any significant hot gas halo. NGC~924 hosts a previously undetected radio point source detected at $>$6$\sigma$ significance in our 610~MHz data but undetected at 235~MHz. 

\subsection{LGG~66 / NGC~978}
Roughly two thirds of our \xmms\ observation of LGG~66 was affected by
background flaring and had to be excluded from further analysis. We detect
diffuse $\sim$0.5~keV thermal emission around the BGE NGC~978, extending to
$\sim$150\arcs\ ($\sim$50~kpc). X-ray point sources are also detected in
group members NGC~974 and NGC~969. No radio emission has previously been
detected from NGC~978, but a radio point source coincident with the galaxy
core is detected at $>$12$\sigma$ significance in our GMRT 610~MHz image.

\subsection{LGG~72 / NGC~1060}
LGG~72 had not been identified as an X-ray bright system prior to our observations. The \xmms\ pointing reveals a disturbed IGM, with X-ray peaks centred on the BGE, NGC~1060, and another bright elliptical, NGC~1066. Diffuse emission fills the field of view, with an arcing 250~kpc ridge linking the two peaks. Based on positions and recession velocities, it appears that NGC~1066 and several close companions form a secondary group which is merging with or falling through the group associated with NGC~1060. The line-of-sight velocity difference between the two ellipticals is $\sim$800\kmps\, considerably greater than the sound speed ($\sim$390-550\kmps) in the $\sim$1-2~keV IGM. The system is therefore an excellent example of one class of objects which has been excluded from previous surveys based on the RASS. Its disturbed morphology, lacking a single central surface brightness peak, prevented its detection in the shallow RASS data, despite a total X-ray brightness (and probable mass) comparable to many well-known groups.

Radio emission from several group members was detected in our GMRT observations, and the 610~MHz data show small-scale bipolar jets associated with NGC~1060. These extend only 10\arcs/4~kpc, perhaps indicating that they have only recently been launched.

In analysing NGC~1060 we have excluded a region 75\arcs-radius circle centred on NGC~1066, but included the X-ray ridge, which may include stripped gas. Temperatures of $\sim$1.9~keV are observed between the two X-ray peaks, and may indicate shock heating. Spherical deprojection is inevitably a poor approximation to the true gas distribution in such a disturbed system, and our results should therefore be treated with caution. The group will be described in more detail in a later paper.

\subsection{LGG~80 / NGC~1167}
The BGE of LGG~80, NGC~1167 is unusually cold gas rich for an early-type galaxy, containing $\sim$1.5$\times$10$^{10}$\Msol\ of \Hi\ in a $\sim$80~kpc radius disk \citep{Struveetal10}. The galaxy also contains 3.3$\times$10$^8$\Msol\ of molecular gas (estimated from CO measurements) and $\sim$1.7$\times$10$^7$\Msol\ of dust \citep{OSullivanetal15}. This material is likely fuelling low-level star formation in spiral-arm-like features in the central $\sim$15~kpc of the galaxy \citep{Gomesetal16}

The galaxy also hosts an compact steep-spectrum (CSS) radio source (B2~0258+35) whose currently active jets extend only $\sim$1\arcs\, and which is detected as a point-like source in our 610~MHz GMRT data, and as a point source with small extensions to east and southwest in our 235~MHz data. However, deep 1.4~GHz and 145~MHz imaging reveals a pair of old radio lobes extending $>$100~kpc from the galaxy \citep{Shulevskietal12, Brienzaetal16} indicating a previous episode of more energetic jet activity. Our X-ray analysis finds no evidence of a hot IGM, while \Hi\ observations show that several other group members are cold-gas rich. It therefore seems likely that accretion of \Hi-rich neighbours is the main process driving the development of the BGE in this group. 

\subsection{LGG~103 / NGC~1453}
LGG~103 has not previously been identified as an X-ray luminous group, but our relatively short \xmms\ observation ($\sim$10~ks after flare removal) traces the $\sim$0.75~keV IGM to a radius of $\sim$105~kpc (350\arcs) around the dominant elliptical NGC~1453. The BGE also hosts a radio point source detected in out GMRT 235 and 610~MHz data and at 1.4~GHz in the NVSS.

\subsection{LGG~117 / NGC~1587}
\citet{Helsdonetal05} used \chandra\ to confirm prior detections of an X-ray luminous IGM in LGG~117, with the BGE, NGC~1587, located at the X-ray peak. We find a very low system temperature ($\sim$0.37~keV, in agreement with Helsdon et al.), but detect diffuse X-ray emission extending at least 135~kpc, confirming that this is an intra-group medium rather than a galaxy halo.

NGC~1587 contains a significant quantity of cold \Hi\ ($\sim$2.5$\times$10$^9$\Msol) and molecular gas \citep[$\sim$2.3$\times$10$^8$\Msol\ based on CO measurements,][]{OSullivanetal15}. It also hosts a radio source which 610~MHz imaging has shown to be extended, with a central point source surrounded by low surface brightness amorphous emission and no clear jets or lobes \citep{Giacintuccietal11}. As noted by Giacintucci et al., AGN activity may have heated the group core, since the temperature profile shows a central temperature peak, and the amorphous radio emission lies entirely within that central bin.

\subsection{LGG~158 / NGC~2563}
The LGG~158 group is X-ray luminous \citep[e.g.,][]{Mulchaeyetal03} with the IGM well centred on the dominant elliptical NGC~2563. We are able to trace the IGM to the edge of the \xmms\ field of view ($\sim$950~kpc), and both \xmms\ and ACIS-I datasets show it to be regular and undisturbed with a cool core clearly visible in both temperature profiles. A previously undetected radio point source in NGC~2563 is detected at $>$6$\sigma$ significance in our GMRT 610~MHz observation, but is undetected at 235~MHz. 

\citet{Morandietal17} present an analysis of the group based on a tiled set of 14 \chandra\ ACIS-I pointings extending out to $\sim$800~kpc (with almost complete coverage to $\sim$450~kpc). These include ObsID~7925 covering the group core, which is included in our analysis. We find our \chandra\ and \xmms\ density profiles to be in good agreement with theirs, but their temperature profile, while showing the same structure we observe, has more extreme values (peak kT$>2$~keV, core kT$\sim$0.5~keV). However, our temperatures agree with theirs to within 1$\sigma$ uncertainties outside 50~kpc. This may be a result of their deprojection extending to larger radii, causing the subtraction of more soft emission in the outer bins. As a result of the higher temperatures, Morandi et al. find somewhat higher entropies than we do, and we note that they also find somewhat lower M$_{500}$ and R$_{500}$ values, based on the $Y_X - M$ relation.

\subsection{LGG~185 / NGC~3078}
LGG~185 is an X-ray faint system. The short ($\sim$8~ks) ACIS-I pointing does detect diffuse $\sim$0.45~keV thermal emission in and around the BGE, NGC~3078, but this only extends to $\sim$100\arcs\ or $\sim$16~kpc radius. The galaxy hosts a core-dominated radio source with extensions to east and west visible in our GMRT 235 and 610~MHz data as well as at 1.5~GHz \citep{WrobelHeeschen84}. 

\subsection{LGG~262 / NGC~4008}
Our \xmms\ observation of LGG~262 was severely affected by background flaring. We excluded all but $\sim$5~ks of data to remove the strongest flares, but low-level flaring continues through the remainder of the observation, seriously impacting the EPIC-pn data. Nonetheless we were able to detect diffuse $\sim$0.4~keV thermal emission extending to 90\arcs\ ($\sim$25~kpc) around the BGE, NGC~4008. The galaxy also hosts a radio point source visible in the NVSS and our GMRT observations.  

\subsection{LGG~276 / NGC~4169}
HCG~61 forms the core of LGG~276. No hot IGM is detected in the available short ($\sim$12~ks) \xmms\ pointing but the AGN in several of the member galaxies, including the BGE NGC~4169, are detected. The BGE is not detected at 235~MHz, but is visible as a point source at 610~MHz and in the FIRST 1.4~GHz survey \citep{Beckeretal95}.

\subsection{LGG~278 / NGC~4261}
Early \rosat\ observations identified the presence of a hot IGM extending $>$40\arcm\ ($>$370~kpc) from the group dominant elliptical, NGC~4261, and the IGM has been studied in some detail using both \chandra\ and \xmms\ data \citep{Crostonetal05b,Crostonetal08,OSullivanetal11c}. NGC~4261 is an FR-I radio galaxy (3C~270) and one of the largest and most radio-luminous sources in the CLoGS sample. \chandra\ revealed X-ray jets corresponding to the inner few kiloparsecs of the radio jets \citep{Gliozzietal03,Zezasetal05}, confirming that the jets are aligned close to the plane of the sky. The jets have swept clear conical regions in the galaxy core \citep{Worralletal10}, while the lobes have inflated large cavities in the IGM, surrounded by rims of compressed gas. Despite the large enthalpy of these cavities and the possibility of shock heating, the X-ray data show that a $\sim$10~kpc radius cool core is still present. Further details of our analysis of the X-ray and radio data for this system are presented in \citep{OSullivanetal11c} and Kolokythas et al. (in prep.).

\subsection{LGG~310 / ESO~507-25}
LGG~310 is an X-ray faint system, our short \chandra\ pointing detecting only slightly extended powerlaw emission coincident with the BGE, ESO~507-25. The galaxy is also detected in the NVSS 1.4~GHz survey and our GMRT 610 and 235~MHz data, with some amorphous extended emission surrounding the radio core at 610~MHz.

\subsection{LGG~338 / NGC~5044}
LGG~338 is the X-ray brightest group in the sky, and as such has a wealth of data available. \rosat\ detected the IGM to at least 26\arcm\ ($\sim$290~kpc) from the BGE, NGC~5044 \citep{Davidetal94}, and hinted at structures in the group core. \chandra\ observations have shown that the core contains a host of filaments and small cavities \citep{Davidetal09,Gastaldelloetal09}, indicating a complex history of AGN jet outbursts and gas motions \citep{Davidetal11}. \xmms\ observations show that the group is sloshing \citep{Gastaldelloetal13,OSullivanetal14a} producing asymmetric temperature, density and abundance structures. The \xmms\ spectral profiles presented in this paper are extracted from the northeast and southwest quadrants described in \citet{OSullivanetal14a} so as to avoid the impact of the asymmetries in the southeast and northwest quadrants. The \chandra\ profiles, which are extracted from the ACIS-S3 chip, only extend a few arcminutes and are less affected by this issue; we therefore use full 360\degree\ azimuthally averaged regions.

Radio observations of NGC~5044 show that current nuclear activity is limited to point-like emission. Evidence of past outbursts is visible in the form of a small 610~MHz extension coincident with one of the brightest X-ray filaments, and a distorted radio filament and detached lobe visible at 235~MHz \citep{Giacintuccietal11}. These morphology of the latter structure appears to have been affected by the sloshing motion of the group, which may also have compressed the radio plasma, causing reacceleration and increasing its radio brightness \citep{OSullivanetal14a}. 

\subsection{LGG~345 / NGC~5084}
The BGE of LGG~345, NGC~5084, is an edge-on S0 with a prominent \Hi-rich disk \citep[e.g.,][]{Pisanoetal11} containing $\sim$10$^{10}$\Msol\ of neutral hydrogen \citep{Koribalskietal04}. Our short \chandra\ pointing shows no evidence of a hot IGM, but does detect powerlaw emission extending $\sim$100\arcs\ ($\sim$10~kpc), with a central point source contributing the majority of the emission. The galaxy is also detected at 1.4~GHz in the NVSS, but not in our GMRT data.

\subsection{LGG~351 / NGC~5153}
LGG~351 is an X-ray faint system, our short \chandra\ observation detecting emission consistent with that expected from LMXBs within $\sim$50\arcs ($\sim$15~kpc) of the BGE, NGC~5153. This peculiar elliptical is interacting with the highly disturbed spiral NGC~5152. Neither galaxy is detected in our GMRT observations, or in the 1.4~GHz VLA surveys. 

\subsection{LGG~363 / NGC~5353}
HCG~68 forms the core of LGG~363, with the BGE NGC~5353 located close to, and likely interacting with, another early-type galaxy NGC~5454. The group is both X-ray luminous \citep[e.g.,][]{Mulchaeyetal03,OsmondPonman04} and cold gas rich, containing \gtsim5$\times$10$^9$\Msol\ of \Hi\ \citep{Borthakuretal10}. The hot IGM is centred on the BGE, extends at least 100~kpc, and is approximately isothermal \citep[e.g.,][]{Finoguenovetal07}. Radio and X-ray point sources are detected in the BGE and a number of subsidiary galaxies, most notably the face-on spiral NGC~5350, located north of the BGE, in whose disk diffuse radio emission is also visible.

\subsection{LGG~393 / NGC~5846}
LGG~393 is another well-known X-ray luminous group centred on the BGE NGC~5846. \chandra\ and \xmms\ observations have show the group to be sloshing \citep{Machaceketal11,Gastaldelloetal13} and as with LGG~338 we extract spectral profiles using only the northeast and southwest quadrants to avoid the asymmetries caused by the gas motion. The BGE hosts an AGN with small-scale jets and lobes \citep{Giacintuccietal11} which have inflated cavities in the centre of the cool core of the group \citep{Machaceketal11}.

\subsection{LGG~402 / NGC~5982}
LGG~402 has not previously been identified as an X-ray bright group. The
BGE, NGC~5982, was detected in the RASS \citep{Beuingetal99} but our \xmms\
observation was required to confirm the presence of an extended IGM. The
temperature profile is centrally peaked, with the scale of the innermost
bin ($\sim$6~kpc) making an unresolved small cool core unlikely. Despite
this, NGC~5982 has a low entropy in its central temperature bin, 14.4$\pm$2.9\kevcmsq, and a
relatively short central cooling time, $\sim$250~Myr. Its AGN is detected as a
radio point source, and cool gas is detected in the galaxy, in the form of
A 3.4$\times$10$^7$~\Msol\ \Hi\ cloud is located 6~kpc east and 200\kmps\
offset from the nucleus \citep{Morgantietal06} and 2.3$\times$10$^4$~\Msol\
of H$\alpha$-emitting ionized gas in the kinematically distinct core
\citep{Sarzietal06}. We will present a more detailed analysis of the galaxy
in a later paper.

NGC~5982 is bracketed by two spiral galaxies, both of which are detected in
our radio observations. NGC~5985, a face-on spiral to the east of NGC~5982,
is noteworthy in that star formation activity in its disk is visible both
in the radio and in our \xmms\ data.

\subsection{LGG~421 / NGC~6658}
No hot IGM is detected in our \xmms\ observation of LGG~421. The BGE NGC~6658, and another member galaxy NGC~6660, are detected as point-like sources, and spectral fitting shows evidence of both thermal and powerlaw emission in the BGE. Neither galaxy hosts a radio source.

\subsection{LGG~473 / NGC~7619}
Although identified as a single group in a number of optical catalogues, X-ray observations show that this is a merging system. \rosat\ observations showing the surface brightness distribution peaked on the dominant ellipticals NGC~7619 and NGC~7626 \citep[e.g.,][]{Mulchaeyetal03}, and \chandra\ mapping confirmed that each galaxy sits at the centre of a cool core, with a ridge of hotter, possibly shock-heated gas between them \citep{Randalletal09}. NGC~7619 is the BGE of LGG~473, and is also located in the brighter and more extended of the two X-ray peaks. However, it only hosts a radio point source, while NGC~7626 is an FR-I radio galaxy with $\sim$100~kpc scale jets and lobes \citep{Giacintuccietal11}. In our X-ray profile analysis we use a 160\degree\ wedge extending west from NGC~7619, so as to avoid inclusion of material associated with NGC~7626 of the potentially shocked gas between the two.

\section{Images}
\label{sec:images}
X-ray, optical and radio images of our groups. The \textit{upper left} panel shows an adaptively smoothed, exposure corrected 0.3-2~keV \xmms\ MOS+pn or \chandra\ ACIS image with the positions of group-member galaxies marked with crosses. The \textit{upper right} panel shows a Digitized Sky Survey optical image with X-ray contours overlaid; the contours are intended to highlight diffuse emission, and emission associated with group member galaxies. The \textit{lower left} panel shows a GMRT radio image (frequency labelled) with group-member galaxies marked (circles are used instead of crosses to avoid obscuring small sources). All three of these panels share a common scale and orientation. The \textit{lower right} panel shows an X-ray image of the group core with radio contours overlaid, showing any radio source associated with the dominant early-type galaxy. Radio contours start at 3$\sigma$ above the r.m.s. noise level and increase in steps of factor 2. Typical r.m.s. noise values are in the ranges 0.05-0.1~mJy~bm$^{-1}$ for 610~MHz and 0.3-0.6~mJy for 235~MHz (see Kolokythas et al., in prep., for more details)

\begin{figure*}
\includegraphics[width=\textwidth]{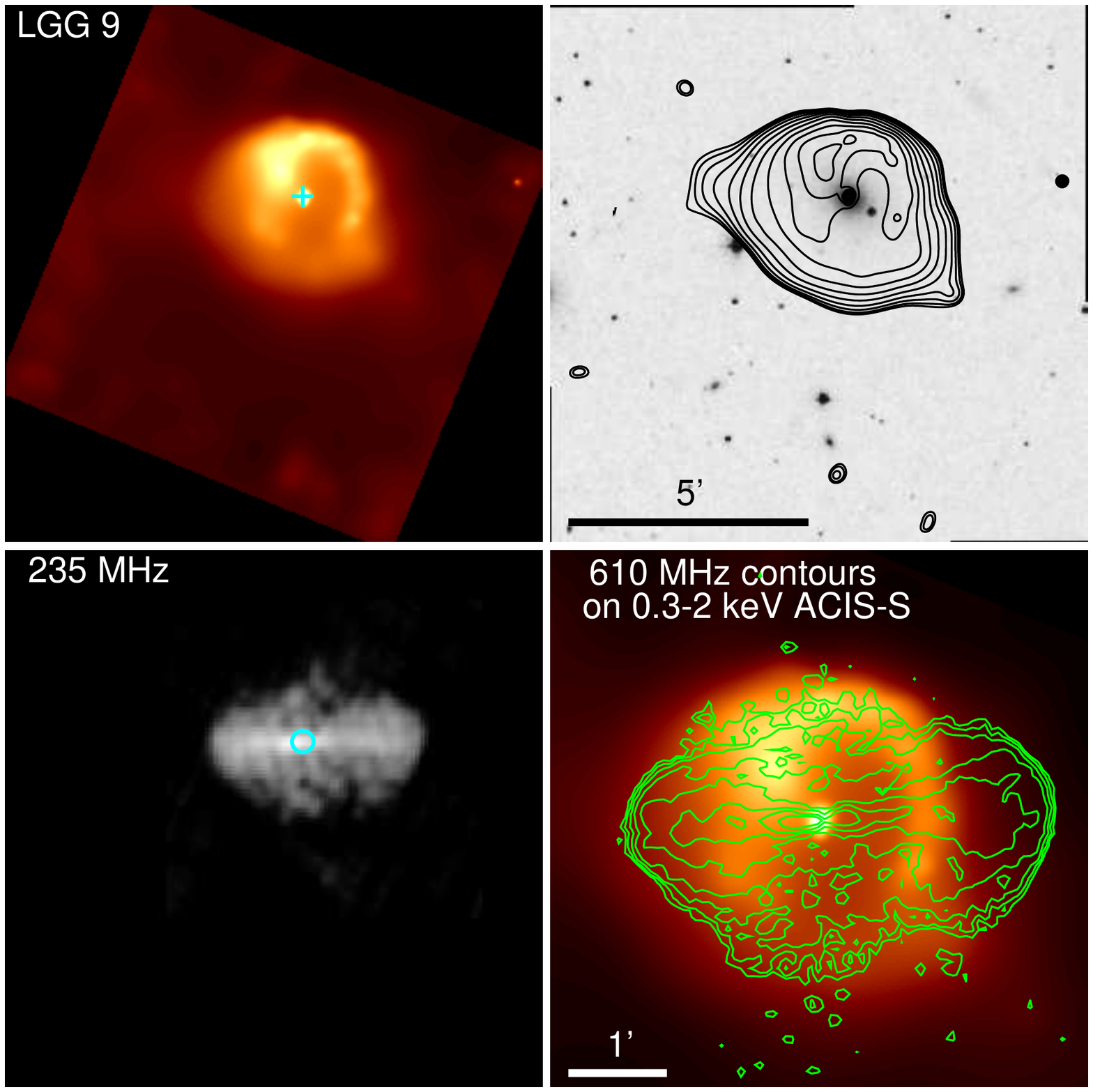}
\caption{LGG 9 / NGC 193. 1\arcm\ = 21.5~kpc.}
\end{figure*}

\clearpage
\begin{figure*}
\includegraphics[width=\textwidth]{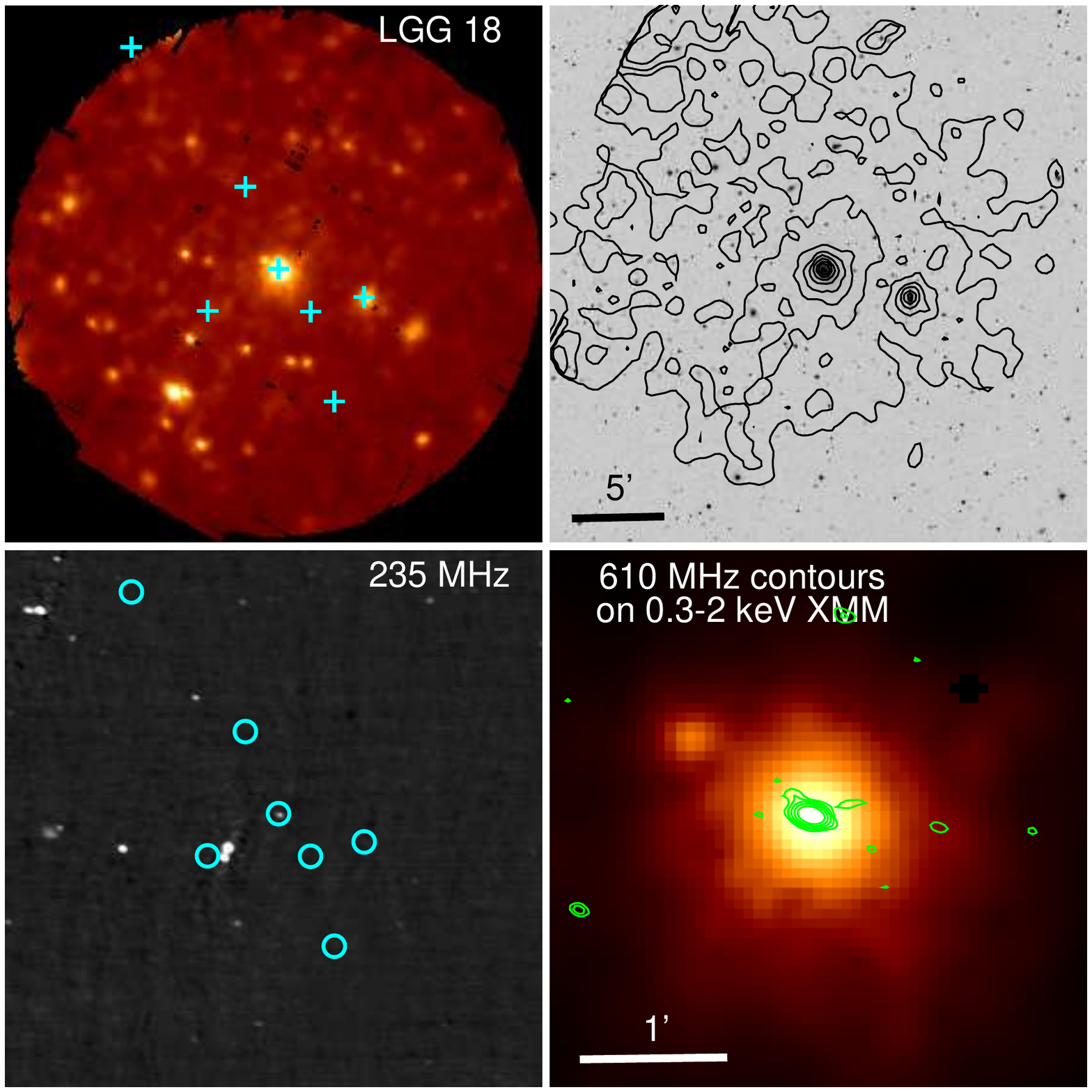}
\caption{LGG 18 / NGC 410. 1\arcm\ = 22.4~kpc.}
\end{figure*}

\clearpage
\begin{figure*}
\includegraphics[width=\textwidth]{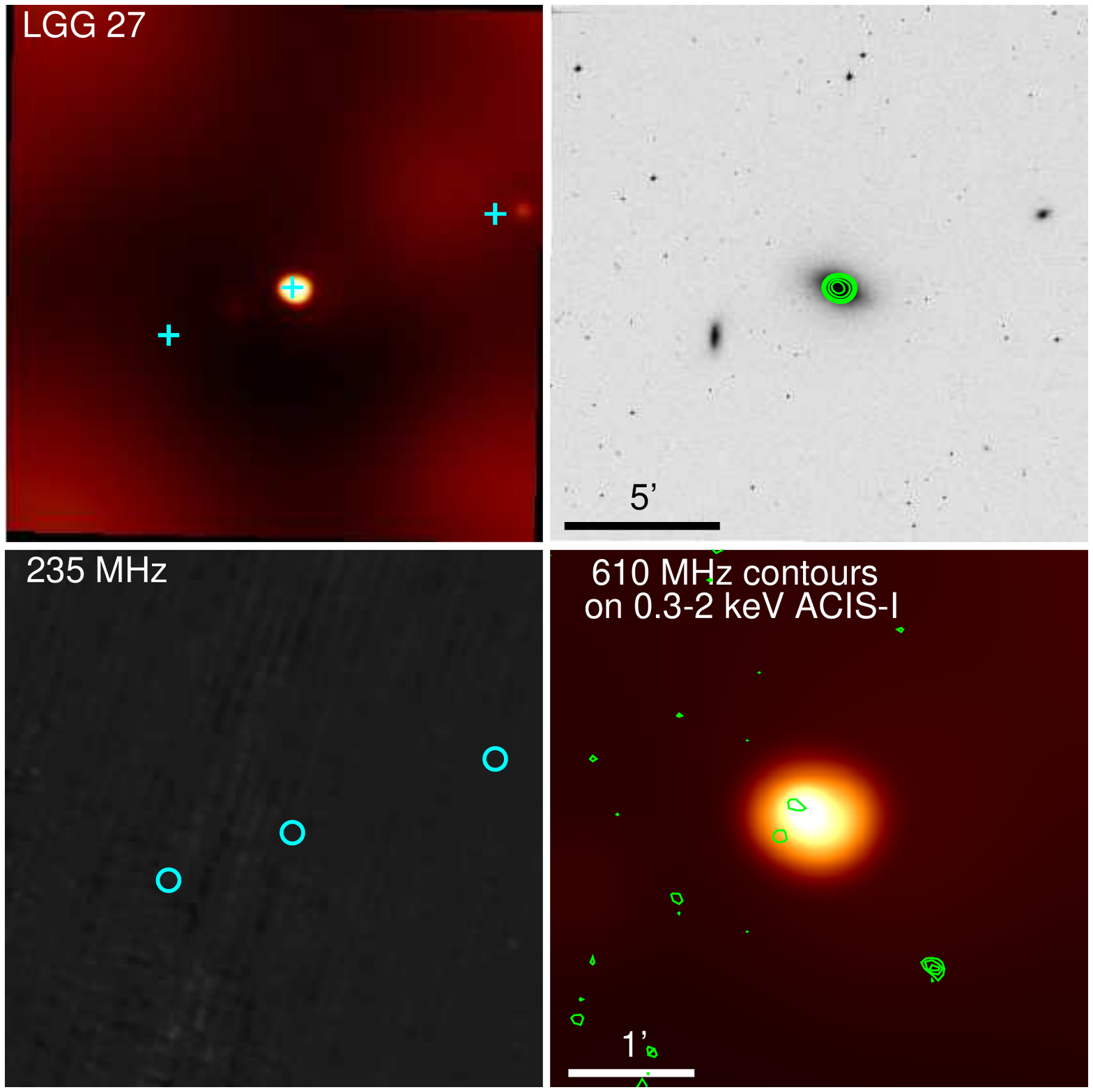}
\caption{LGG 27 / NGC 584. 1\arcm\ = 7.3~kpc.}
\end{figure*}

\clearpage
\begin{figure*}
\includegraphics[width=\textwidth]{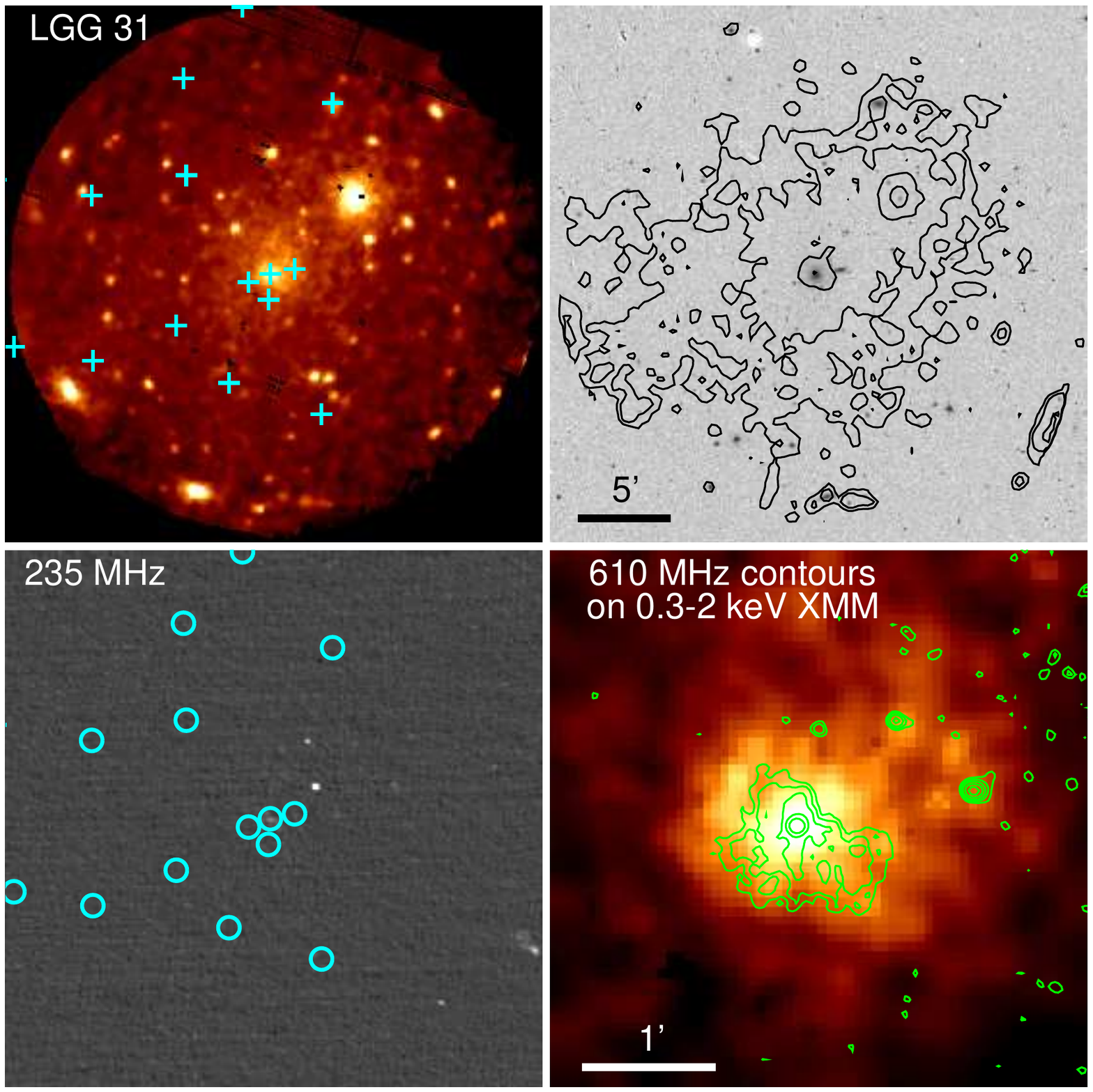}
\caption{LGG 31 / NGC 667. 1\arcm\ = 22.7~kpc. Note that the diffuse source on the upper right is a background cluster.}
\end{figure*}

\clearpage
\begin{figure*}
\includegraphics[width=\textwidth]{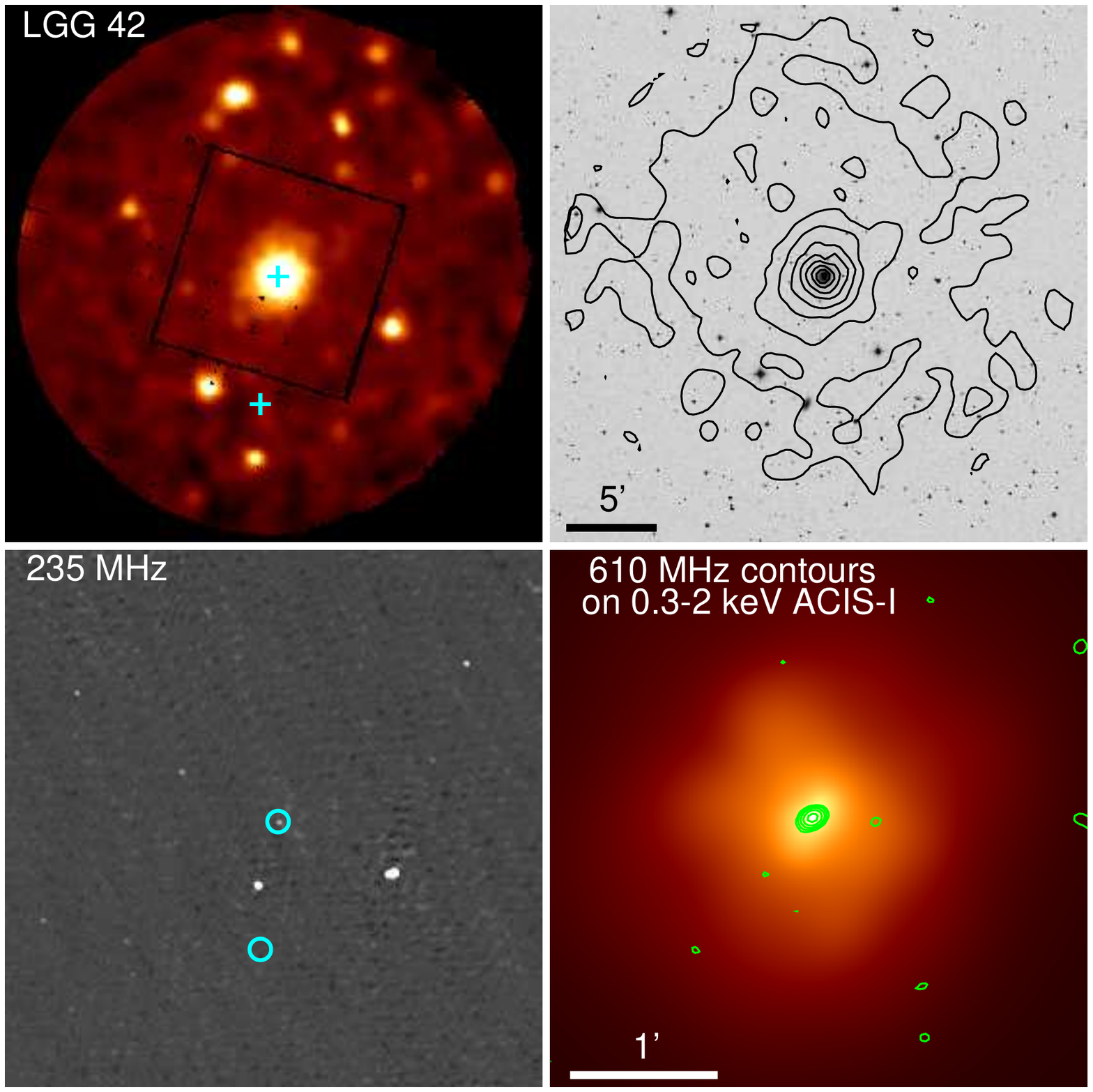}
\caption{LGG 42 / NGC 777. 1\arcm\ = 21.2~kpc.}
\end{figure*}

\clearpage
\begin{figure*}
\includegraphics[width=\textwidth]{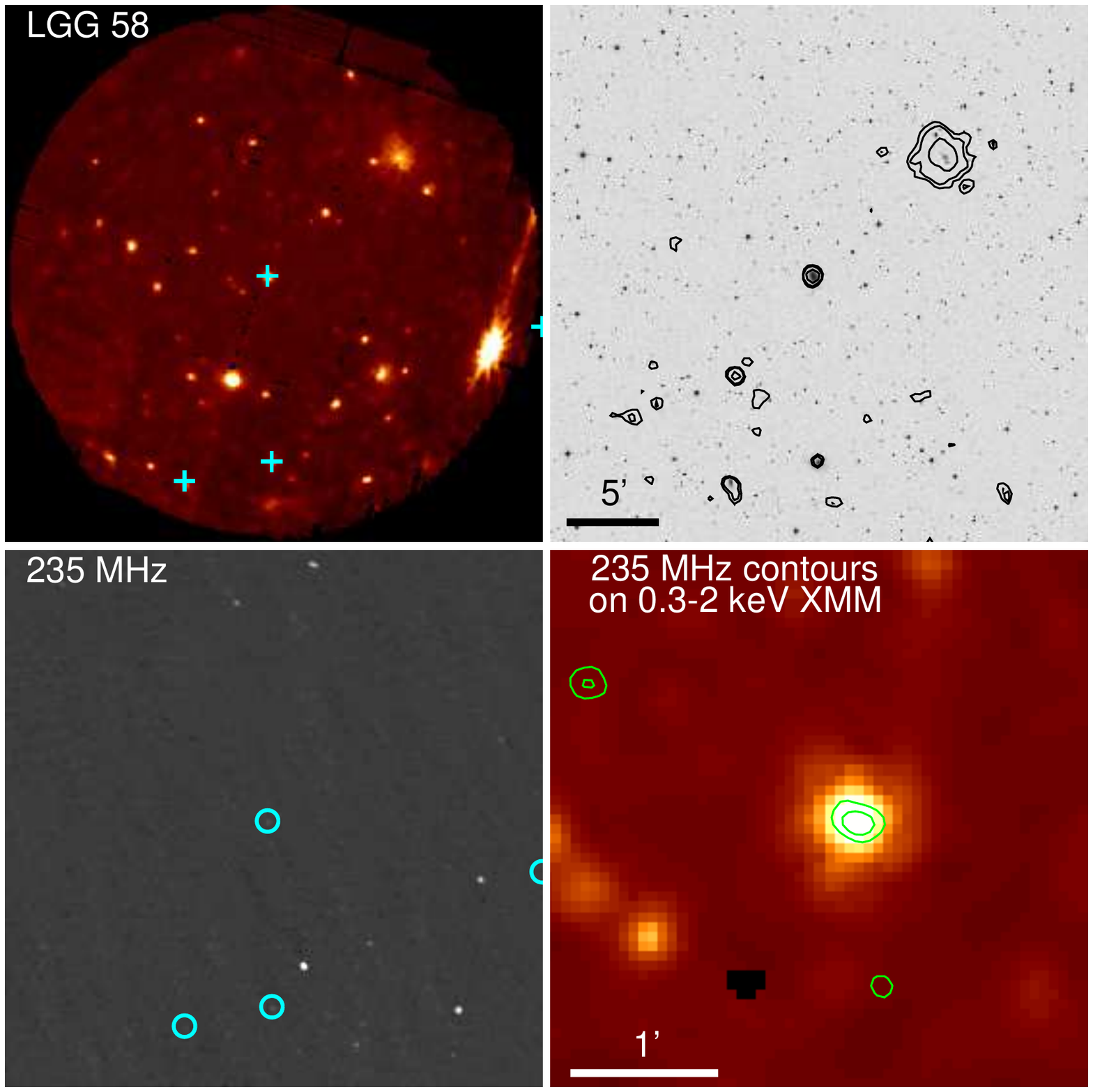}
\caption{LGG 58 / NGC 940. 1\arcm\ = 21.5~kpc. Note that the diffuse source on the upper right is a background group.}
\end{figure*}

\clearpage
\begin{figure*}
\includegraphics[width=\textwidth]{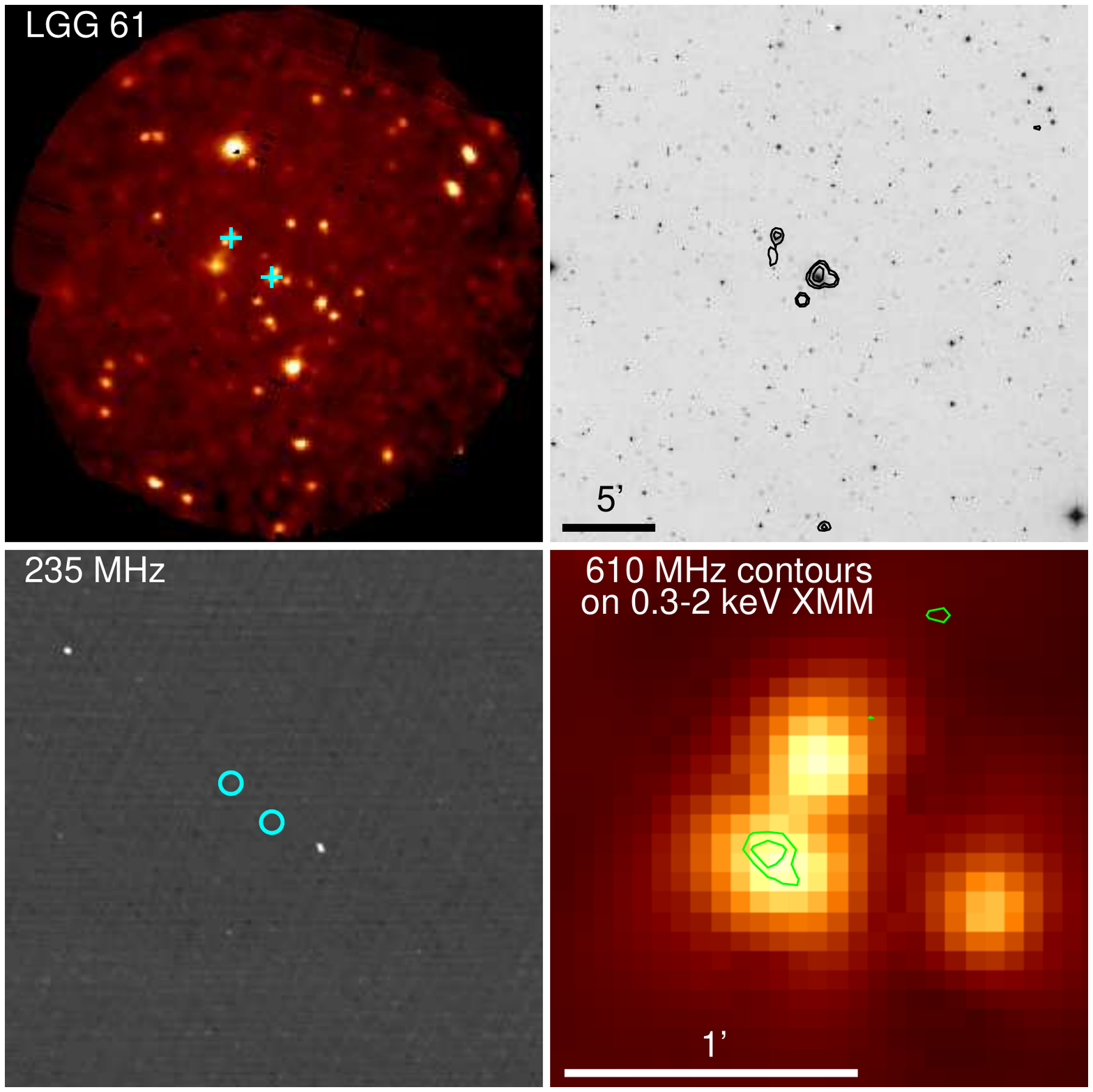}
\caption{LGG 61 / NGC 924. 1\arcm\ = 18.6~kpc.}
\end{figure*}

\clearpage
\begin{figure*}
\includegraphics[width=\textwidth]{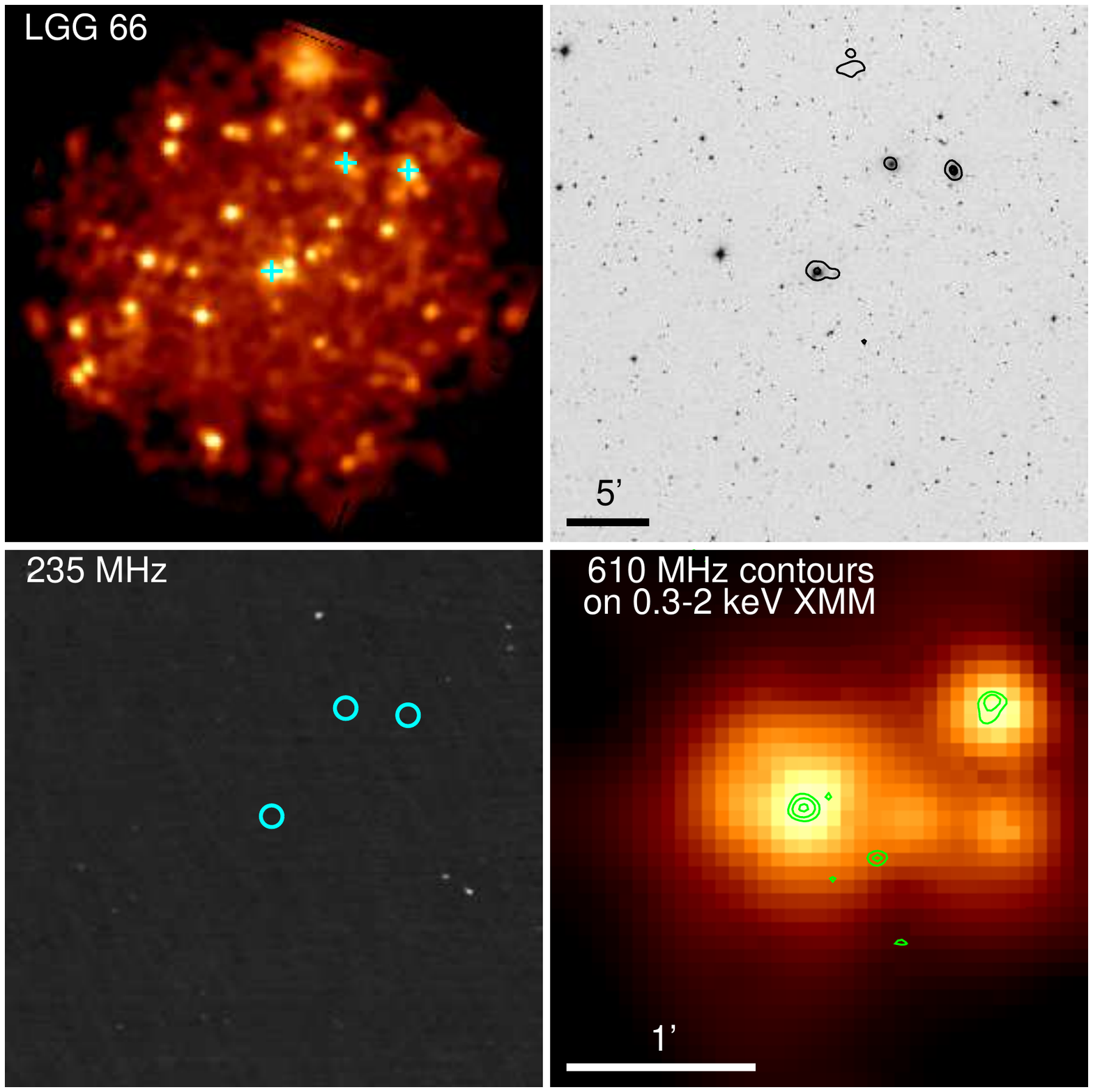}
\caption{LGG 66 / NGC 978. 1\arcm\ = 20.0~kpc.}
\end{figure*}

\clearpage
\begin{figure*}
\includegraphics[width=\textwidth]{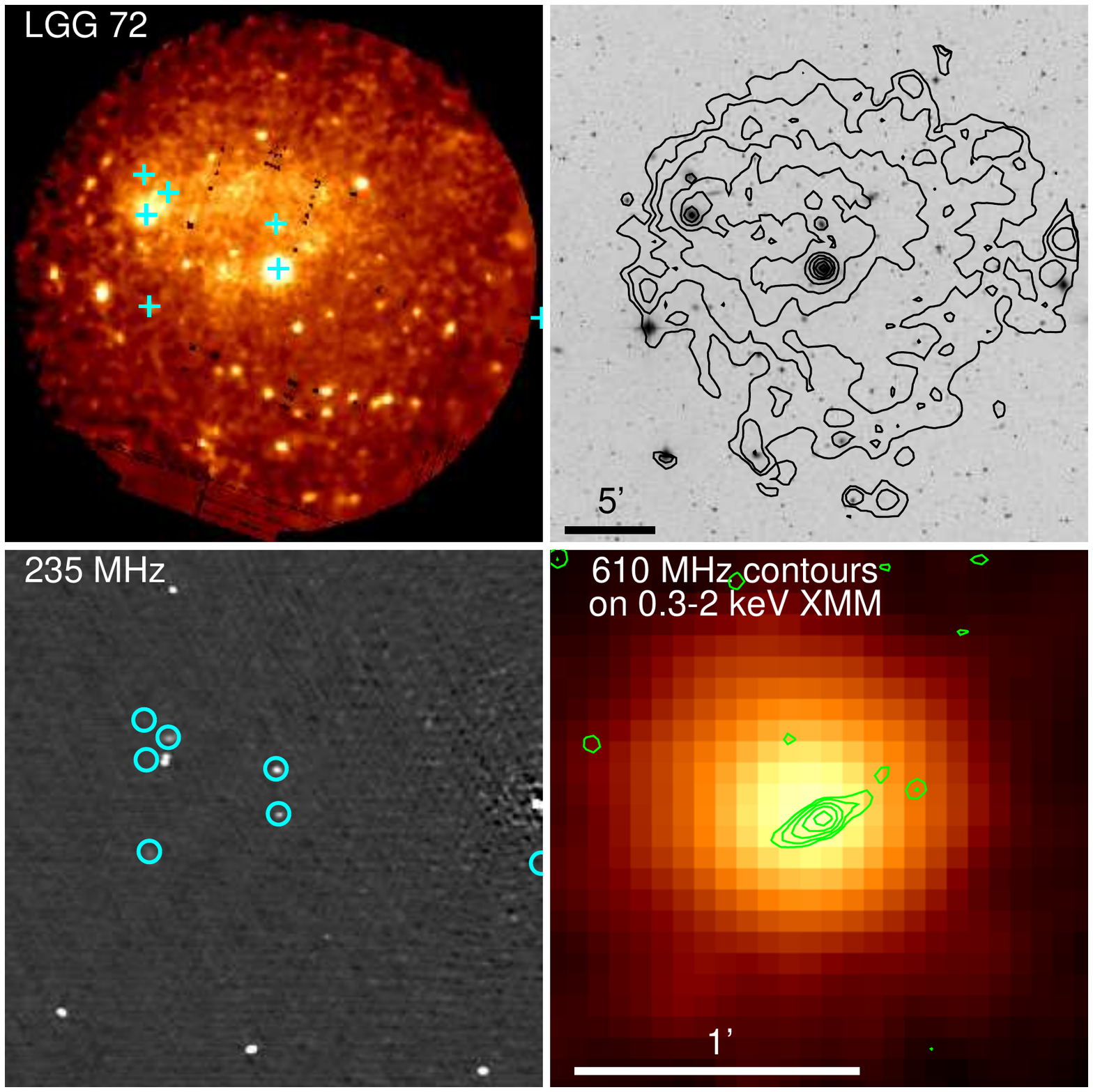}
\caption{LGG 72 / NGC 1060. 1\arcm\ = 22.1~kpc. This is a merging system, with BGE NGC~1060 in the centre of the images, and the dominant galaxy of the secondary core, NGC~1066, on the upper left.}
\end{figure*}

\clearpage
\begin{figure*}
\includegraphics[width=\textwidth]{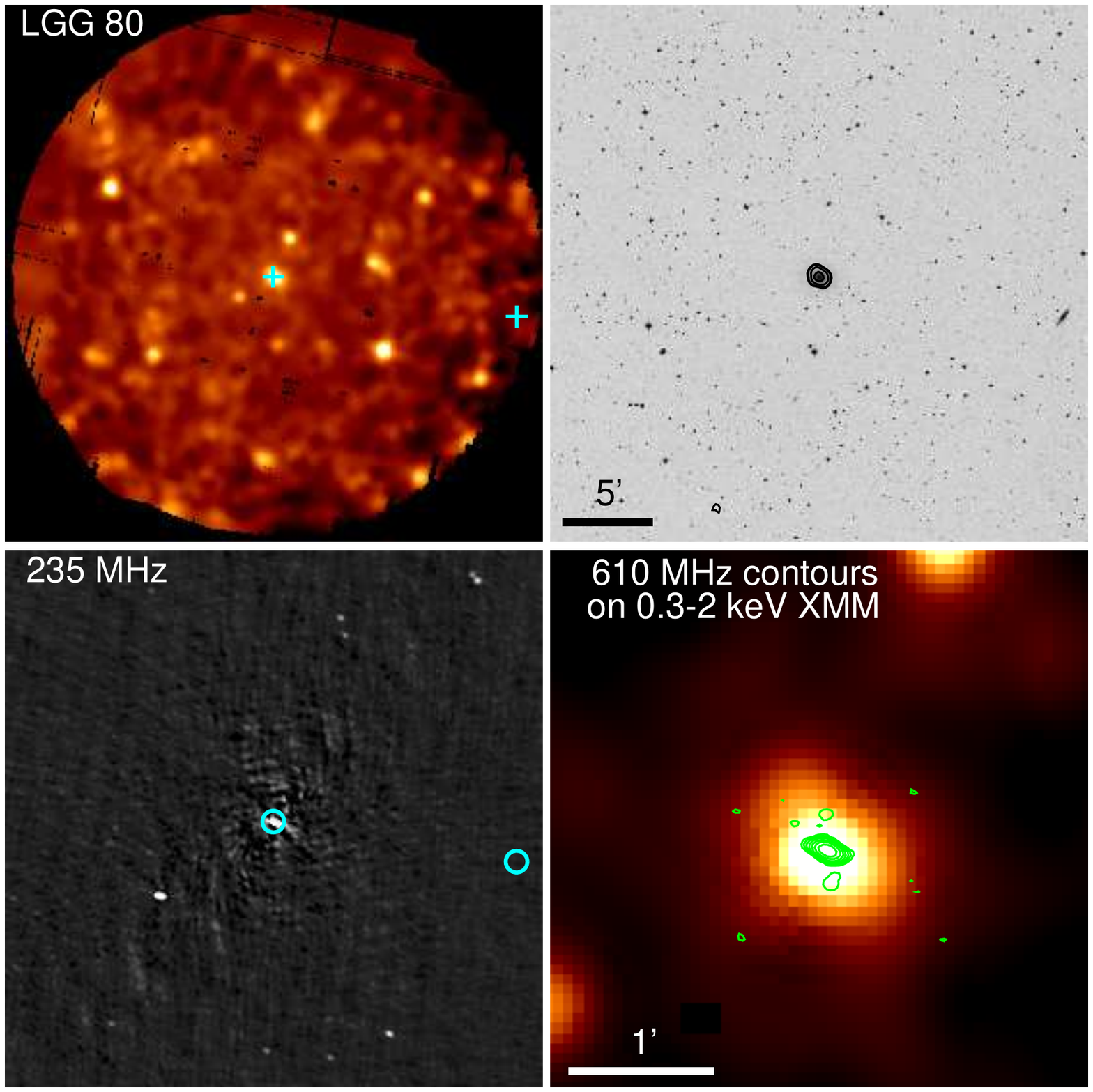}
\caption{LGG 80 / NGC 1167. 1\arcm\ = 20.9~kpc.}
\end{figure*}

\clearpage
\begin{figure*}
\includegraphics[width=\textwidth]{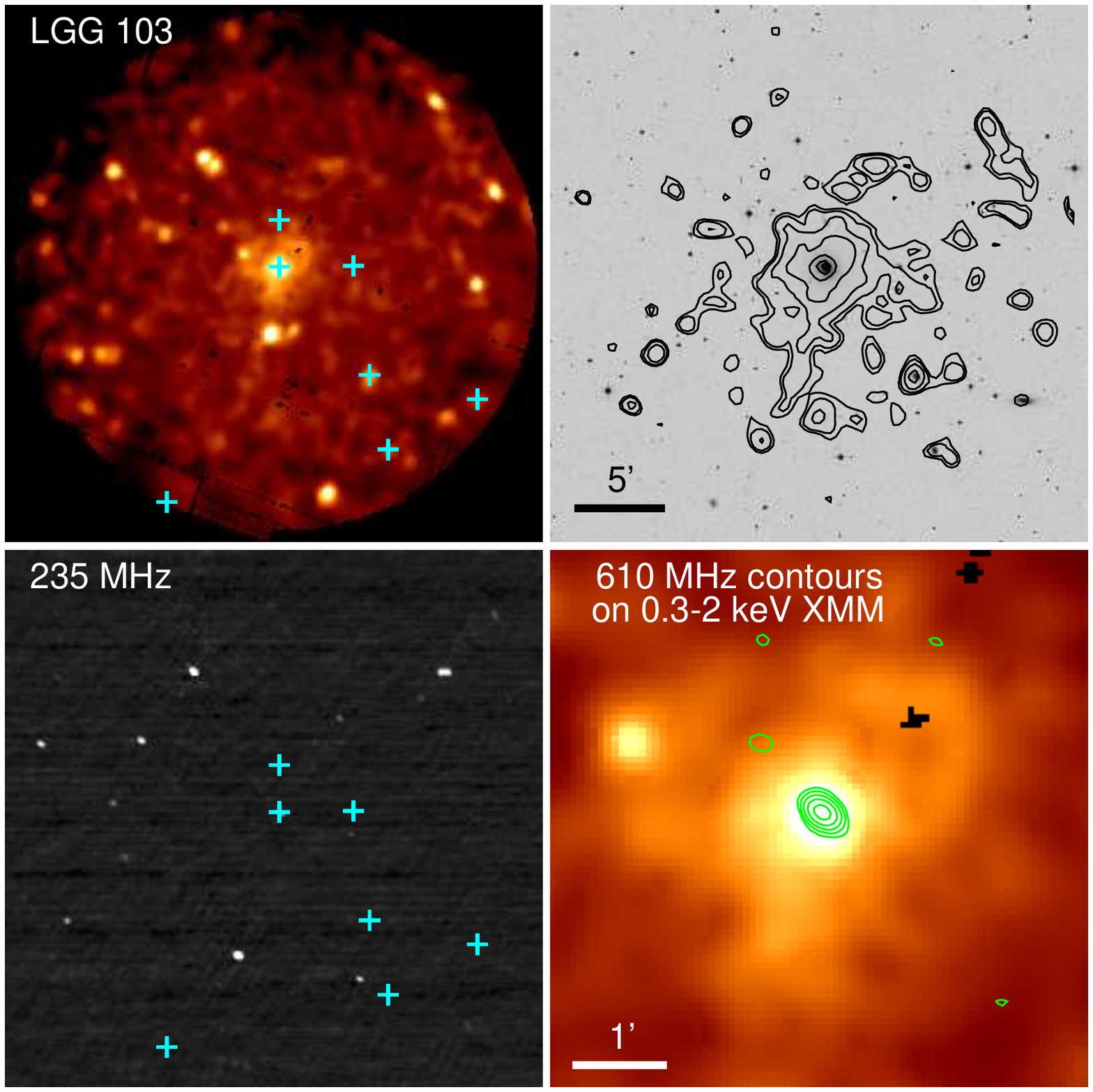}
\caption{LGG 103 / NGC 1453. 1\arcm\ = 18.3~kpc.}
\end{figure*}

\clearpage
\begin{figure*}
\includegraphics[width=\textwidth]{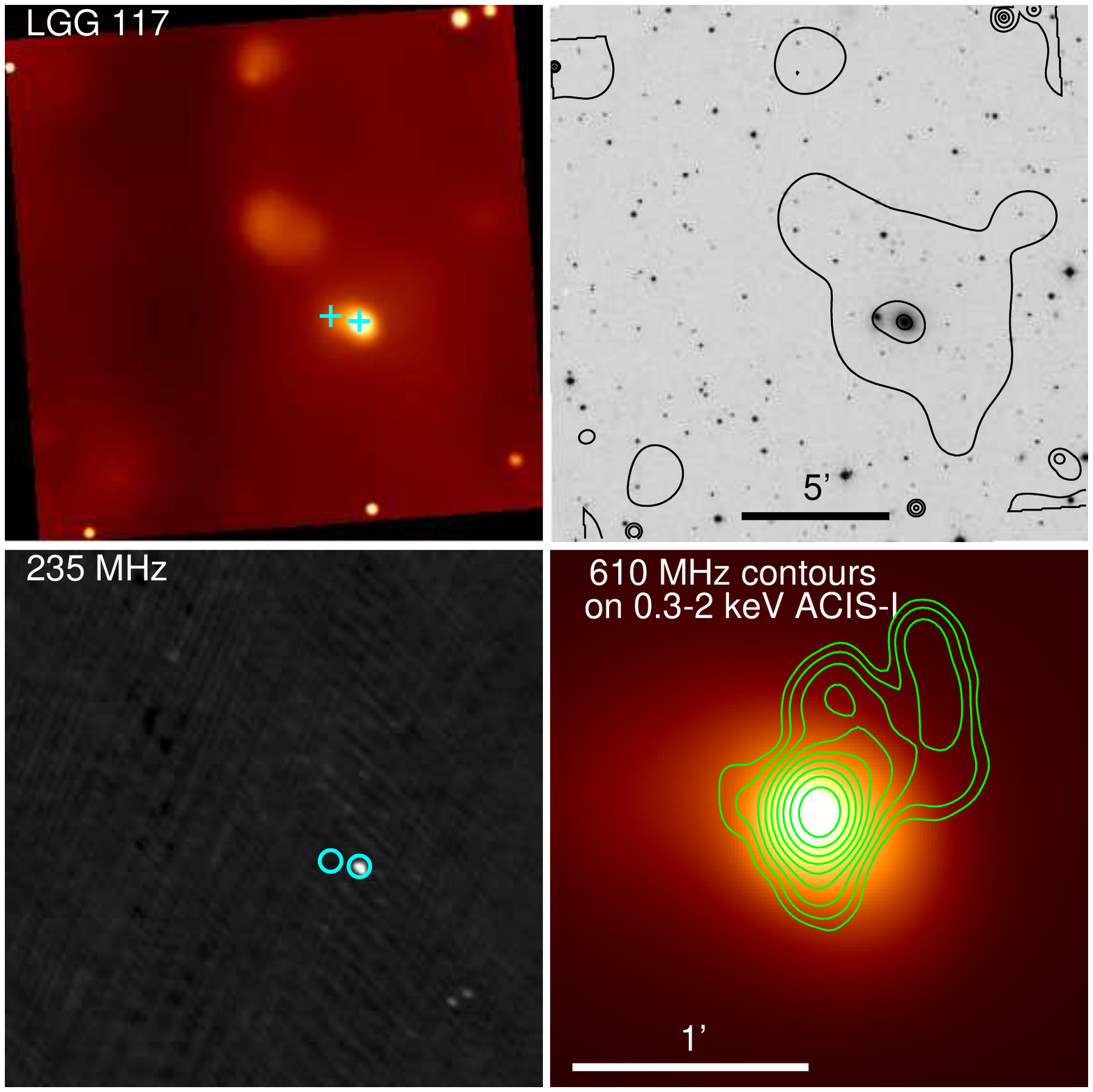}
\caption{LGG 117 / NGC 1587. 1\arcm\ = 14.8~kpc.}
\end{figure*}

\clearpage
\begin{figure*}
\includegraphics[width=\textwidth]{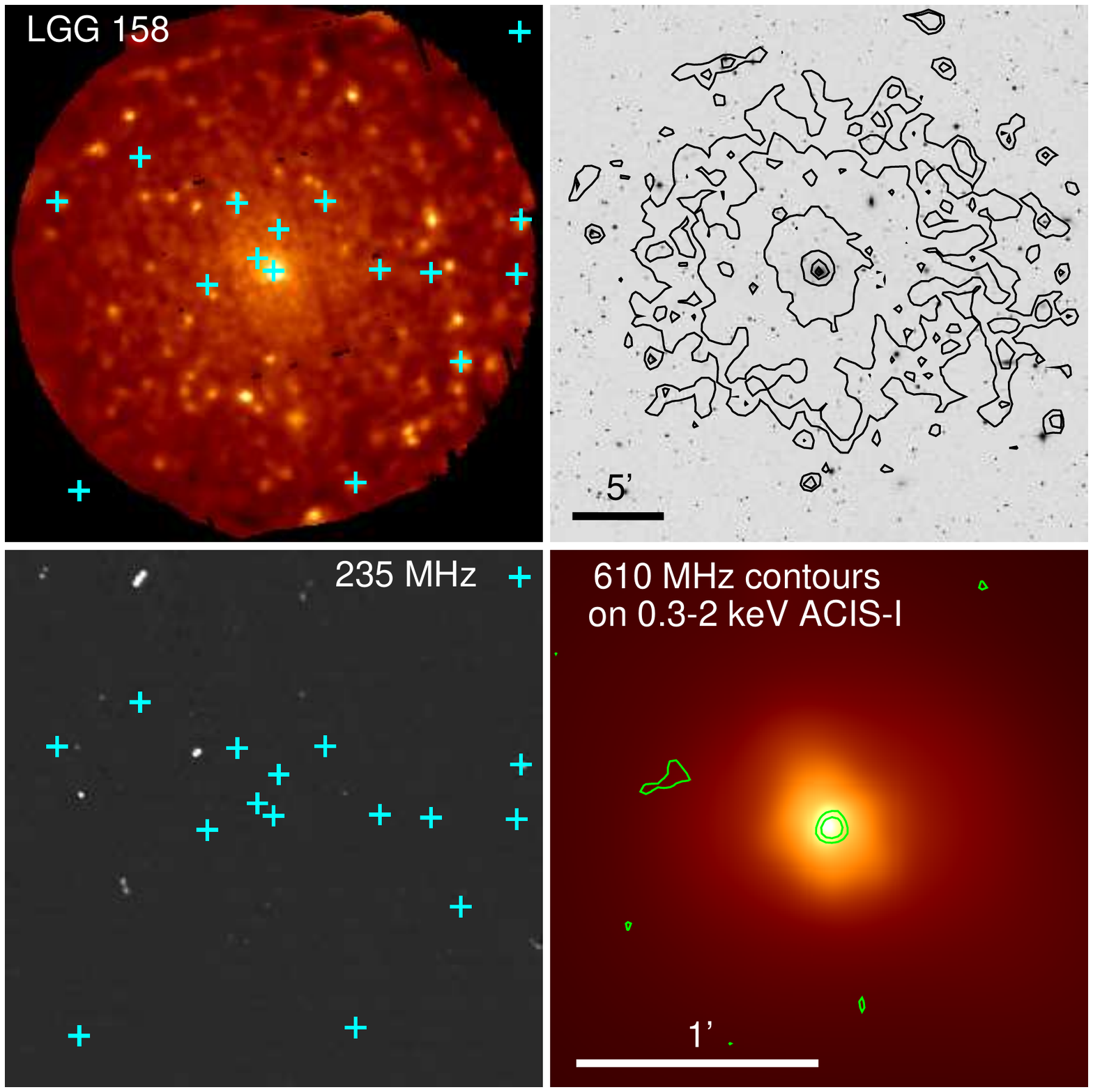}
\caption{LGG 158 / NGC 2563. 1\arcm\ = 18.9~kpc.}
\end{figure*}

\clearpage
\begin{figure*}
\includegraphics[width=\textwidth]{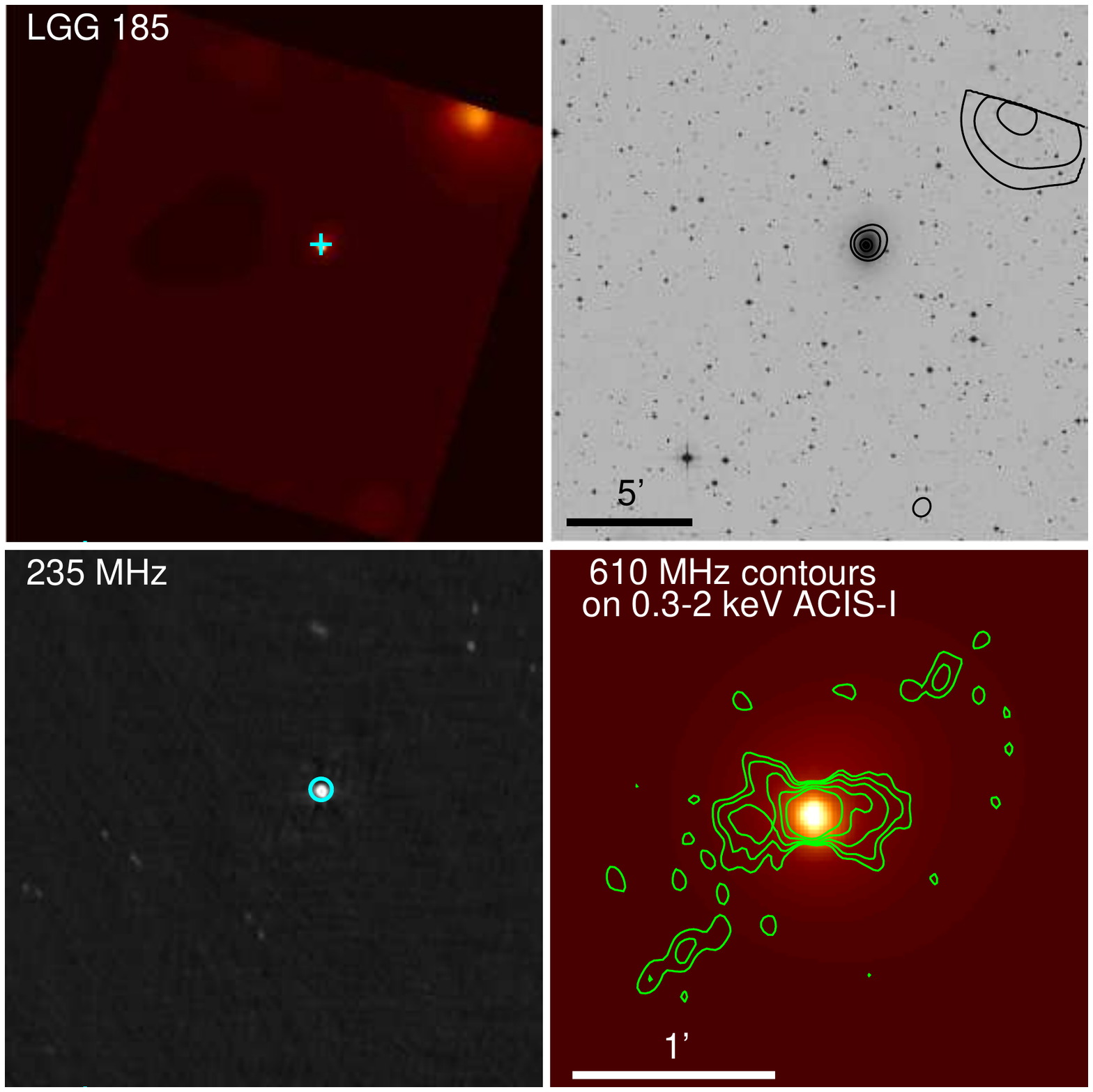}
\caption{LGG 185 / NGC 3078. 1\arcm\ = 9.9~kpc. Note that the diffuse X-ray source on the upper right is background cluster MCXC~J0958.0-2650 at redshift$z$=0.145.}
\end{figure*}

\clearpage
\begin{figure*}
\includegraphics[width=\textwidth]{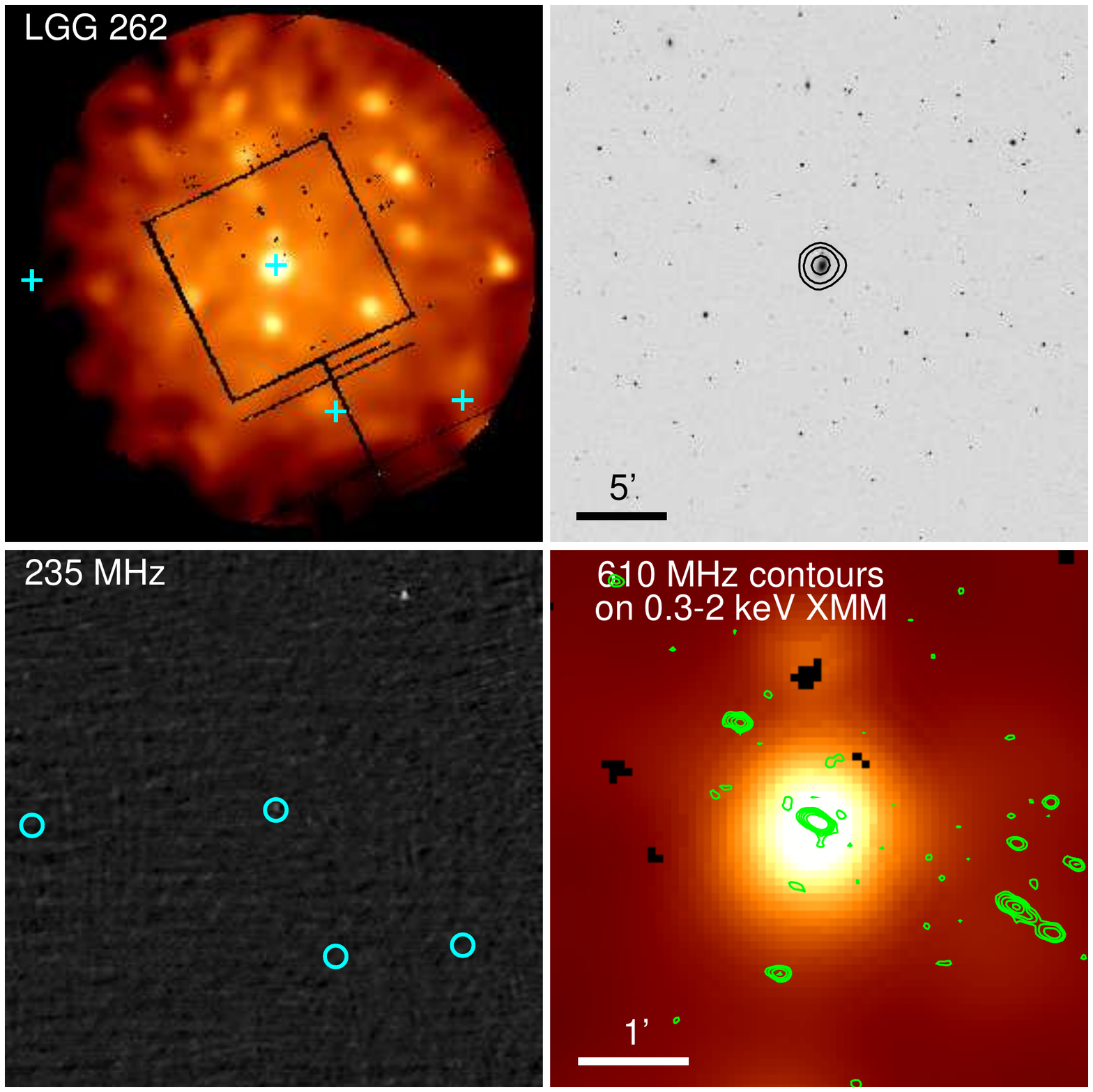}
\caption{LGG 262 / NGC 4008. 1\arcm\ = 15.7~kpc.}
\end{figure*}

\clearpage
\begin{figure*}
\includegraphics[width=\textwidth]{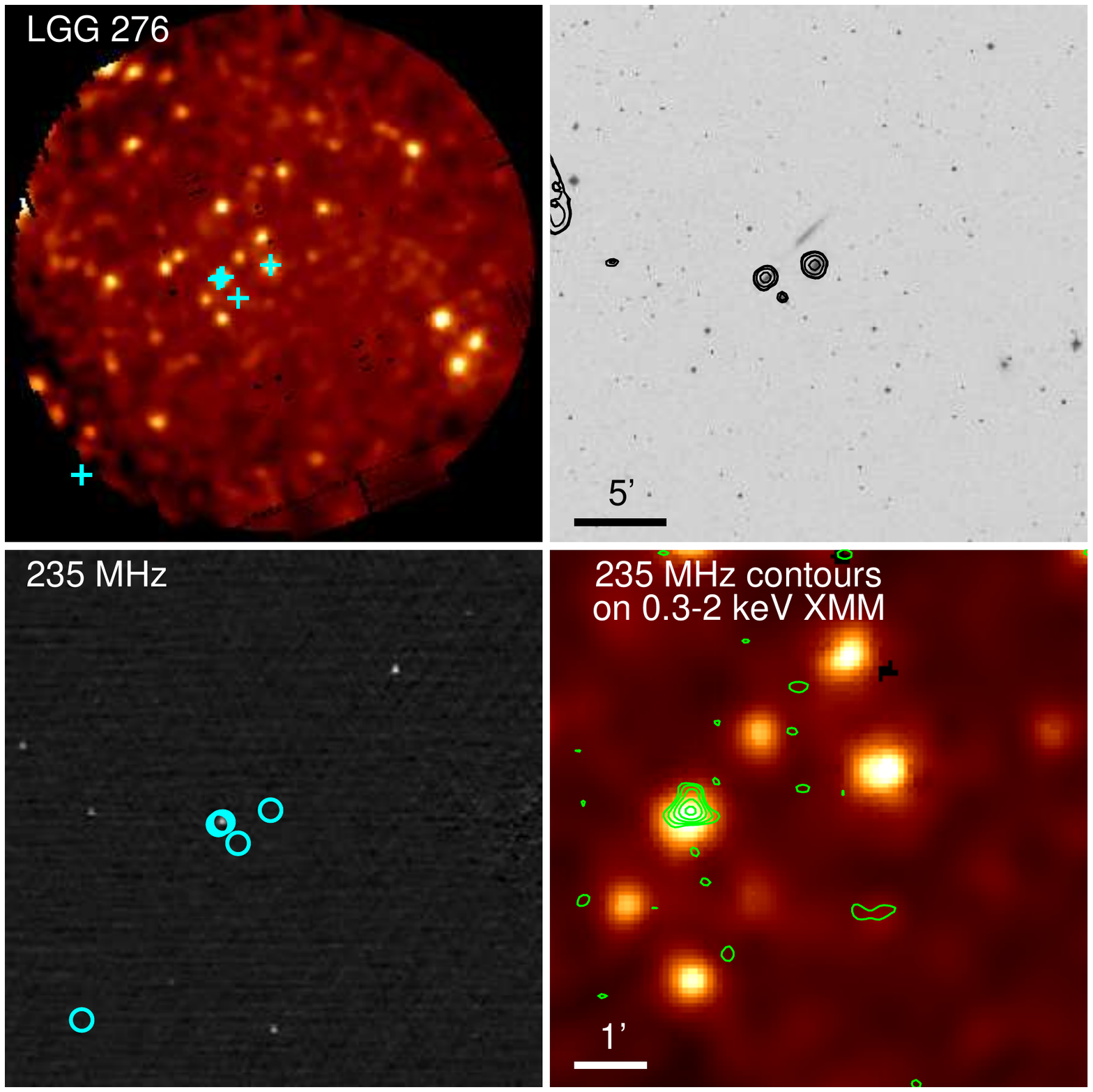}
\caption{LGG 276 / NGC 4169. 1\arcm\ = 13.1~kpc}
\end{figure*}

\clearpage
\begin{figure*}
\includegraphics[width=\textwidth]{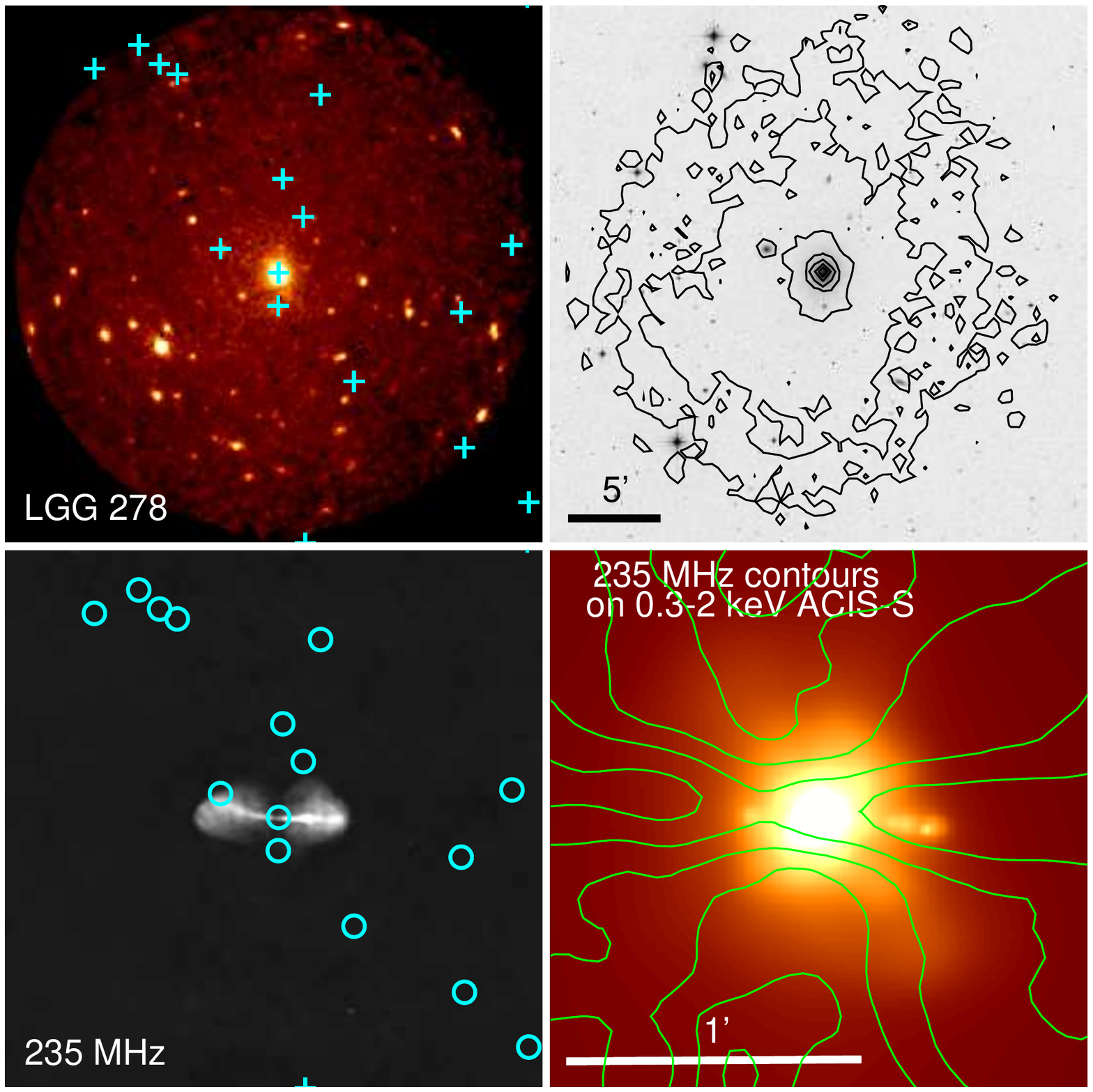}
\caption{LGG 278 / NGC 4261. 1\arcm\ = 9.3~kpc.}
\end{figure*}

\clearpage
\begin{figure*}
\includegraphics[width=\textwidth]{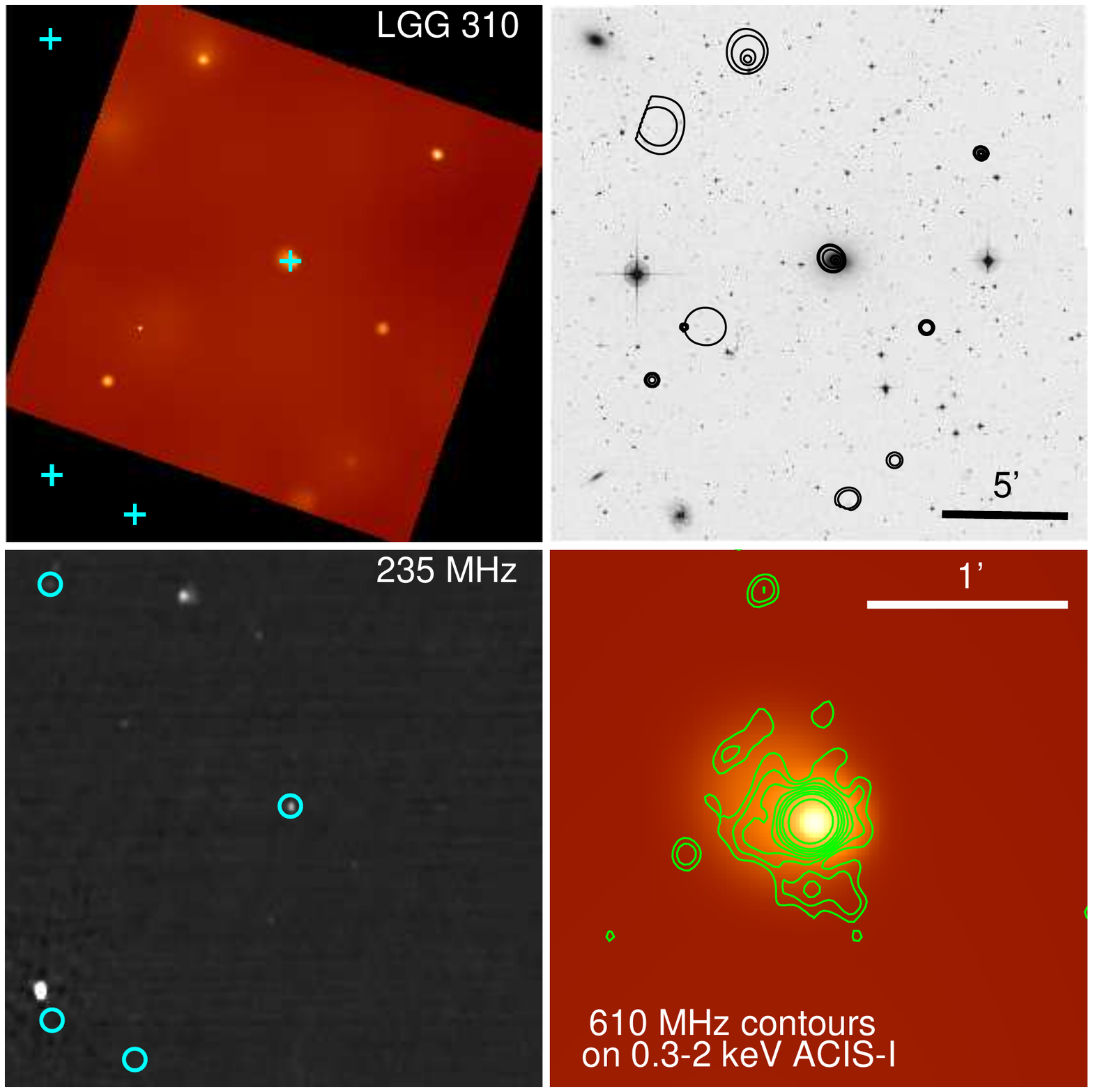}
\caption{LGG 310 / ESO~507-25. 1\arcm\ = 13.1~kpc.}
\end{figure*}

\clearpage
\begin{figure*}
\includegraphics[width=\textwidth]{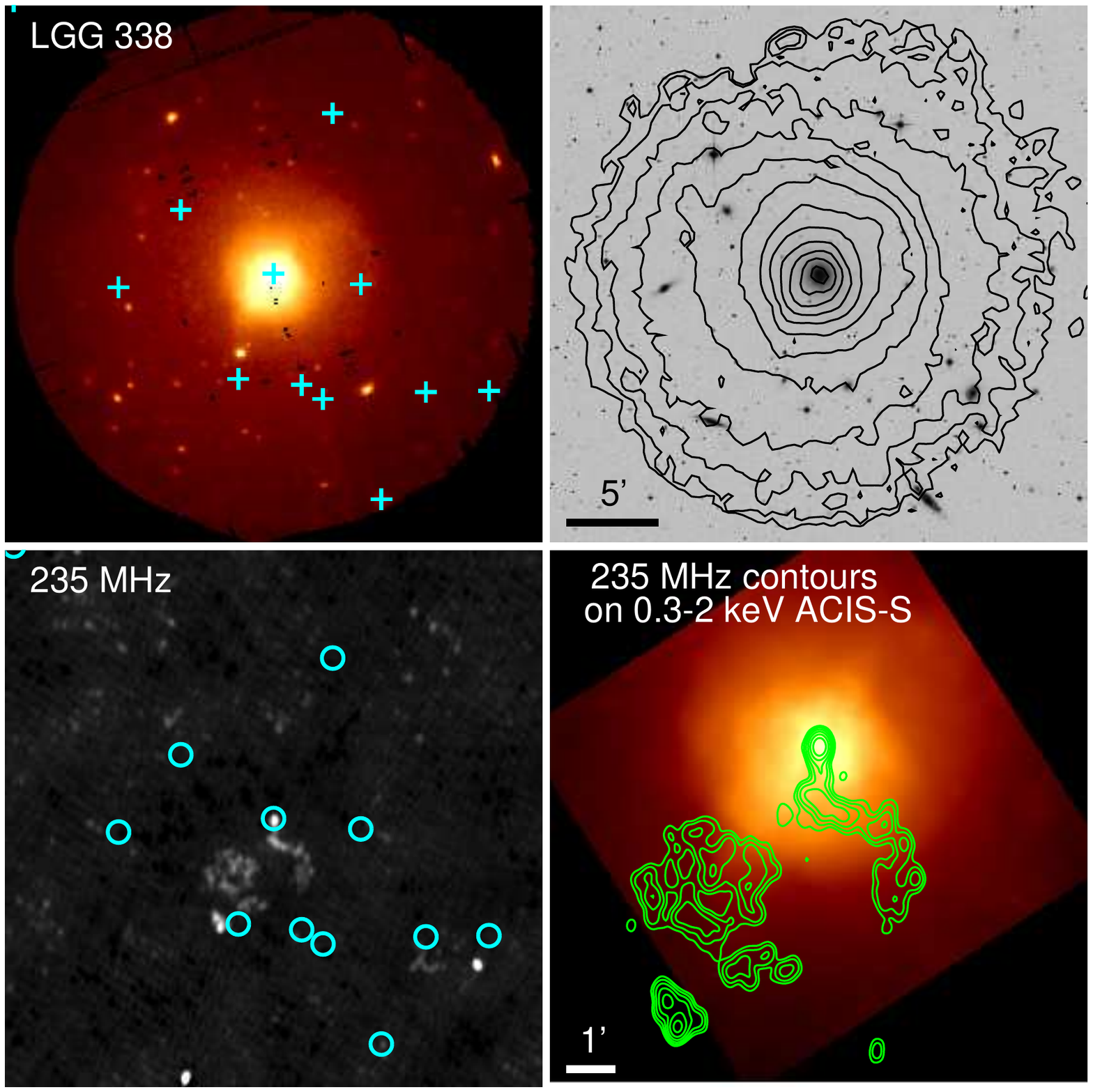}
\caption{LGG 338 / NGC 5044. 1\arcm\ = 11.0~kpc.}
\end{figure*}

\clearpage
\begin{figure*}
\includegraphics[width=\textwidth]{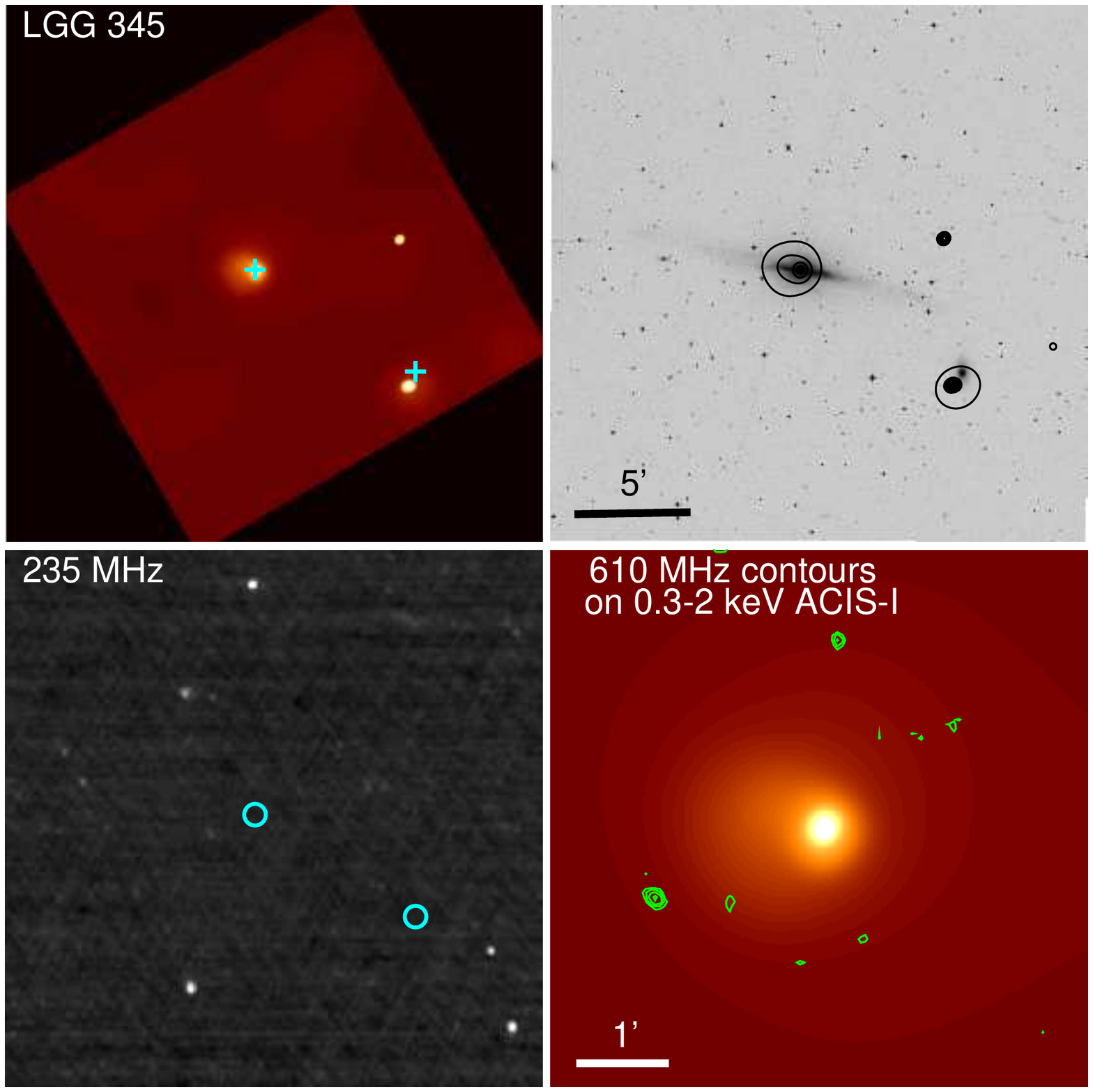}
\caption{LGG 345 / NGC 5084. 1\arcm\ = 6.7~kpc.}
\end{figure*}

\clearpage
\begin{figure*}
\includegraphics[width=\textwidth]{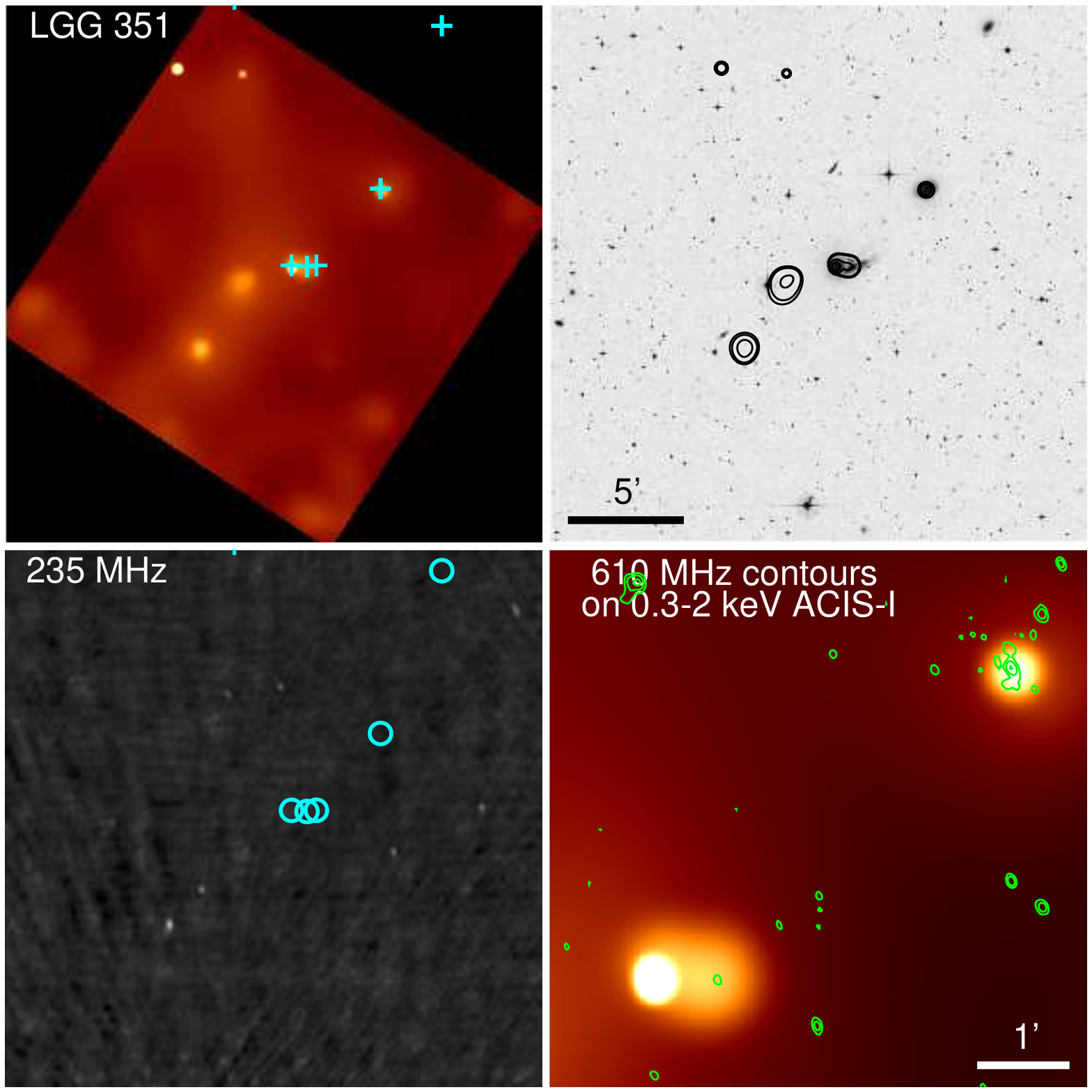}
\caption{LGG 351 / NGC 5153. 1\arcm\ = 17.5~kpc.}
\end{figure*}

\clearpage
\begin{figure*}
\includegraphics[width=\textwidth]{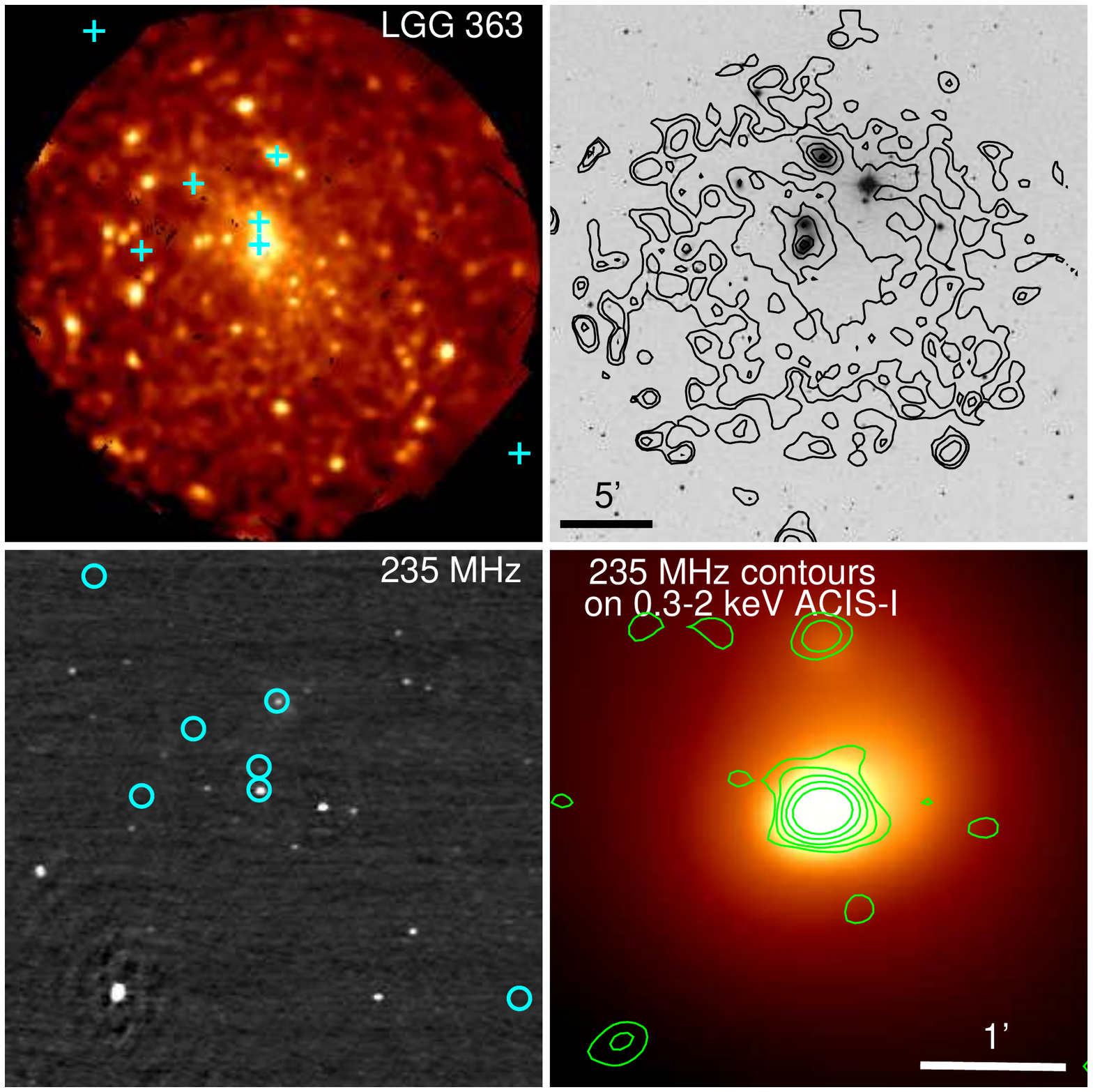}
\caption{LGG 363 / NGC 5353. 1\arcm\ = 10.2~kpc.}
\end{figure*}

\clearpage
\begin{figure*}
\includegraphics[width=\textwidth]{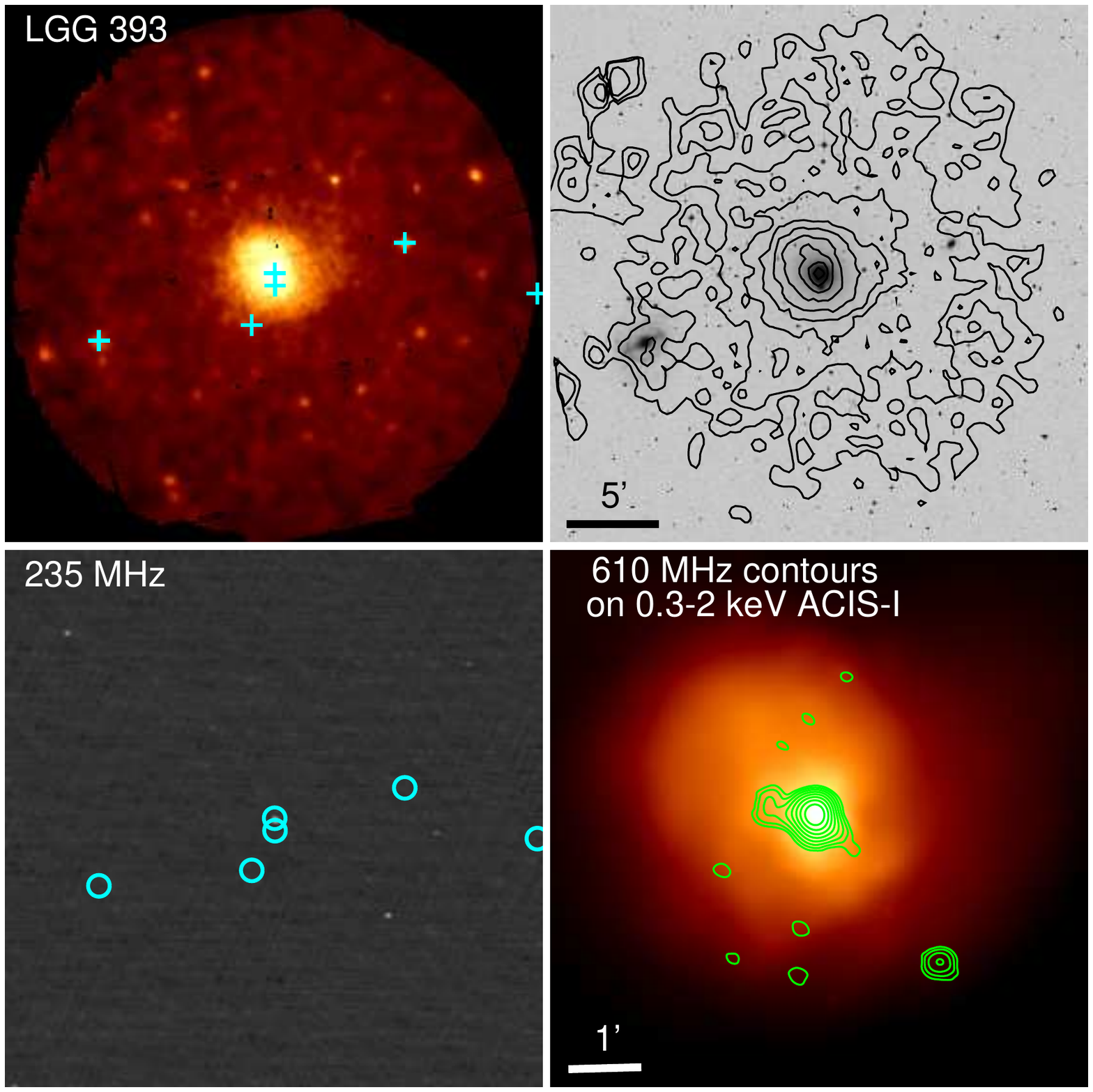}
\caption{LGG 393 / NGC 5846. 1\arcm\ = 7.6~kpc.}
\end{figure*}

\clearpage
\begin{figure*}
\includegraphics[width=\textwidth]{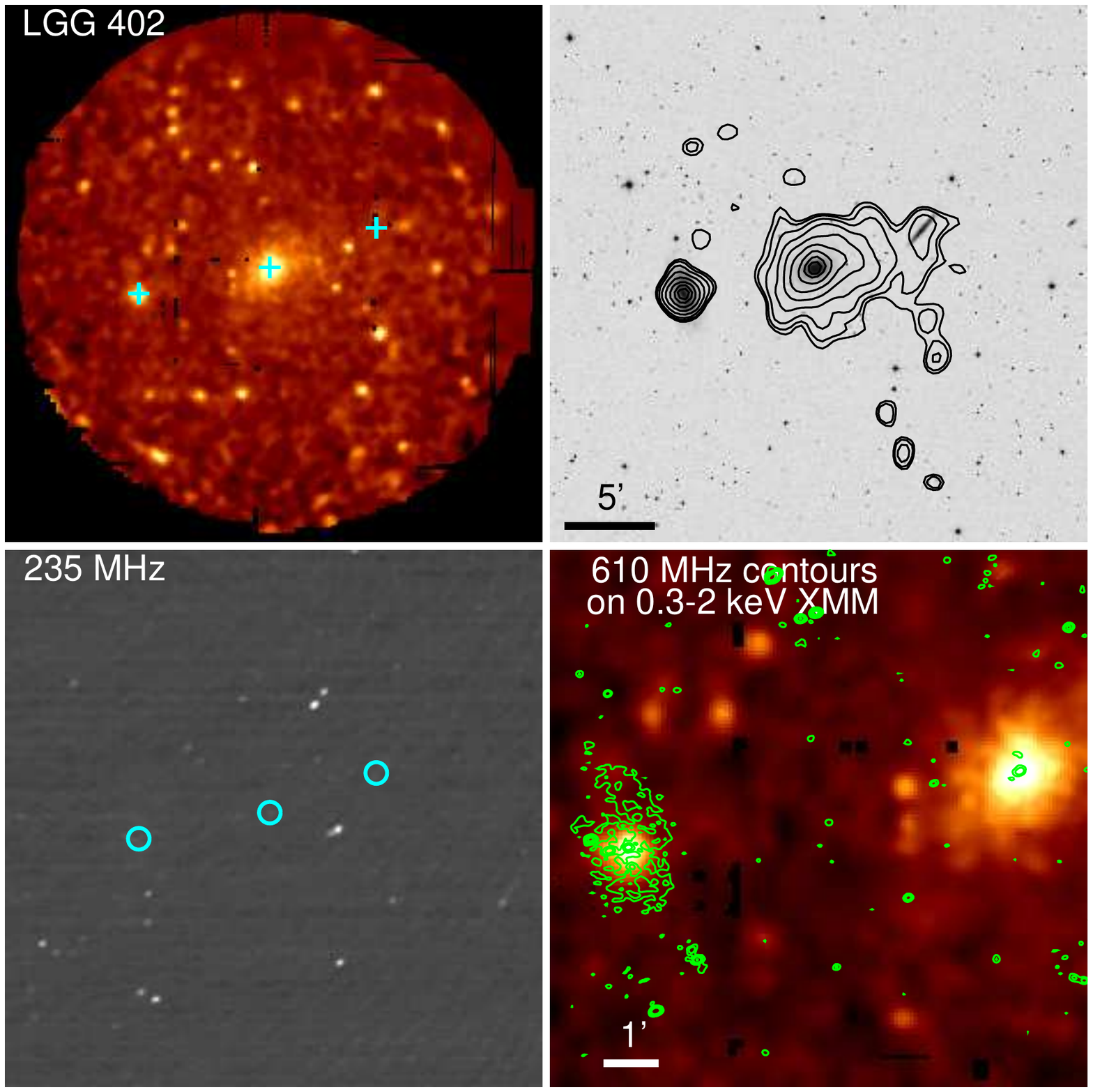}
\caption{LGG 402 / NGC 5982. 1\arcm\ = 12.8~kpc. The extended X-ray emission is centred on the dominant elliptical, NGC~5982, in the centre of the image, but the disk of the face-on spiral NGC~5985, to the left, is visible as extended 610~MHz radio emission.}
\end{figure*}

\clearpage
\begin{figure*}
\includegraphics[width=\textwidth]{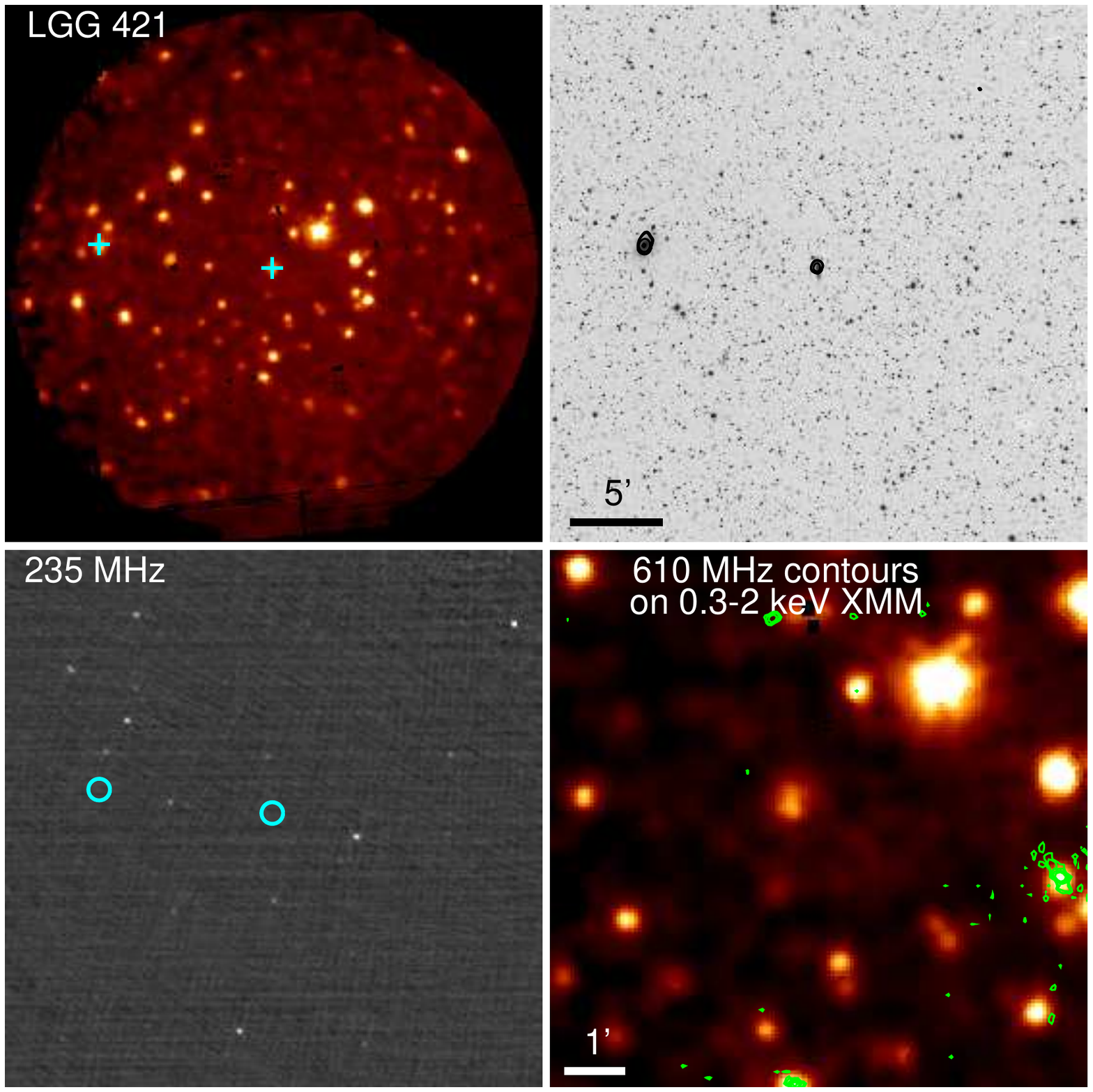}
\caption{LGG 421 / NGC 6658. 1\arcm\ = 18.3~kpc.}
\end{figure*}

\clearpage
\begin{figure*}
\includegraphics[width=\textwidth]{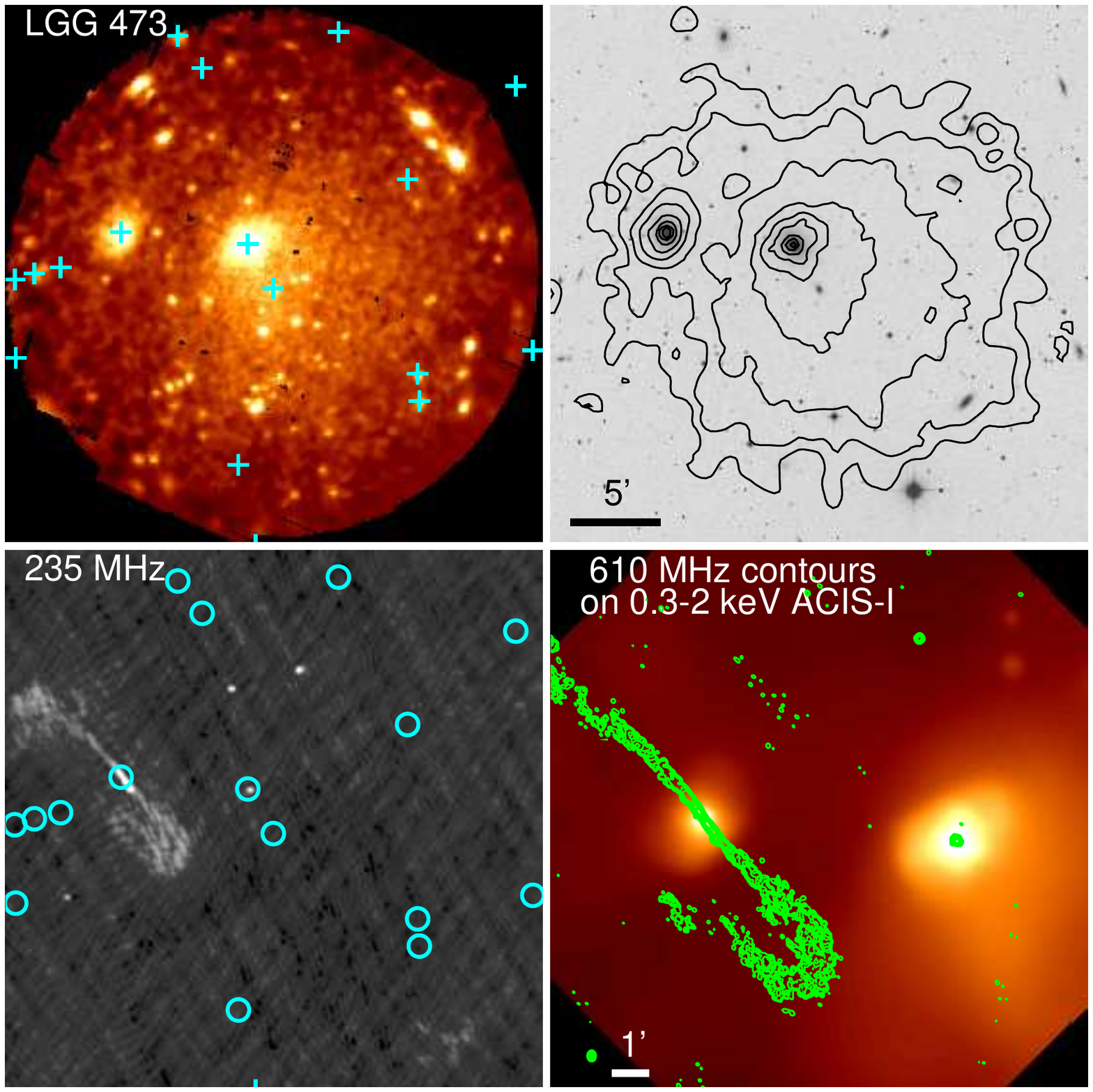}
\caption{LGG 473 / NGC 7619. 1\arcm\ = 15.7~kpc. This is a merging system, with the subsidiary X-ray peak centred on NGC~7619, which hosts a large double-lobed radio source. NGC~7626 hosts a radio point source.}
\end{figure*}

\bsp
\label{lastpage}
\end{document}